\documentclass[rmp,twocolumn,floats]{revtex4}
\usepackage{graphics,graphicx,epsfig}
\usepackage{amssymb,color}
\usepackage{epsf,epstopdf,wrapfig}

\newcommand{\beq}{\begin{equation}}
\newcommand{\eeq}{\end{equation}}
\newcommand{\beqn}{\begin{eqnarray}}
\newcommand{\eeqn}{\end{eqnarray}}
\newcommand {\e}[1]{\mathrm{~#1}}    
		
\begin{document}

\title{From statistical mechanics to information theory: understanding biophysical information-processing systems.}

\author{Ga\v{s}per Tka\v{c}ik\footnote{gtkacik@sas.upenn.edu}  }

\affiliation{Department of Physics and Astronomy, University of Pennsylvania, Philadelphia, Pennsylvania 19104--6396}

\date{\today}

\begin{abstract}
These are  notes for a set of 7 two-hour lectures  given at the 2010 Summer School on Quantitative Evolutionary and Comparative Genomics at OIST, Okinawa, Japan. The emphasis is on understanding how biological systems process information. We take a physicist's approach of looking for simple  phenomenological descriptions that can address the questions of biological function without  necessarily modeling all (mostly unknown) microscopic details; the example that is developed throughout the notes is transcriptional regulation in genetic regulatory networks. We present tools from information theory and statistical physics that can be used to analyze noisy nonlinear biological networks, and build generative and predictive models of regulatory processes. 
\end{abstract}

\maketitle
\tableofcontents

\section*{Introduction}
In cell biology, neuroscience, as well as the study of collective behavior of organisms, networks of interacting agents or elements orchestrate the cellular, organismal or group response to the changes in the \emph{environment} or the \emph{internal conditions} of the system. For example, in cells, signaling proteins on the membrane can detect  external chemicals and respond by chemically modifying other intracellular proteins, leading to a cascade of activity that can end with up- or down-regulation of the appropriate genes. Similarly, a genetic regulatory network comprises a set of regulatory proteins, called transcription factors, that find and bind  special non-coding regions on the DNA, thus causing a change in the expression levels of the regulated genes. In the nervous system, signals are propagated as ``digital'' action potentials: each neuron in the human brain receives synaptic input from other neurons and integrates it into its own decision to spike or not. Finally, flocks of birds, aggregating collections of single-celled amoeba (\emph{Dictyostelium}), schools of fish and even groups of people can exhibit collective behaviors that are  not necessarily trivially understood from the properties of single group members.

As physicists, we often see collective behaviors as emerging from interactions among basic ``simple'' elements. Yet in biology, even the building blocks of information processing networks are not simple. Proteins can have many conformational states that are hard to predict from the amino-acid sequence. The integration of information at the enhancer site in metazoan gene regulation -- that is, how levels of regulatory proteins together determine the expression level of the regulated gene -- is still mostly unknown. And in the nervous system, examples in which single units change their information-processing properties through adaptation to sensory statistics or where the network dynamically modifies its inter-neural connections as a consequence of learning, have been studied extensively, but are still poorly understood.

Despite this complexity in network building blocks and the fact that processes in biological networks can occur on many (not necessarily well-separated) timescales, we can still hope to find phenomenological or coarse-grained descriptions. There is no underlying theory of biological networks in the sense known to hard physical science. But our view is that while they operate and are essentially constrained by basic physics (some of the examples we will see later), the biological networks are also subject to evolutionary pressures for \emph{function}. This is a crucial distinction in comparison with inanimate systems, and there is hope that the intuition of a ``network X being driven to perform function Y well'' might generate a predictive theory for biological network X. For example, in the metabolic pathway of \emph{Escherichia coli}, the function that ``given the level of nutrients, the bacteria should maximize the growth rate'' can be mathematically formulated and actually leads to verifiable predictions about the network architecture \cite{iberra2002}.

In this Introduction, we motivate the lectures by asking three questions \cite{phd}:
\begin{enumerate}
\item \emph{When considering  biological networks that} process information, \emph{how might one quantify the network function in a mathematically concise way? Is it possible to derive network properties by optimizing for such function, as is the case with metabolic networks? Are there general principles that underlie information processing in living systems?}
\item \emph{What kinds of measurements can we perform on biological information processing networks and, having these measurements, how can they be analyzed? } 
\item \emph{How is the analysis of such networks different from the typical analyses of collective behaviors in physics? Which concepts and tools from physics can be borrowed to dissect and understand biological networks? }
\end{enumerate}

We'll use transcriptional regulation as a concrete example to illustrate these questions throughout the lecture notes. Analogies in neuroscience and other fields will be commented on.

These notes are organized as follows: Section \ref{lec1} provides an introduction to biological information processing networks by defining the basic terminology and approaches, and illustrating why these systems are interesting to study; Section \ref{lec2} discusses the response properties of single network elements, e.g. genes or neurons; Section \ref{lec3} discusses the role, types and origins of noise in biological networks; Section \ref{lec4} lays the groundwork for information theoretic approach, defining quantities such as entropy, mutual-  and multi-information, synergy, redundancy etc; Section \ref{lec5} illustrates how tools from information theory can be used to infer models of transcriptional regulation, i.e.  transcription factor -- DNA interaction; Section \ref{lec6} proceeds to extend the tools to analyze simultaneous interactions of more than pairs of elements in network; and finally, Section \ref{lec7} proposes a new information-theoretic principle that  can perhaps explain the design of several developmental transcriptional regulatory networks.

\section{Biological information processing networks: Noisy nonlinear dynamical systems}
\label{lec1}

The primary source for this section is Ref \cite{enc}, which provides a more complete bibliography and a review of the study of biological networks.

Let us start by describing several biological networks and reviewing some of their general properties:

{\bf Transcriptional regulatory networks.} In their genomes, organisms contain from several hundred to several tens of thousands of genes. The expression levels of these genes are primarily regulated by a set of proteins known as transcription factors (TFs), which are encoded by a few up to about $~10\%$ of genes in the genome. TFs bind specifically to short regulatory sequences of $\sim 10 - 20$ nucleotides in length, also known as \emph{binding sites}, on the DNA, thereby modifying the expression levels of regulated genes. TFs can cross- and self-regulate, opening up a possibility of feedback regulation. They are usually present in nuclei in small, nanomolar range concentrations (for a nucleus with several $\e{\mu m}$ radius, these concentrations  correspond to several hundred to thousands of TF molecules per nucleus). The timescales of such regulation span from minutes to hours. Some known examples of such regulation are the Lac operon in \emph{Escherichia coli}, the $\lambda$ repressor switch, many examples in yeast, genes of early development in \emph{Drosophila melanogaster} (such as bicoid, hunchback, even skipped), Hox genes etc.

{\bf Signaling networks.} 
``Sensory proteins,'' such as G-protein-coupled receptors, sense various extracellular molecules. They respond to incident photons, like rhodopsin, and bind neurotransmitters and environmental molecules to which we respond by sense of smell. Similarly, in bacteria, histidine kinase proteins (one part of two component signaling systems) are also membrane-bound proteins that detect their specific ligands. Upon binding their ligands these proteins change confirmation and cause a chain of phosphorylation / dephosphorylation reactions that chemically modify their target proteins (e.g. in bacterial two component signaling systems, the targets are the so-called ``response regulator'' proteins), thus altering their activity. These proteins can be present in thousands per cell, and the chemical reactions are much faster than in the case of transcriptional regulation, with equilibration times on the order of milliseconds to seconds. The reaction specificity is thought to occur via molecular `lock-and-key' like recognition mechanisms, but there exist cases of both unwanted crosstalk and intentional signal integration, where the same molecular targets are modified by various upstream enzymes.

\begin{figure}
\includegraphics[width =  \linewidth]{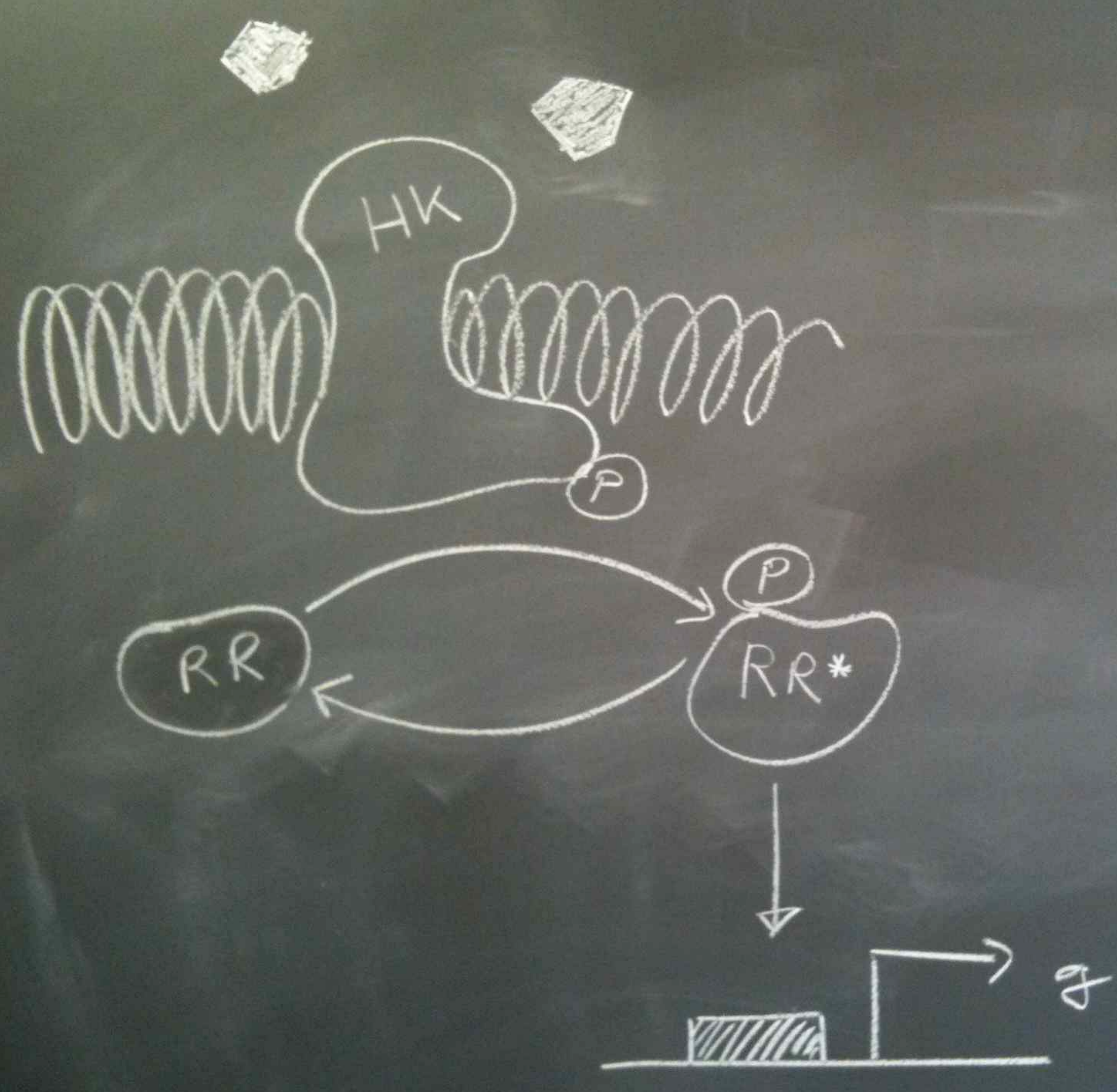}
\caption{A 2 component signaling system used by bacteria to transduce information from the environment into the cell. A receptor protein, \emph{histidine kinase} or HK, is embedded into the membrane and able to bind ligands (shaded pentagons). Upon binding, HK undergoes a conformational change and transfers a phosphate group (P) to a specific diffusable protein, called a \emph{response regulator} (RR). An activated form of this protein, $RR^*$, can act as a transcription factor and regulate target genes (such as gene $g$ in this figure). }
\label{f-2comp}
\end{figure}

{\bf Neural networks.}
Neurons transmit signals by propagating stereotyped voltage pulses, or \emph{action potentials}, across their membranes \cite{spikes}. These excitations are driven by ionic currents that flow through special proteins embedded in the membrane, called \emph{ion channels}. The speed of propagation along the fibre is on the order of meters per second, and the timing precision of spikes can below a millisecond. Most neurons are all-or-nothing devices: upon receiving input from other neurons  through its dendrites (which can number into tens of thousands, in the mammalian cortex), a neuron with some probability either produces a spike and sends it down its axon, or not. The axon  synapses upon thousands of other neuron's dendrites. The complexity of neural network processing stems from the fact that synapses are state- and history-dependent, as their transmission probabilities can be adjusted on long timescales by chemical modification (this is responsible for learning). Neurons too have complex internal states and exhibit many interesting computational capabilities, such as nonlinear input summation, adaptation, resonance properties, and different regimes of operation dependent on the precisely controlled ion channel composition.

These networks share a number of common features:
\begin{itemize}
\item Biological networks are {\bf dynamical systems}. The relevant timescales are the time on which the input fluctuates, the timescale on which single elements respond (neural spiking, protein decay rates) and potentially the timescale on which the network itself changes its properties (learning in neural networks, change in signaling protein concentrations in signaling networks). Networks can be (self-)tuned to special operating points (e.g. dynamical criticality), where new, ``emergent''  timescales might appear in the system. The networks often contain positive or negative {\bf feedback loops}.
\item The wiring in the network is {\bf specific}. Specificity can be achieved by spatial organization (neural networks, chromosomal organization, metazoan gene regulation) and selective establishment or death of connections (neural networks), or by molecular mechanisms of recognition (TF-DNA interaction, signaling enzymes). 
\item The network dynamics is {\bf noisy}. This is a consequence of the stochasticity in single molecular events at low concentrations of the relevant molecules. Neuronal spikes are not completely reproducible even if the same stimulus is played to the neuron over and over again, because fluctuations in opening and closing of a finite number of ion channels in the membrane are not negligible. The nanomolar concentrations of TFs in the cell mean that the precise timing when a TF finds and binds a regulatory site on the DNA is a random variable which results in stochastic gene activation. We discuss the importance of noise below.
\item The network elements are {\bf nonlinear.} There are saturation effects, as in when all enzymes of the signaling pathway are operating at capacity, or a gene is fully activated. Neurons themselves are excitable systems, that either don't respond or respond fully with a spike. Functionally, nonlinearities enable the systems to make ``decisions'' (e.g. by thresholding) and to re-represent their inputs in nontrivial ways (e.g. not by simple linear, rotation-like transformations).
\end{itemize} 

\subsection{Frameworks for describing biological networks}

\begin{figure}
\includegraphics[width =  \linewidth]{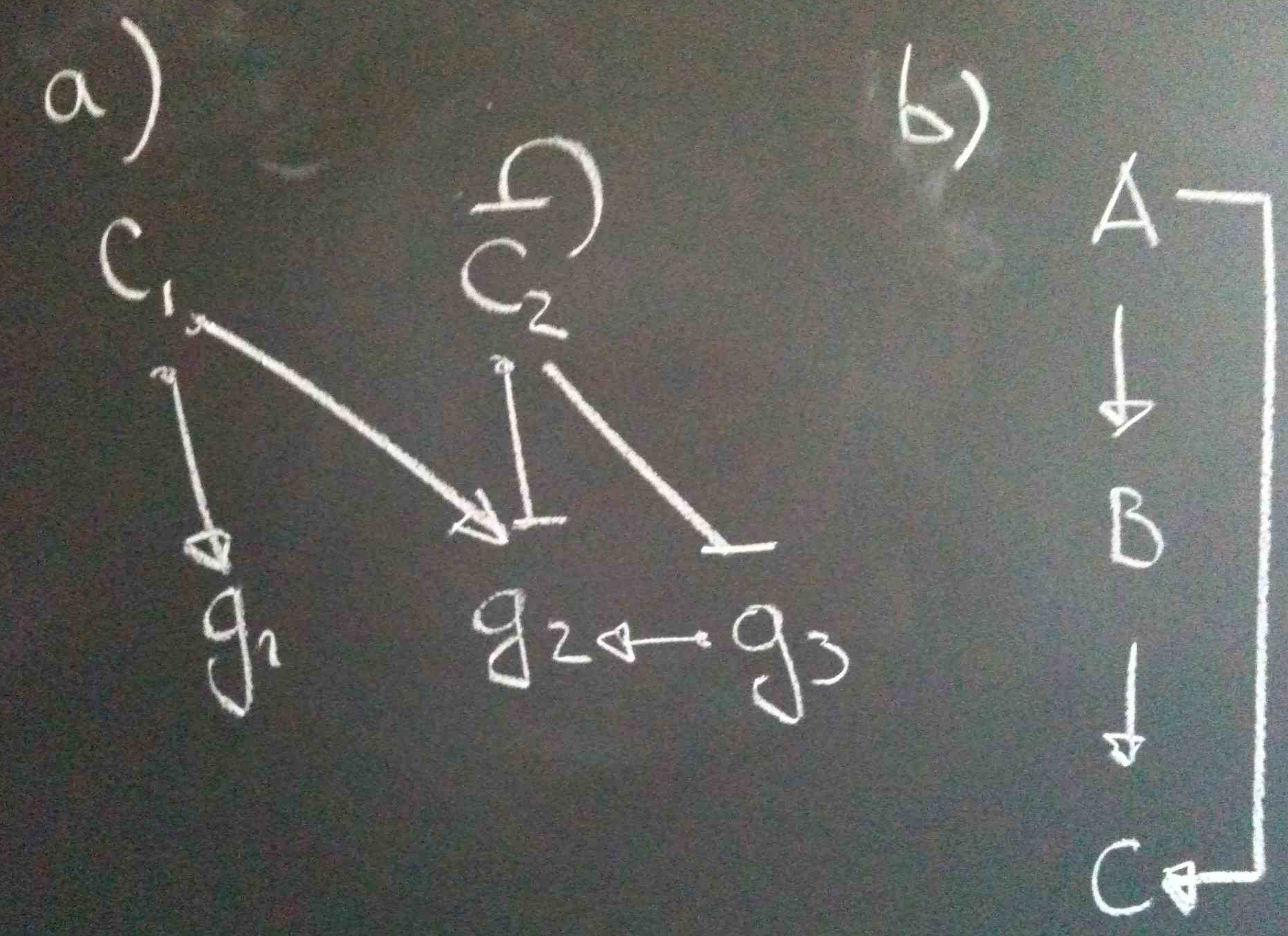}
\caption{Graph representation of simple regulatory networks. {\bf a)} Transcription factors $c_1$ and $c_2$ regulate genes $g_1$, $g_2$ and $g_3$; arrows represent activation, blunted lines represent repression. $c_1$ activates $g_1$ and $g_2$, while $c_2$ auto-represses itself and genes $g_2$ and $g_3$. $g_3$ laterally activates $g_2$. {\bf b)} Feed-forward loop, one of the motifs found to be overrepresented in the transcriptional regulatory network of \emph{E coli}: gene $A$ activates $B$ which activates $C$, but $A$ also directly activates $C$.}
\label{f-motifs}
\end{figure}

Various formal levels of description have been used to analyze biological networks, and different approaches emphasize some aspects enumerated above at the expense of the others:

{\bf Topological models} discuss networks in terms of wiring diagrams. The focus is on summarizing experimental data about the patterns of interconnection using a (possibly directed) graph. In genetic regulation, a particular kind of arrow in a wiring diagram might imply that gene A is activating gene B, while another kind might imply repression. Topological models are concerned with statistical properties of such graphs, in particular, in how they differ from various simple models of random  graphs. For instance, global statistical properties, such as node-degree distributions, clustering coefficients, mean path lengths etc have been studied for metabolic and regulatory networks, and the network of interconnections of all 302 neurons of \emph{C elegans}\footnote{Interestingly, the full wiring diagram of all the neurons in a relatively simple animal has not really brought us closer to understanding how the functions emerge in this model nervous system.}. The advantage of this method is that reasonably complete regulatory network diagrams exist, and their properties can be compared to other networks, including engineered ones (e.g. transport, internet routing etc). A parallel line of inquiry has shown that certain local graph connectivity features, known as \emph{motifs}, are overrepresented in, for example, transcriptional regulatory networks compared to randomized network ensembles \cite{shenorr2002}; see Fig.~\ref{f-motifs}. One should bear in mind, however, that the connectivity graph is a drastic oversimplification of the true network: without knowing {\bf (i)} the kinetic parameters on the arrows, {\bf (ii)} how the regulatory arrows converging onto a single node combine in their effects, and {\bf (iii)} what are the internal states of each of the nodes in the regulatory graph (e.g. often the same node represents both mRNA and its protein product, both of which have their own dynamics with delays), one can not predict from the connectivity alone how the network will behave, although certain classes of dynamical behavior can be excluded.

Interesting findings pertaining to biological networks described by connectivity graphs have been: {\bf (i)} their scale free degree distribution, with the probability of a node to be connected to $k$ neighbors being $P(k)\sim k^{-\gamma}$ (with $\gamma\sim 2-3$) \cite{barabasi2004}, and the consequent identification of \emph{hub} nodes (often essential proteins); {\bf (ii)} robustness to breakup of the connected component with respect to removal of most nodes, but fragility with respect to removal of hubs; {\bf (iii)} ``small world'' architecture, with short mean path lengths between pairs of nodes, and high clustering coefficients \cite{strogatz2001}; {\bf (iv)} hierarchical yet clustered nature (no preferred size of the cluster, consistent with scale-free property) \cite{ravasz2002}; {\bf (v)} ``dissociative structure'', in which hubs are often \emph{not} connected to each other (in contrast to social networks, where friends with lots of friends are often friends among each other). 

{\bf Boolean models.}
Here, the network is represented as a collection of boolean variables $\{\sigma_i\}$, which can take ``on'' ($\sigma=1$) and ``off'' ($\sigma=0$, sometimes $\sigma=-1$) values, and a set of dynamical update rules, $\sigma_i(t+1)=f(\sigma_j(t))$, that evolve the system forward in time in discrete steps. The variables might be thought of as two-state genes (or neurons), and the update rules are combinatorial (Boolean) expressions involving input TFs on the promoters (or  dendritic inputs in neurons). Generalizations to more than 2 states have been used, e.g. in modeling \emph{Drosophila} developmental gene network \cite{sanchez2001}. This method scales well to  simulations of many genes and emphasizes the dynamic and nonlinear nature of the network, but can introduce serious artifacts due to synchronous update rule. In neuroscience, the Hopfield network is of the similar form, with asynchronous update, and provides a clean theoretical model of associative memory that is able to recall a stored binary pattern  from any closely matching subpattern \cite{hopfield1982}. In that case the states of $N$ neurons are given by $\{\sigma_i\}$, the update rule is downhill descent $\sigma_i \leftarrow \mathrm{sgn}(\sum_j J_{ij} \sigma_j + h_i)$ evaluated under the network wiring parameters $\{h_i, J_{ij}\}$, the input pattern is the initial state of the network  $\{\sigma_i\}_0$, and the final state is the attractor of the dynamics. The memories are $K$ binary patterns $\vec{\xi}^\mu$, $\mu=1\dots K$, stored into $J_{ij}$ by the prescription $J_{ij}=K^{-1}\sum_\mu \xi_i^\mu \xi_j^\mu$.

Interesting applications of Boolean network modeling include the description of the \emph{Drosophila} gap-gene system in which 4 gap genes respond to 3 maternal TF inputs with spatially varying profiles; this model could describe correctly the expression patterns in a number of known mutants \cite{sanchez2001}. Another interesting case involved describing the budding yeast cell cycle driven by a network of 11 interacting nodes \cite{li2004}. In this network the topology induces a robust sequence of state transitions (growth, DNA duplication, pause, division), triggered by a ``cell-size'' checkpoint; robustness here refers to the fact that the dynamical trajectory through the states remains unchanged upon perturbation of update rules, small changes in topology and  the parameters.

{\bf Dynamical systems.}
In the field of dynamical systems analysis, one usually assumes that each network element can have a certain degree of ``activity'' (expression level for a gene, fraction of activated proteins of a given type in a signaling network, firing rate of a neuron). These activities $g_i$ are treated as continuous variables undergoing network dynamics. As a simple example one could write:
\begin{equation}
\frac{dg_i}{dt}=-\frac{1}{\tau}g_i + F\left(\sum_j J_{ij} g_j + I_i\right); \label{dsys}
\end{equation}
here, the first term describes relaxation towards equilibrium with timescale $\tau$, while the second ``activation'' term adds up contributions from other network elements weighted by a connection matrix $J_{ij}$, and a possible direct input $I_i$ to element $i$. $F$ is a nonlinear function, often saturating (or sigmoid).  In the case of gene regulation, $g_i$ could be the expression level of gene $i$, $\tau$ the protein decay rate, $J_{ij}$ the contributions of transcription factor $j$ to the expression of gene $i$, $I_i$ the direct external influence on $i$ by e.g. induction, and $F$ would be the promotor input/output function, discussed later. Of course more complex and realistic examples are possible, potentially including spatial effects due to, for example, diffusion of TF molecules in the cells. In the study of dynamical systems one focuses on the search for universality classes of dynamical behaviors, and looks for limit cycles and steady states in dynamics and their sensitivity to network control parameters [such as $J_{ij}$ in Eq (\ref{dsys})]. 

Well-known applications of this approach involve modeling the chemotaxis module of \emph{Escherichia coli} \cite{bray1993}, a model of cell cycle control in fission yeast (approximately 10 proteins and 30 rate constants) \cite{Novak1997}, and the circadian clock in mammals (a system of about 20 rate equations) that exhibits autonomous oscillations and reproduces the entrainment of oscillations through light-induced gene expression \cite{Goldbeter2003}. This approach is popular also for modeling of smaller and simpler (sometimes synthetic) systems, such as the repressillator \cite{Elowitz2000} or the Min oscillator in bacteria.

{\bf Probabilistic models and stochastic dynamics.}
Probabilistic models try to capture the noise inherent in biological processes and experimental protocols. For a set of (for example) binary variables $\{\sigma_i\}$ that denote the simultaneous expression levels of genes (or firing / silence states of the neurons), one class of probabilistic models tries to find good approximations for $P(\{\sigma_i\}|\mathcal{C})$, that is, the probability distribution that all genes (or neurons) are in a particular configuration $\{\sigma_i\}$ given the external conditions (or stimulus) $\mathcal{C}$. This is then compared to experimental data, which is a sample of many measurements of $\{\sigma_i\}$ across cells or time. $P(\{\sigma_i\}|\mathcal{C})$ gives a generative model from which synthetic data that resembles real measurements can be drawn. Various approaches differ in what forms for $P$ are assumed. We will mention Bayesian network inference and maximum entropy models in Section \ref{lec6}.
A second class of probabilistic models attempts to generalize dynamical equations, e.g. Eq (\ref{dsys}). If the noise is not too big, the simplest way is to write:
\begin{equation}
\frac{dg_i}{dt}=-\frac{1}{\tau}g_i + F\left(\sum_j J_{ij} g_j + I_i\right) + \xi_i(t); \label{dsysn}
\end{equation}
where $\langle \xi(t)\xi(t')\rangle=2T(\{g_j\})\delta(t-t')$ is the Langevin noise strength, i.e. a random force whose variance $T$ might depend on the state of the system (braces denote averaging over many realizations of the noise time series). Techniques from statistical mechanics and diffusion can then be used not only to calculate the \emph{mean} dynamics $\bar{g}_i(t)$ of this system, but also the evolution of the ``noise'' or the fluctuation around the mean, characterized by the covariance matrix of all $g_i$, $C_{ij}(t,t')=\langle \delta g_i(t)\delta g_j(t')\rangle$.
When concentrations of constituents are low (or when we are interested in single neural spikes), the continuous description with $g_i$ must break down and the state space is replaced by $\{n_i\}$, the integer counts of individual molecules such as DNA or TFs. One switches to describing the full evolution of the probability distribution using the Master equation \cite{vanKampen_07}:
\begin{equation}
\frac{dP(\{n_i\}|t)}{dt}=\hat{L}P(\{n_i\}|t),
\end{equation}
where $\hat{L}$ is some linear evolution operator.

Finally, there are techniques like the Stochastic Simulation Algorithm (SSA) of Gillespie \cite{gillespie2007} and generalizations to spatially extended systems, where the discrete stochastic dynamics can exactly be simulated at the expense of slow execution speeds. One of the pioneering studies with SSA was the simulation of the $\lambda$ lysis / lysogeny switch \cite{McAdams1997}, where the simulation tried to reproduce the fraction of lysogenic phages as a function of multiplicity of infection.

\subsection{A simple regulatory element formulated in different frameworks}
Let's start by discussing the simplest possible example of genetic regulation; we will develop this example as we progress through the lectures.

\begin{figure}
\includegraphics[width =  \linewidth]{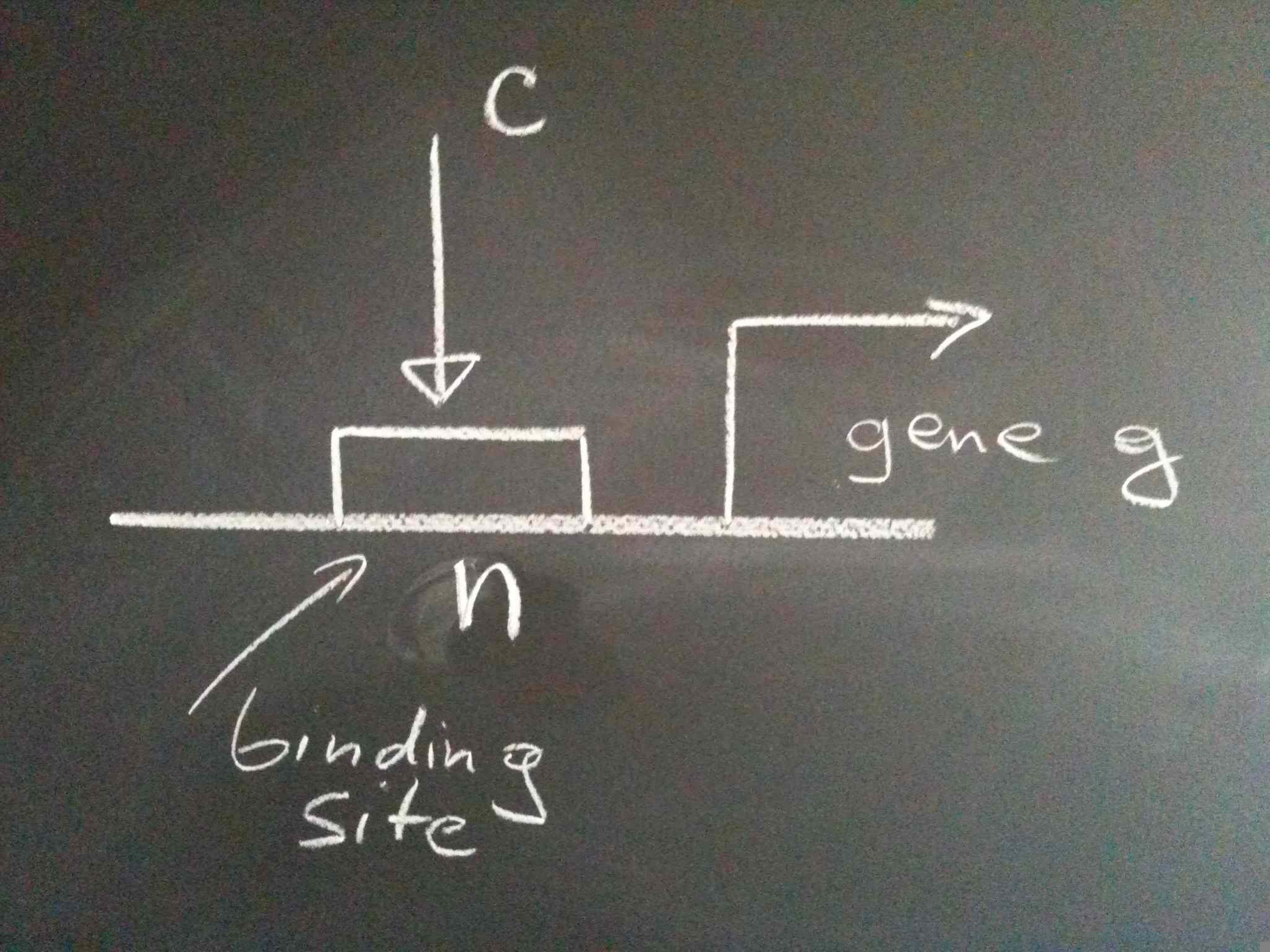}
\caption{The simplest regulatory graph, where an input transcription factor at concentration $c$ regulates the output expression level $g$ by binding to a binding site $n$, which can be empty or occupied. Since $c$ acts as an activator, an occupied site results in transcription and translation of $g$.}
\label{f-sscheme}
\end{figure}

Let the transcription factors (TFs) be present at concentration $c$ in the cell. On the DNA, there is a single specific binding site that can be occupied or empty; we'll denote this occupancy with $n(t)$. When the site is occupied, the regulated gene will get transcribed into mRNA, which is later translated into proteins whose count we denote by $g(t)$, at the combined rate that we denote by $R$. The proteins are degraded with the characteristic time $\tau$.  In this case, our TF thus acts as an activator, see Fig.~\ref{f-sscheme}. Here and afterwards we will refer to the transcription factor $c$ as an \emph{input}, and the regulated gene product $g$ as \emph{output}.

This model discards a lot of molecular complexity: there is no explicit treatment of diffusion of TFs, no non-specific binding,  no separate treatment of mRNA and protein, no chromatin opening / closing etc; in addition, we group many multi-stage molecular processes (such as TF binding, RNAP assembly, processive transcription etc) into single steps. Thus, our model is a gross but tractable oversimplification. As an illustration, let us formulate it in all of the mathematical frameworks discussed above.

The topological diagram for this example is simple $c\rightarrow g$. In the case of Boolean network models, the state space of this network is $\{c,g\}\in [0,1]^2$, indicating that both the transcription factor $c$ and the regulated gene $g$ can be ``on'' or ``off.'' The update rules are trivial: $c(t+1)=c(t), g(t+1)=c(t)$.

Treating now concentrations $c$ and $g$ as continuous, we can describe the same regulatory process by the set of differential equations:
\begin{eqnarray}
\frac{dn}{dt} &=& k_+ c(t)(1-n) - k_-n	\label{occ}\\
\frac{dg}{dt}&=&-\frac{1}{\tau}g + R n. \label{prot}
\end{eqnarray}
Equation (\ref{occ}) is an equation for occupancy $n$, which is a number between 0 and 1. Nominally, the site can only be fully empty or occupied, but in this approximation, we treat it as a continuous variable that can be interpreted as a ``probability of the site being bound.'' $k_+c$ is the TF-concentration-dependent on-rate, and $k_-$ is the first-order off-rate. Often, it is assumed that there is a separation of time scales: the first equation for occupancy equilibrates much faster than $\tau$, meaning that the mean occupancy 
\begin{equation}
\bar{n}(t) = \frac{k_+ c(t)}{k_+c(t) + k_-}
\end{equation}
can be inserted into Eq (\ref{prot}) to get
\begin{equation}
\frac{dg}{dt}=-\frac{1}{\tau}g(t) + R\frac{k_+ c(t)}{k_+c(t) + k_-}. \label{prot1}
\end{equation}
In this simple case without feedback, the approach to the equilibrium at fixed $c$ is exponential with the rate $\tau$, and the steady state is simple: $\bar{g} = R\tau \bar{n}$. Equation (\ref{prot1}) has precisely the form of Eq (\ref{dsys}) with $F$ having a sigmoidal shape. We discuss in the next section how the particular sigmoidal regulation functions are connected to equilibrium statistical mechanics of this system, how noise can be added by an introduction of the Langevin force into Eq (\ref{prot}), and why the assumption of fast equilibration of $n$ strongly influences the noise. 

Finally, if we wanted to fully capture the noise in this model, the object of our inquiry would be $P_n(g|t,c)$: the time-dependent joint probability of observing $g$ molecules of the resulting gene and the state of the binding site being $n=0,1$ (empty, occupied), given some concentration of the input $c$. One can marginalize over $n$ to get the evolution of probability of observing $g$ output molecules: $P(g|t,c)=\sum_{n=0,1}P_n(g|t,c)$. Writing down the Master equation and for simplicity suppressing the parameters $(c,t)$ on which all terms $P(g|t,c)$ are conditioned, we find:
\begin{eqnarray}
\frac{dP_0(g)}{dt} &=& \frac{g+1}{\tau}P_0(g+1)+ k_-P_1(g) - (k_+ c + \frac{g}{\tau})P_0(g) \nonumber\\
\frac{dP_1(g)}{dt} &=& \frac{g+1}{\tau}P_1(g+1) +RP_1(g-1)+k_+cP_0(g) -\nonumber\\
&-& (k_- + \frac{g}{\tau}+R)P_1(g); \label{mastereq}
\end{eqnarray}
the reader should recognize degradation-related terms (proportional to $1/\tau$), the protein production terms (prefixed with $R$ and present only in the case when the gene is on, i.e. $n=1$) and the switching terms of the promoter containing $k_+ c$ and $k_-$, which couple the $n=0$ to $n=1$ states.  In this simple case, the equilibrium distribution can be solved by zeroing out the left-hand side of Eq (\ref{mastereq}). This yields an infinite dimensional system in $g$ that can be truncated at some $g_{\rm max}\gg R\tau$; we would end up with a homogenous linear system that can be supplemented by a normalization condition $\sum_{n=0,1}\sum_{g=0}^{g_{\rm max}}=1$, which can be inverted and solved for steady state $P_n(g)$. More sophisticated methods are available when the number of genes grows and they are interacting \cite{alex1}. Note that in this example we treated the $g$ as discrete, but $c$ is still a continuous input parameter (not a variable whose distributions we are also interested in).

Finally, let's mention the Gillespie algorithm. For this algorithm we start with enumerating all reactions and their rates:
\begin{eqnarray}
k_+ &:& c+n\rightarrow cn	 \nonumber\\
k_- &:& cn \rightarrow c + n \nonumber\\
R&:& cn \rightarrow cn + g \nonumber \\
\tau^{-1}&:& g \rightarrow \oslash \label{cgillespie}
\end{eqnarray}
One initializes the state of the system as a vector $(c,n,cn, g)$ of integer counts of molecular species (here $cn$ denotes a molecular complex of a c molecule bound to the promoter; there can only be 0 or 1 $n$ and $cn$, and you can quickly convince yourself that $n=1-cn$). Then the probability per unit time of each of the 4 reactions is the product of the rate constant and the number of reactants properly normalized by the relevant volume. The algorithm randomly draws the next reaction consistent with the probabilities per unit time, updates the state of the system and repeats. This algorithm is exact for well-mixed systems, but {\bf (i)} can be slow in case there are fast and slow reactions in the system; {\bf (ii)} one needs to sample many simulation runs to accumulate the noise statistics. 

From the presented example it is clear that the fully stochastic dynamical description can be relatively complicated even for a very simple system. To proceed and be able to connect to data, we will in these lectures drop the time dependence and only focus on the steady state, while emphasizing the nonlinear and noisy nature of the system. Sections \ref{lec2} and \ref{lec3} therefore use the Langevin description to treat noise in a dynamical system, and Section \ref{lec6} presents a probabilistic maximum entropy model of an interacting network. Our assumption to only study the steady state will preclude us from discussing network phenomena that are intrinsically dynamic, e.g. the cell cycle clock, the circadian clock, or the detailed nature of excitability in neurons and \emph{Dictystelium} cell cultures. But for many biologically realistic cases, such as in developmental biology, or in many experimental settings, such as measuring the gene response to  constant levels of inducer, the steady state approach is useful.
\section{Network building blocks as input / output devices}
\label{lec2}
In this lecture we will first explore simple thermodynamical models of gene regulation, by studying how the concentration of a transcription factor relates to promotor occupancy and thus to the expression level of the regulated gene. We will then look at one possible model for combinatorial regulation and introduce the notion of the input/output relation. The section finishes by briefly surveying the experimental measurements of input/output relations.

\subsection{Simple models of regulation: Hill functions}
Let's first revisit our simple description of gene regulation from Eqs (\ref{occ},\ref{prot}). We found that in steady state the occupancy of the promoter is $\bar{n}(c)=k_+ c / (k_+ c + k_-)$, where $k_+$ and $k_-$ are on- and off rates, respectively. 

Since this is an equilibrium system, we can ask for the equivalent statistical mechanics description. Suppose we have a site $n$ that can be occupied or full. In case it is occupied, there is a binding energy $E$ favoring the occupied state, relative to the reference energy 0 in the unbound state. But in order to occupy the state, one needs to remove one molecule of TF  from the solution. The chemical potential of TFs, or the free energy cost of removing a single molecule of TF from the solution, is $\mu=k_B T \log c$, where $c$ is the TF concentration, measured in some dimensionless units of choice. 
In statistical physics we can calculate every equilibrium property of the system if we know how to compute the \emph{partition sum}, which is $Z=\sum_i e^{-\beta(E_i - \mu n_i)}$, where the sum is taken over all possible states of the system (in our case binding site empty and binding site occupied), $E_i$ is the energy of the system is the state $i$, and $n_i$ is the number of molecules in the system in the state $i$. 

In our case of a single binding site, the partition sum is over the empty ($n=0$) and occupied ($n=1$) state:
\begin{equation}
Z=e^{-\beta (E-\mu)}+1,
\end{equation}
where $\beta=1/(k_BT)$, $T$ is the temperature in Kelvin and $k_B$ is the Boltzmann constant.  The probability that the site is occupied is then
\begin{equation}
P(n=1) = \frac{1}{Z}e^{-\beta (E-\mu)}.
\end{equation}
Inserting the definition of $\mu$, we get
\begin{equation}
P(n=1) = \frac{c}{c+K_d},
\end{equation}
where we write $K_d=\exp(\beta E)$. But $\bar{n} = 1\cdot P(n=1) + 0\cdot P(n=0) = P(n=1)$, so by comparing with Eq (\ref{occ}) we can make the identification
\begin{equation}
K_d = e^{\beta E}= \frac{k_-}{k_+}, \label{db}
\end{equation}
which connects our statistical mechanics and dynamical pictures. Note that $k_-$ is measured in units of inverse time, $\e{s}^{-1}$, $k_+$ is measured in units of $s^{-1}\times [\mathrm{conc}]^{-1}$ (but by convention we here measure concentration in dimensionless units, as in $\mu = k_BT\log c$), so $K_d$ has units of concentration.

Suppose we make the model somewhat more complicated: let us have two binding sites, which together will constitute a system with 4 possible states of occupancy: both sites empty, either one occupied, and both occupied, which we'll write compactly as $(00,01,10,11)$. Let's also assume that there is cooperativity in the system -- if both sites are occupied, then there will be an additional favorable energetic contribution of $\epsilon$ to the total energy of the state $(11)$. Finally, when promoters can have multiple internal states, we need to decide which state is the ``active'' state, when the gene is being transcribed\footnote{In general, each internal promoter state could have its own transcription rate, but often one state is picked as having the maximal transcription rate, and the other states represent the gene being ``off'' or expressing at some small, ``leaky'' rate of gene expression.}; here we pick the state $(11)$ as the active state.

\begin{figure}
\includegraphics[width =  \linewidth]{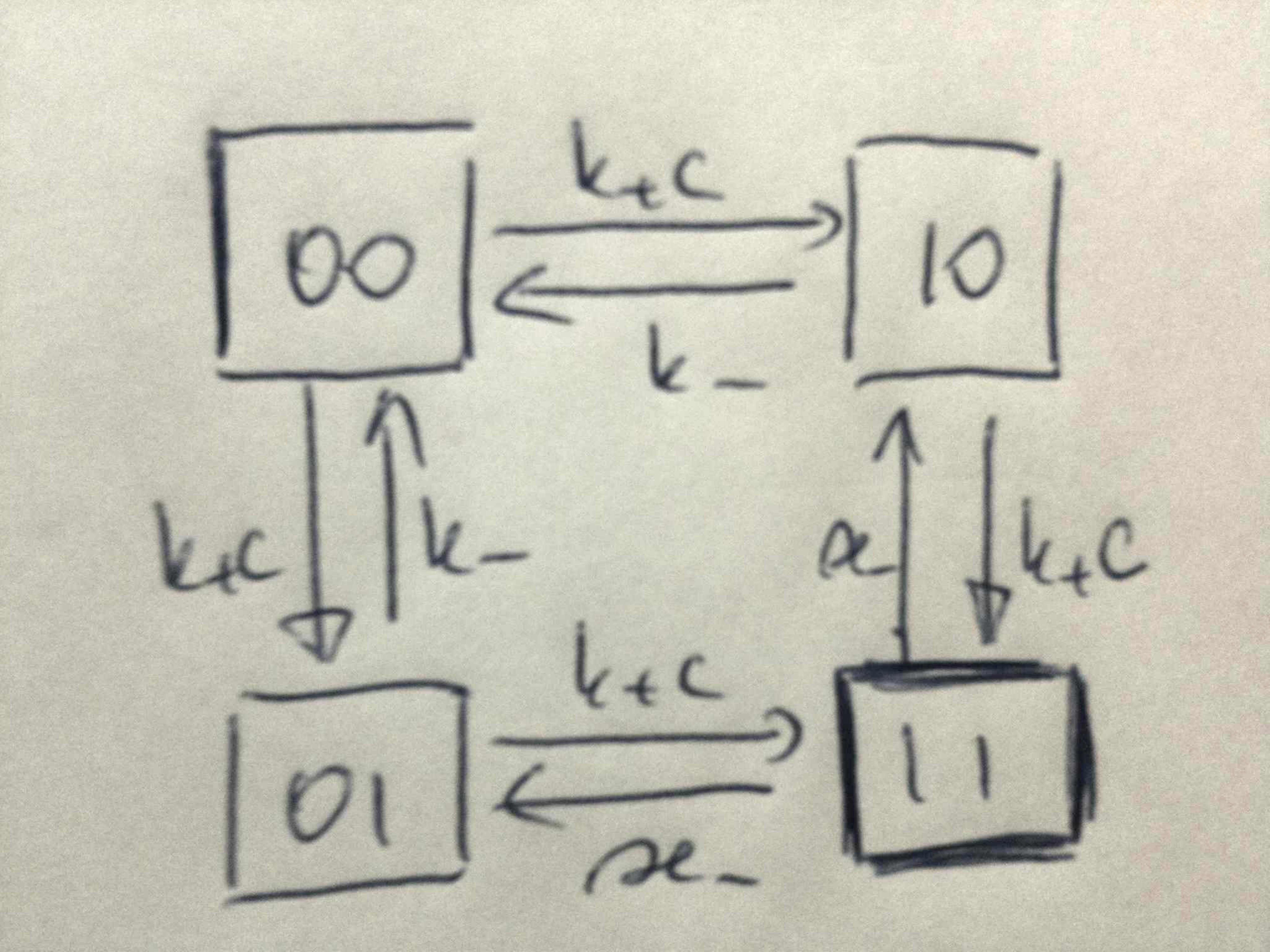}
\caption{The transitions in the model with 2 binding sites and 4 occupancy states, $(00,01,10,11)$. The binding of an additional molecule of TF happens at a rate $k_+c$, whereas unbinding rates are state dependent: a singly occupied promoter returns to the non-occupied state with a rate $k_-$, but the doubly occupied promoter loses a molecule of TF with the  rate $\kappa_-$. This difference is due to cooperativity, where the binding of one molecule stabilizes the binding of the other, and this makes the unbinding rates state dependent. The ``active'' state is $(11)$ in the lower right corner. We leave it as an exercise for the reader to write down the dynamical equations $dn_{00}/dt=\dots$, $dn_{01}/dt=\dots$ etc, observe that $n_{00}+n_{01}+n_{10}+n_{11}=1$, and compute the steady state activation if cooperativity is strong, $\bar{n}_{11}$. As in the case of a single binding site, this expression can be connected to the thermodynamic result of Eq (\ref{p11}). }
\label{f-2state}
\end{figure}

The probability of being active is then
\begin{equation}
P(11) = \frac{e^{-2E-\epsilon+2\mu}}{e^{-2E-\epsilon+2\mu} + 2e^{-E+\mu}+ 1}, \label{p11}
\end{equation}
where we use the units where $\beta=1$, that is, we express the energies and chemical potential in thermal units of $k_BT$.
If the cooperativity is strong, i.e. the additional gain in energy $\epsilon$ is larger than the favorable energy of putting a molecule of TF out of the solution onto the binding site, $\epsilon \ll \mu-E$, we can drop the middle term of the denominator in  Eq (\ref{p11}) and simplify it into:
\begin{equation}
P(11) = \frac{c^2}{c^2 + K_d^2}, \label{hill2}
\end{equation}
with $K_d=\exp[\beta (E+\epsilon/2)]$, where again we have used the definition of chemical potential $\mu$. This problem with 2 binding sites and 4 states of occupancy also has a complementary dynamical picture, which is already quite complicated, see Fig.~\ref{f-2state}.

\begin{figure}
\includegraphics[width =  \linewidth]{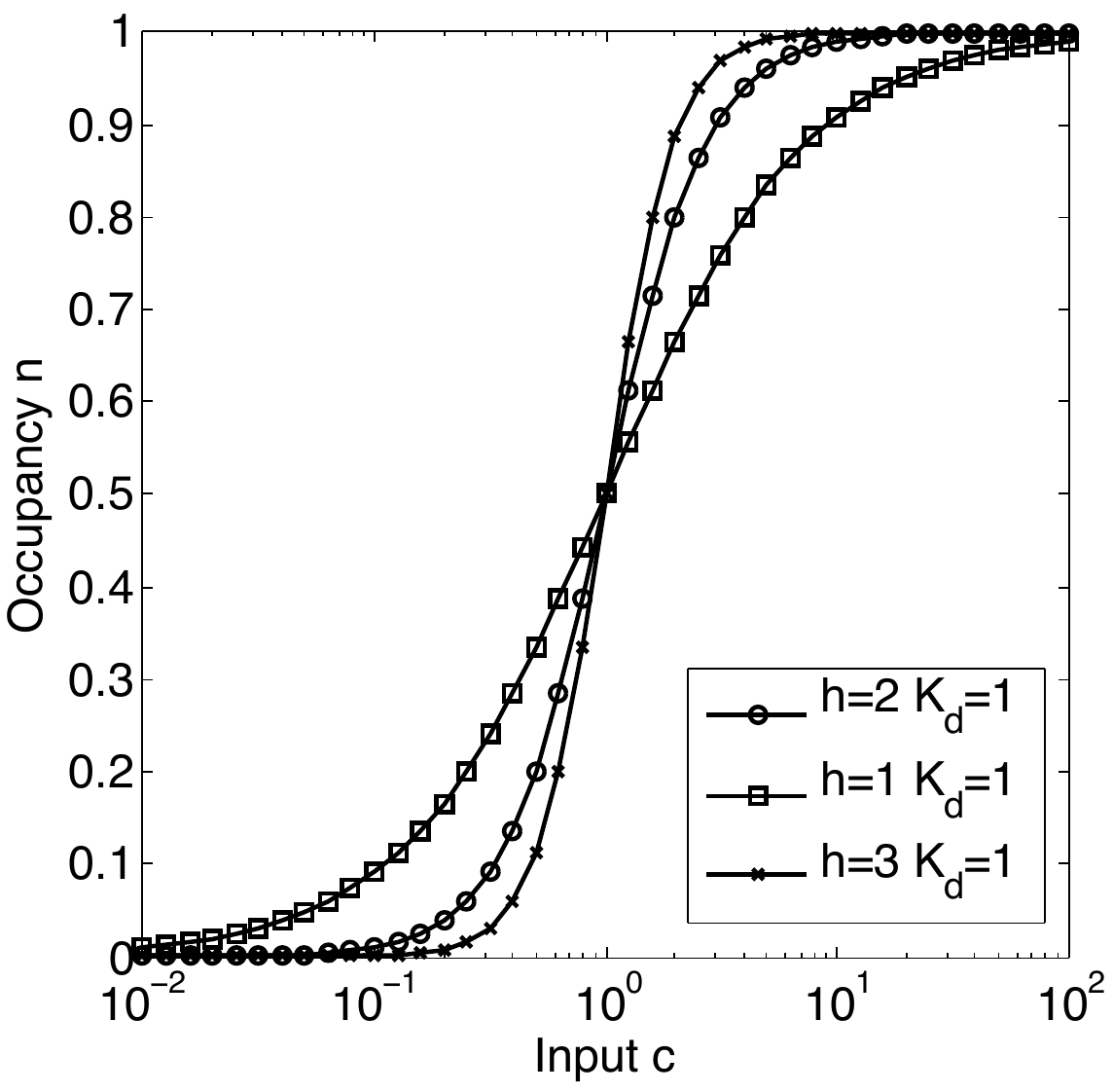}
\caption{Three Hill regulatory functions with different slopes (Hill coefficients $h$), as in the legend. All functions have $K_d=1$. Input TF concentration is customarily plotted on logarithmic horizontal axis, while the average promotor occupancy $\bar{n}$ is on the vertical axis. The output gene expression is in steady state $\bar{g}=(R\tau)\bar{n}(c)$, i.e. proportional to occupancy. The slope of $\bar{n}(c)$ on the log-log plot at half-induction ($c=K_d$) is related to the Hill coefficient, $d(\log \bar{n})/d(\log c)|_{K_d}=h/2$.}
\label{f-hill}
\end{figure}

Readers used to molecular biology models of gene regulation will recognize sigmoidal functions in Eqs (\ref{occ},\ref{hill2}), also known as Hill functions, with a general form (see Fig.~\ref{f-hill}):
\begin{equation}
\bar{n}(c) = \frac{c^h}{c^h+K_d^h}, \label{hillg}
\end{equation}
where the dissociation constant $K_d$ is interpreted as the concentration at which the promoter is half induced, and $h$ is known as the cooperativity or Hill coefficient, usually interpreted as the ``number of binding sites''\footnote{In case where there is self-activation of the gene, i.e. gene $g$ can activate its own transcription in addition to being activated by the input $c$, that conclusion is incorrect.}.  Here we have shown how such phenomenological curves arise from simple statistical mechanics models of gene regulation with cooperative interactions. For repressors, one can show that $\bar{n}(c) = K_d^h/(c^h+K_d^h)$.

Before proceeding, let's inspect more closely the relation between the dynamical rates and the binding energy for a single site: $k_-/k_+=\exp(\beta E)$. As we have shown in Eq~(\ref{db}), this equality is required by detailed balance if thermodynamic and kinetic pictures are to match. Molecularly -- and we will devote the entire lecture of Section \ref{lec5} to this -- the energy of binding $E$ in the case of transcription factor -- DNA interaction depends on the sequence. So if we were to vary the sequence and binding energy $E$ would change, which of the two rates, $k_-$ or $k_+$ would vary as a result? In general one cannot answer this question without knowing in detail the sequence of molecular transitions that happen at the binding site. However, there is a useful limit, called the \emph{diffusion-limited on rate}, that is often applicable. In this regime, the limit to how quickly the TF can bind is given by the speed at which it can diffuse to the binding site. It has been shown that if TF diffuses with diffusion constant $D$ and is trying to bind a site with linear dimension $a$, the fastest on rate is $k_+\approx 4\pi D a$, for spherical TF and binding site\footnote{If assumptions about geometry are relaxed, the prefactor $4\pi$ will change.}\cite{bergreview}. In the diffusion-limited approach, if the binding site is empty, as soon as a TF diffuses into a region of size $a$ around the binding site, it will immediately bind. Then, all dependence on binding energy $E$ will be absorbed into the off rate $k_-$. Intuitively we can understand this by imagining that once the TF is bound in an energetically favorable configuration, it has to wait for a random thermal kick of typical size $k_BT$ to unbind, and the probability of that kick being able to overcome the binding energy barrier $E$ is $\sim \exp(E/k_B T)$ (remember, the more negative the binding energy, the stronger the binding).
%
%
\subsection{Phenomenological models of regulation}
In the previous chapter we have shown how thermodynamic and kinetic models are connected for simple cases of gene regulation where a single transcription factor binds cooperatively to different numbers of binding sites. In many cases, however, several transcription factors together regulate a single gene. How can such situations be addressed from a theoretical perspective?

Here we briefly discuss three possibilities, their drawbacks and benefits.

{\bf First,} if one knows the detailed molecular map of the states and their transitions, it is possible to write down the diagram as in Fig.~\ref{f-2state} and compute the steady-state occupancies; alternatively, this can be done in the thermodynamic picture if one knows the binding energies and cooperativities. For simple cases, such as a single activator and repressor jointly controlling a gene and having an energetic interaction, this is feasible and has been done for, e.g., the \emph{lac} operon in \emph{E coli}, as we discuss later. As we advance towards more and more complex organisms, however, this approach loses its appeal: for metazoan enhancers, for instance, we don't even know the microscopic states (much less their energies), and so cannot write down the partition function. While correct, this approach does not scale to more complex regulatory strategies well.

{\bf Second,} we can decide for an ad hoc approach. Here, we write down a model probability that the gene is on as a function of its TF concentrations, but do not worry whether this probability is actually derivable from some statistical mechanical system, or alternatively, if there is a realistic dynamical scheme that would generate this model probability. Often, such approaches borrow the intuition from simple thermodynamics results and combine them into a more complicated regulatory scheme. For example, if the gene $g$ is activated by TF $A$, present at concentration $c_A$, and repressed by TF $B$, present at concentration $c_B$, one could postulate (without deriving) that the occupancy of the promoter is
\begin{equation}
\bar{n}(c_A,c_B)=\frac{c_A^{h_A}}{c_A^{h_A}+K_A^{h_A}}\cdot \frac{K_B^{h_B}}{c_B^{h_B}+K_B^{h_B}}. \label{andreg}
\end{equation}
This expression assumes that molecules of $A$ bind independently to $h_A$ sites with dissociation constant $K_A$, and molecules of $B$ bind to $h_B$ sites with dissociation constant $K_B$;  importantly, we also assume that the joint regulation is \emph{and}-like, meaning that gene $g$ will only be activated when both $A$ is bound and B is \emph{not} bound [that's why there is a product in Eq (\ref{andreg})]. More complex schemes like this can clearly be derived, and while they will not necessarily correspond to any possible thermodynamic system, they might be useful \emph{phenomenological} models that can be fit to the data. 

{\bf Third,} we can pick a real thermodynamic model that is flexible enough to encompass many possible combinatorial strategies of gene regulation but will have a small enough number of parameters to connect to available data. As in the previous case, this model might not correspond on a molecular level to the events on the promoter, and would thus also qualify as a phenomenological model. It would, however, have the advantage of being more easily interpretable and understandable within the context of statistical physics. One such model is the so-called Monod-Wyman-Changeaux model, which we discuss below.
\subsection{A model of combinatorial regulation: MWC model}
In this section we'll discuss a thermodynamic model that can easily be extended to include combinatorial regulation. The model has been motivated by the work on allosteric transitions and was used to explain hemoglobin function \cite{mwc}. When applied to the case of gene regulation, the central idea is that as a whole, the promoter can be in two states, ``on'' (1) and ``off'' (0). Remember that in our previous examples we had to declare one of the combinatorial states as the ``active'' state; here, this distinction is built into the model by assumption.

\begin{figure}
\includegraphics[width =  \linewidth]{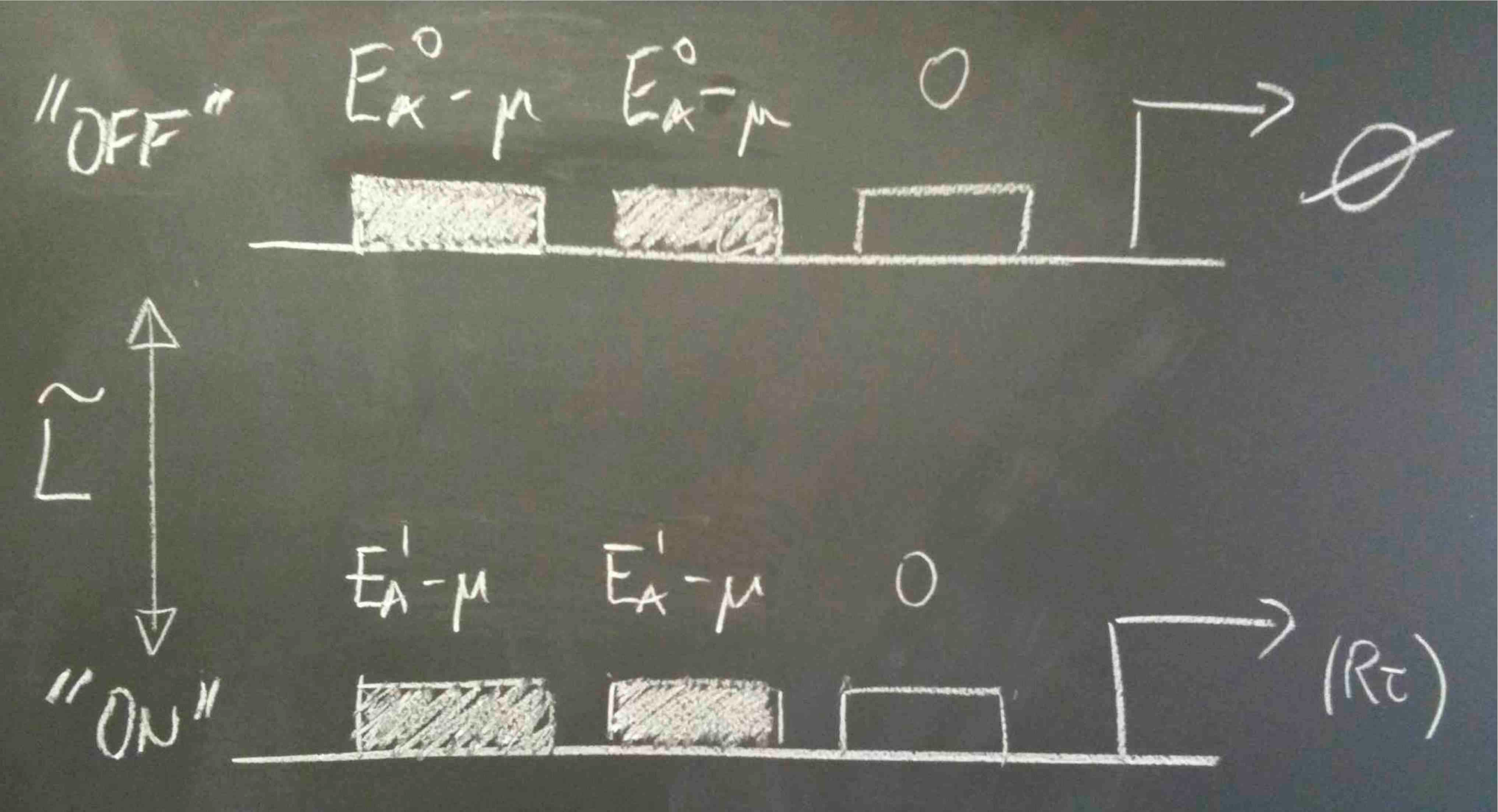}
\caption{A schematic diagram of MWC model. Two possible states of the promoter, ``on'' and ``off'', are separated by an energy barrier of $\tilde{L}$. There are 3 binding sites for the transcription factor in this example, to which TFs bind independently; their binding energy, however, depends on the state of the promoter. Here, 2 of the 3 sites are occupied, and are contributing $E_A^{0,1}-\mu$ each to the total free energy. If the promoter is ``off,'' there is no transcription, if it is ``on,'' transcription proceeds at rate $R$ and gives rise to $R\tau$ molecules of output in steady state at full induction.}
\label{f-mwc}
\end{figure}

The regulatory region has $n_A$ binding sites for transcription factor $A$. These sites can be bound in both the active and inactive state, and molecules of $A$ always bind independently, see Fig.~\ref{f-mwc}. However, the binding energy for each molecule of $A$ to its binding site is state dependent, i.e. $E_A^0$ when the whole promoter is ``off'' vs $E_A^1$ when it is ``on.'' Let's work out the thermodynamics of this system. For each of the two states, we can write down the free energies of $k$ molecules of type $A$ bound:
\begin{eqnarray}
F_0&=&k(E_A^0 - \mu) + \tilde{L}	\\
F_1&=&k(E_A^1 - \mu), 	
\end{eqnarray}
where $\mu=\log c$ (we are writing everything in units of $k_BT$ and dimensionless concentration again), and $\tilde{L}$ is how favoured the ``off'' state is against ``on'' state even with no TF molecules bound. The partition function is then
\begin{equation}
Z=\sum_{k=0}^n {n\choose k}e^{-k(E_A^0-\mu)+\tilde{L}} +\sum_{k=0}^n {n\choose k}e^{-k(E_A^1-\mu)}.
\end{equation}
Recognizing that the sums are simply binomial expansions\footnote{Note that $\sum_{k=0}^n{n \choose k}r^k=(1+r)^n$.}, we get for the probability of the ``on'' state (proportional to the expression of the gene):
\begin{eqnarray}
P(\mathrm{on}) &=& \frac{(1+e^{-E_A^1+\mu})^n}{(1+e^{-E_A^1+\mu})^n + (1+e^{-E_A^0+\mu})^ne^{\tilde{L}} } \label{mwce}\\
&=&\frac{(1+c/K_A^1)^n}{(1+c/K_A^1)^n+L(1+c/K_A^0)^n}. \label{mwcclass}
\end{eqnarray}
Equation (\ref{mwcclass}) is written in the standard form, with the identifications $K_A^1=\exp(\beta E_A^1)$, $K_A^0=\exp(\beta E_A^0)$ and $L=\exp(\tilde{L})$. 

A regulatory impact of transcription factor $A$ onto the regulated gene is described by quantities $(n,K_A^0,K_A^1)$ in the MWC model. There is one additional parameter $L$, the offset (or ``leak'') favoring the ``off'' state. Note that the parameters $K_A^{0,1}$ of the MWC model are not directly comparable to Hill model parameter $K_d$; however, we can make the identification in the regime where $c/K_A^0\ll 1$ and $c/K_A^1\gg 1$ . Then the term $(1+c/K_A^0)^n$ in Eq~(\ref{mwcclass}) can be approximated with 1, and $(1+c/K_A^1)^n\approx (c/K_A^1)^n$. Equation~(\ref{mwcclass}) then reduces to
\begin{equation}
P(\mathrm{on})=\frac{c^n}{c^n + L(K_A^1)^n},
\end{equation}
and we can identify the parameter $n$ in the MWC model with the Hill coefficient $h$, and the dissociation constant of the Hill model, $K_d$, with $K_d=L^{1/n}K_A^1$.

In general, for a single gene, the MWC model is not much different from Hill, producing sigmoidal curves that don't necessarily cover the whole range from 0 to 1 in induction as the input changes over a wide range. However, in the limit where $c/K_A^0\ll 1$, we can easily generalize MWC to regulation by several transcription factors. To see how, rewrite Eq~(\ref{mwce}) as
\begin{equation}
P(\mathrm{on}) = \frac{1}{1+e^{F(c)}}, \label{fenergy}
\end{equation}
where $F(c) = -n \log(1+c/K_A^1) + \tilde{L}$. In this picture, the binding and unbinding of transcription factors simply shifts the free energy of ``on'' vs ``off'' state. We can easily see that if $K$ transcription factors $\mu = A, B, \dots$  with concentrations $c_\mu$ regulate the expression of a gene, we can retain Eq~(\ref{fenergy}), but write
\begin{equation}
F(\{c_\mu\})=-\sum_{\mu} n_{\mu}\log\left(1+\frac{c_\mu}{K_\mu}\right) + \tilde{L};
\end{equation}
it is easy to check that positive $n_\mu$ represent activating influences, while flipping the sign of $\mu$ makes that gene $\mu$ repress the expression of $g$ \cite{awpre}.
\subsection{Input/output relations and surfaces}
These considerations lead to a useful abstraction that phenomenologically describes the behavior of single genetic regulatory elements, and summarizes how the input TF levels get integrated on the promoter into the activity of the regulated gene. We need to plot the curve (or surface in case of combinatorial regulation) of gene activity as a function of all relevant input TF concentrations, $\bar{g}=\bar{g}(\{c_\mu\})$: this is a nonlinear mapping that combines all inputs into a given output. In simple organisms (prokaryotes), the molecular understanding of these relations is probably close to correct: the promoter activity is just the occupancy of the promoter by the RNA polymerase and thus is proportional to the rate of making new transcripts. For higher organisms, the molecular picture is likely incorrect. Nevertheless, input/output surfaces are  useful abstractions for thinking about regulation, and concrete models such as MWC can compactly summarize experimental data.
\begin{figure}
\includegraphics[width =  \linewidth]{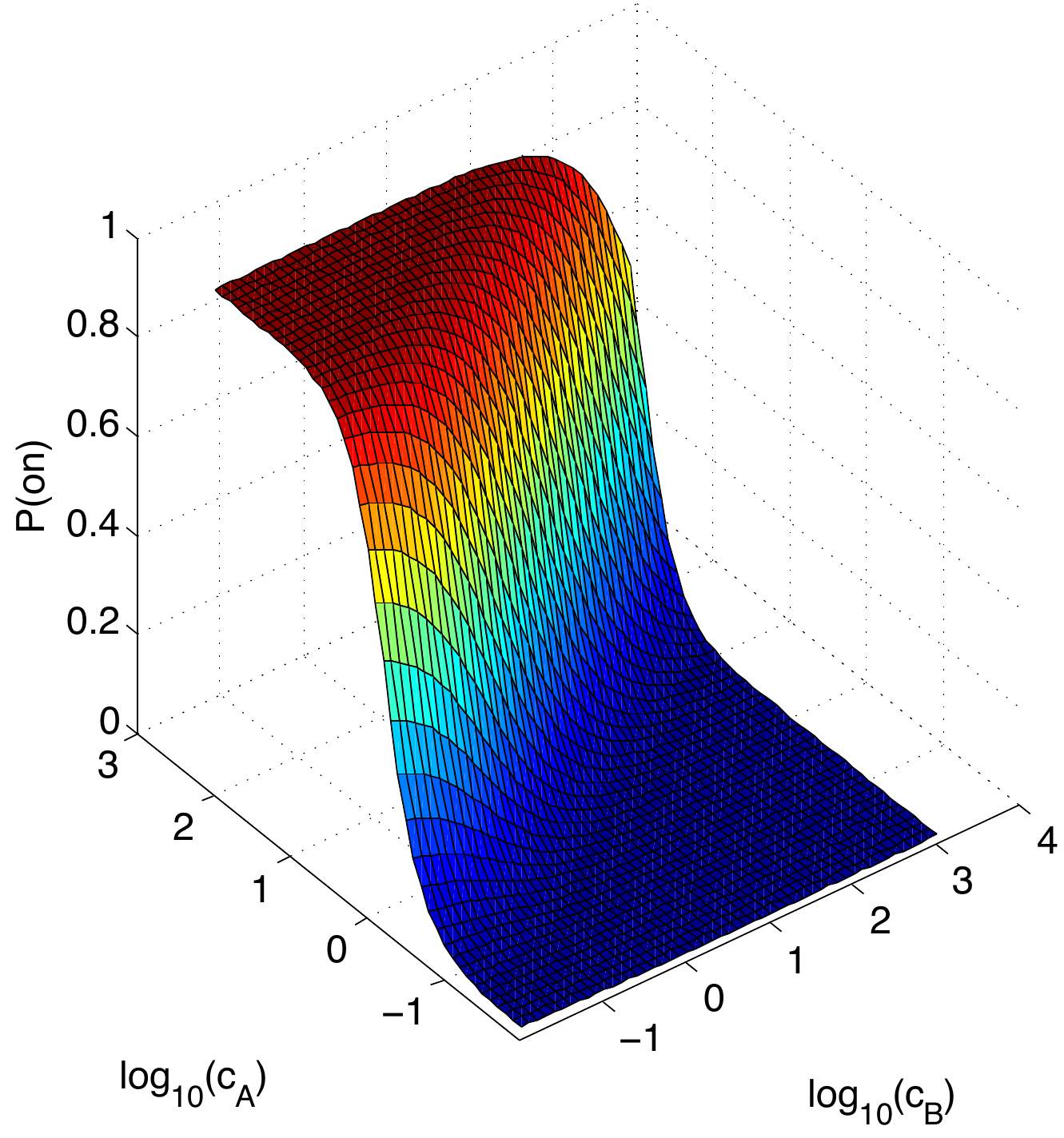}
\caption{A regulatory surface for a gene with two inputs, an activator A present at concentration $c_A$ and a repressor B with concentration $c_B$. The model is MWC, with $\tilde{L}=5$, $n_A=2$, $n_B=-1$ (negative since it is a repressor), $K_A=0.5$ and $K_B=0.7$.  }
\label{f-surf}
\end{figure}
\subsection{Experiments}
Are there quantitiative measurements of input/output relations in gene regulation?

\begin{figure}
\includegraphics[width =  \linewidth]{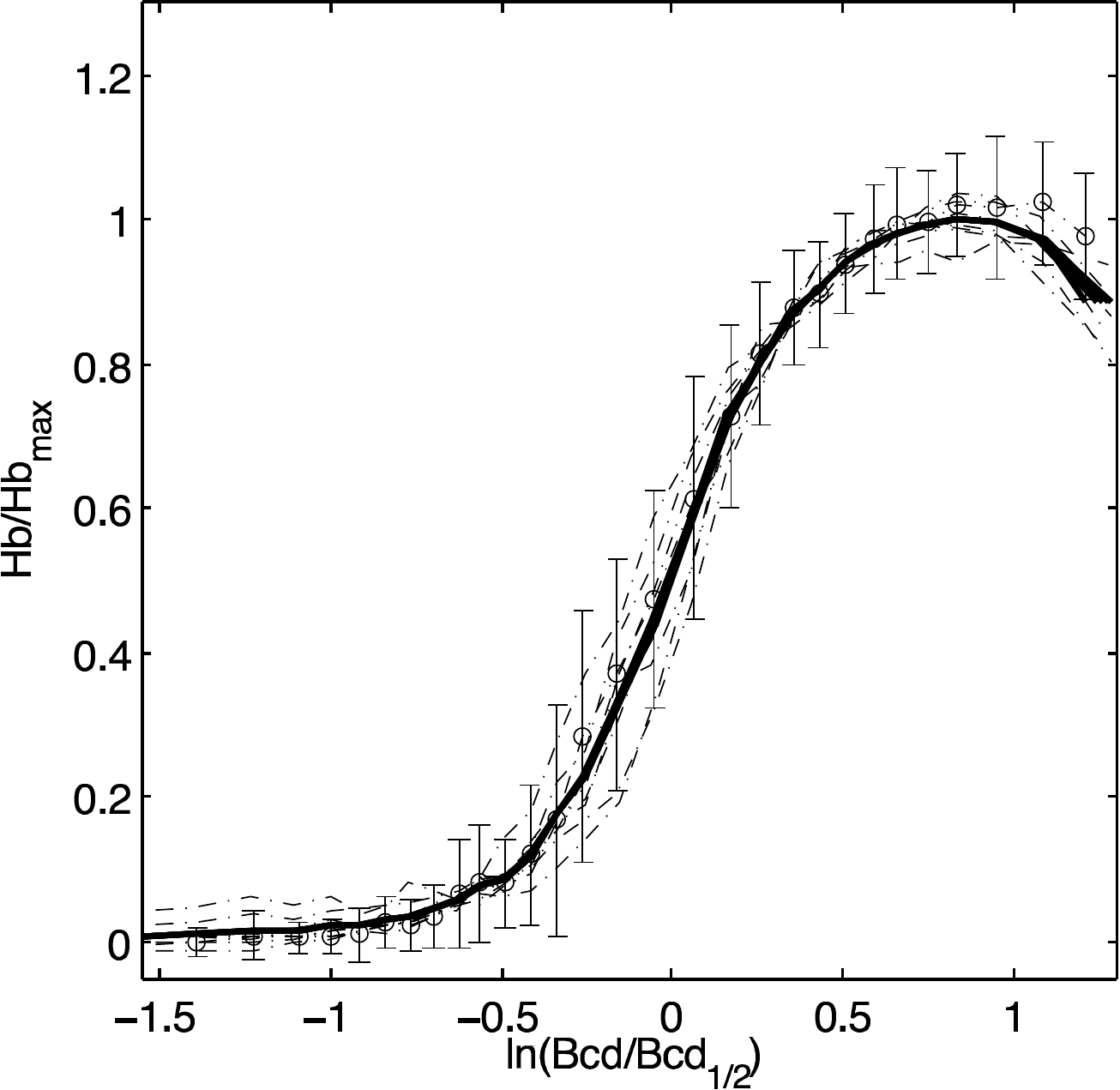}
\caption{The input/output relation between 2 genes involved in the early embryonic development in \emph{Drosophila melanogaster} \cite{thomas,plosone}. On the horizontal axis, the concentration of the input transcription factor called \emph{bicoid}, on the vertical axis the expression level of the target of bicoid regulation, called \emph{hunchback}. The data has been normalized so that at full mean induction hunchback reaches the value 1. Bicoid is expressed in units of its dissociation constant denoted by Bcd$_{1/2}$: when the concentration of bicoid is Bcd$_{1/2}$, hunchback is expressed at half-maximal levels.  Dashed curves show the measurements in different embryos, and the error bars show the noise in hunchback expression that we'll study in the next lecture. As we will see later, development is a very convenient system for extracting input/output relations because the physiologically relevant range of inputs is naturally laid down in the form of a spatial gradient across the embryo, and the output hunchback level for each input bicoid concentration can be read out simultaneously under the microscope in a single field of view. }
\label{f-noise2}
\end{figure}

For systems with a single input and a single output, it is commonplace to measure such relations and characterize them  with Hill-like functions.  One example of a single input, single output function is shown in Fig.~\ref{f-noise2}. This relation has been obtained in quantitative measurements using immunostaining methods and microscopy in the fruit fly embryos \cite{thomas, plosone}; the degree of reproducibility is evident from the match between measurements on multiple embryos. This curve can be fit very well by a Hill-like activation, Eq~(\ref{hillg}), with the fitted Hill coefficient of 5. The slight dip at high induction is a consequence of the fact that in the embryo, hunchback (the output) is regulated by other transcription factors, not only  the input (bicoid)\footnote{Hunchback is also thought to self-activate in addition to being activated by bicoid.}. The situation where the factors other than the explicitly measured input influence the expression level of the target gene is common in  higher organisms -- the list of all inputs is not necessarily even known, and the extracted input/output relations are of necessity phenomenological, reflecting the direct influence of the observed input regulator, but also the indirect influences through unobserved intermediaries.

\begin{figure}
\includegraphics[width =  \linewidth]{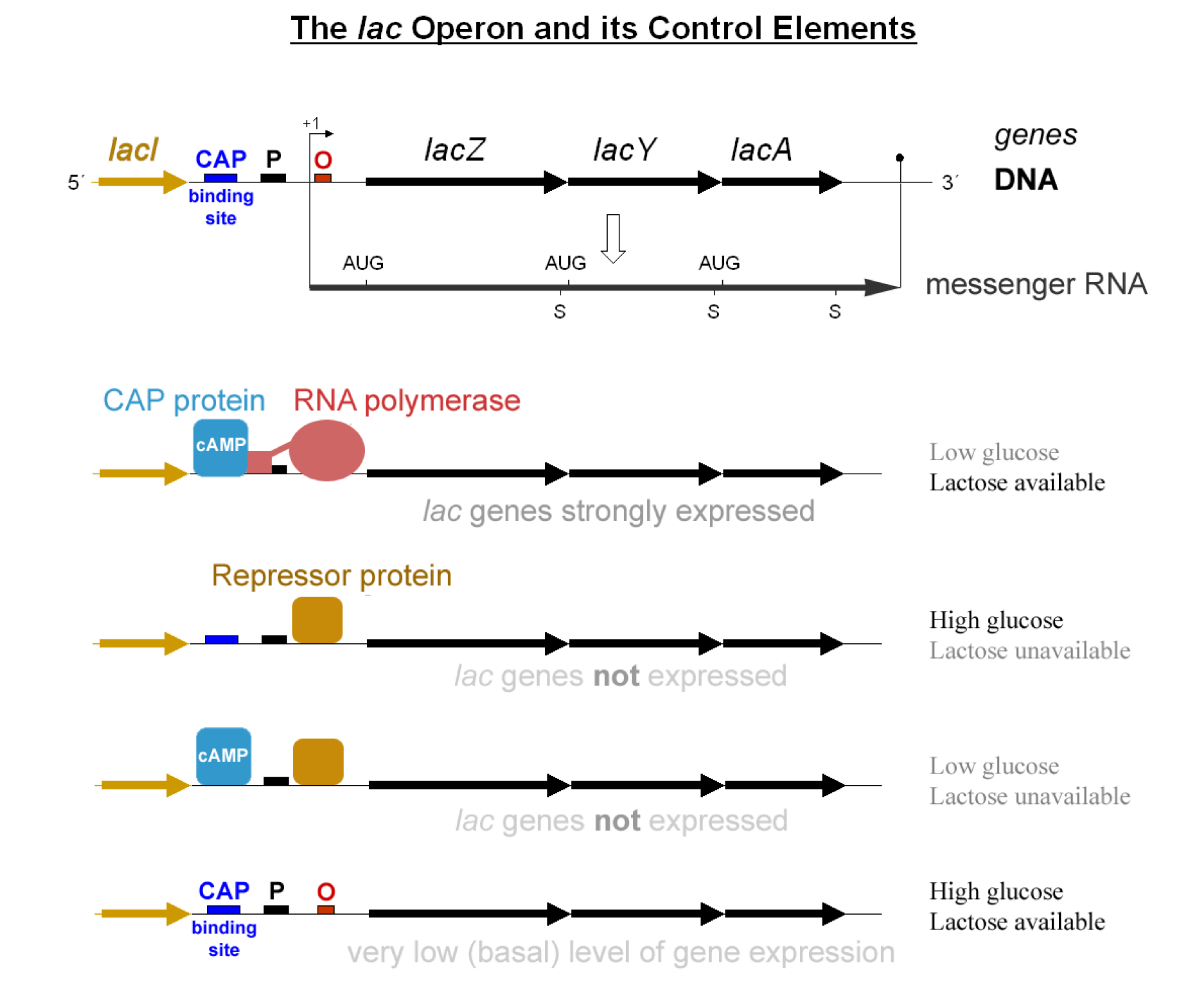}
\caption{A schematic diagram of the lac operon in \emph{Escherichia coli}. Genes for utilizing lactose are regulated jointly by the lac repressor (lacI gene and the associated lacR protein), and by the CAP (also known as CRP) activator. The bacterium expresses lactose genes strongly only when lactose is available, and the preferred sugar, glucose, is not. When glucose is also available, induction is low, but when lactose is unavailable, the genes in the operon are completely shut off. Image from Wikimedia commons.}
\label{f-lac}
\end{figure}

Perhaps the best worked out example is that of the lac operon in \emph{Escherichia coli} \cite{monod}, see Fig~\ref{f-lac}. The work of Refs~\cite{setty, kuhlman} has constructed a thermodynamic model of joint regulation of lac by its two transcription factors, CRP (activator) and LacR (repressor). In a series of experiments the concentrations of IPTG (an inducer for LacR) and cAMP (an inducer for CRP) has been varied and the 2D input/output surface has been mapped out. Making this surface consistent with known molecular facts about the regulatory proteins (such as the cooperativity of IPTG-LacR interaction) has required an in depth understanding of the system, including the thermodynamics of DNA looping and finding other mechanisms that influenced the intra-cellular concentrations of cAMP (such as an enzyme called PDE that degrades cAMP). This body of work demonstrated how thermodynamical models of regulation, information about relevant molecular properties of the regulatory proteins, and quantitative experiments can generate real understanding in a (simple) biological system. On the other hand, the same work highlighted practical problems connected with inferring the input/output relations when input is externally experimentally controlled: part of the mystery of making the models consistent with the measurements was the discrepancy between externally delivered concentrations of the inducer, and its actual intracellular concentration (\emph{E coli} possesses alternative mechanisms that affect these concentrations). In the end, understanding even a simple bacterial system such as the lac operon proved quite challenging. In the case of \emph{Drosophila} hunchback / bicoid regulation, similar problems can be avoided, because the input (bicoid) gradient is established naturally during early morphogenesis, and no experimental (possibly physiologically irrelevant)  interference with the system is needed.

Finally, in metazoan regulation our knowledge is much more limited. In enhancer regions that control the expression levels of genes possibly far away on the DNA, transcription factor binding sites form clusters and combinatorial regulation is abundant. Several kinds of TFs might bind, each to possibly more than one specific  binding site in the enhancer. These binding sites surprisingly appear to be less specific (i.e. have shorter recognition sequence lengths) than in bacteria. Genetic studies have shown that in constructs in which some of these sites have been permuted, the biological function is retained, while other tweaks disrupt the function; it is not clear what is the appropriate ``grammar'' that separates viable from non-functioning binding site configurations. Attempts are being made, however, to both fit simple thermodynamical models of regulation to the data in a predictive fashion \cite{siggia, segal}, and to map out the input/output surfaces of genes involved in early fruit fly patterning \cite{arnosti}, when gap genes respond to spatially varying concentrations of the morphogens (maternally deposited transcription factors). These studies have revealed a strong role for genetic cross-repression among the gap genes, as well as spatial effects in positioning of the binding sites that might be due to the packing and regulatory role of chromatin that can make the genes (in)accessible for expression.
\subsection{Relation to neuroscience}
Neurons, especially in the sensory periphery, can also be viewed as input/output devices \cite{spikes}. In the retina, the so-called retinal ganglion cells are sensitive to \emph{features} in the visual space: each neuron observes a small visual angle and fires when the spatio-temporal pattern of light in that visual angle matches the feature that the neuron is looking for. The output of the neuron can be viewed as a scalar, the probability of spiking $r$. In a typical experimental paradigm, visual stimuli can be precisely repeated many times and played to the neuron, so the probability of firing at each point in time, $r(t)$,  is accessible as empirical probability, or firing rate, computed across many aligned stimulus presentation repeats. 

The input characterization is more problematic. In principle, one projects a movie, that is, a set of image frames of $N\times N$ pixels each, refreshed every $t_0$ milliseconds, onto the neuron. Even if the neuron is only locally sensitive in time, i.e. its spiking at the current moment only depends on $T$ previous movie frames, the dimensionality of the input space is huge. Therefore, naively charting out an input/output relation, $r=r(N\times N \times T\;\mathrm{ parameters})$, is infeasible. 

It turns out, however, that the features that neurons look for in this huge dimensional input space are often simple. The neuron's sensitivity can be described by a spatio-temporal linear filter $\mathcal{L}$, local in space and time, also known as the \emph{receptive field}. A good model of the typical sensory neuron is that the neuron first takes a convolution of the filter with the movie, $s(t)=\int dt'\mathcal{L}(t')*\mathrm{movie}(t-t')$. The resulting$s(t)$ is now a scalar that gets mapped through the point-wise (instantaneous) input/output relation into the spiking rate $r(t)$. 

This is the traditional phenomenological model of a neural response, called the LN (linear-nonlinear) model. The linear part describes ``feature extraction'' and reduces a stimulus of high-dimensionality to a single scalar projection. The nonlinear part maps that projection through a nonlinearity to generate probability of spiking, $r(t)=r(s(t))$. Some neurons are sensitive to more than one feature, and thus have, e.g., two linear filters $\mathcal{L}_1$, $\mathcal{L}_2$, generating two projections $s_1, s_2$ at each instant in time. The input/output function is then a surface, or a 2D nonlinearity in neuroscience jargon, $r=r(s_1,s_2)$. A large effort in neuroscience is expended on deriving methods of experimentally probing, and inferring the linear filters and non-linearities. As an additional complication, the neural behavior is adaptive, meaning that the properties of both filters and non-linearities can be dynamically tuned on slow timescales to reflect the changes in the statistics of neural inputs (e.g. the change in the movie properties, such as average luminosity or contrast).
\section{The input / output devices are noisy}
\label{lec3}
\subsection{Encoders, decoders, and noise}
Information flow, in biological as well as engineered networks, is limited due to noise. 
To gain intuition about the influence of noise, we consider an information transmission system as a ``black box'' that takes some input signal $c$, transforms (encodes) it and transmits it.  The output signal $g$ is delivered to a readout device -- a decoder -- that then tries  to determine which input was sent. If there were no noise, each input would be uniquely mapped to some output, and this mapping would be fully specified by a one-to-one input/output relation $\bar{g}=\bar{g}(c)$. 

When noise is present, there is no such one-to-one map in general, and we must take into account the possibility that given an input symbol $c$, the system output is not uniquely determined. Instead, there exists a distribution over $g$, $P(g|c)$, that tells us how likely we are to receive a particular $g$ at the output if the symbol $c$ was transmitted. This distribution is also known as the \emph{encoding distribution}. A listener at the output could then use the Bayes' rule:
\begin{equation}
P(c|g) = \frac{P(g|c)P(c)}{\sum_c P(g|c)P(c)},
\end{equation}
to construct the \emph{decoding distribution}, that is, an inverse mapping for the likelihood of each input signal $c$, given that $g$ was received. Two things are worthy of note: {\bf (i)} the decoding party needs to know $P(c)$,  the distribution of inputs that are being sent (for instance, if $c$ are the letters of the English alphabet, one needs to know the letter frequencies in the written language to decode optimally); {\bf (ii)} $P(c|g)$ is not the final decoding result -- the decoding party does not want a distribution over possible $c$ that were sent, but instead want the specific $c$ that was ``most likely'' sent. Decoding thus requires an additional rule for choosing the best guess of $c$ from $P(c|g)$, and there are various optimal choices for this rule, depending on how and which errors in decoding are penalized. For example, if $c$ were continuous, we might want to choose a decoding rule that picks the best guess $\hat{c}$ out of the decoding distribution $P(c|g)$, such that the estimated L2 norm between the true transmitted $c$ and decoded $\hat{c}$ is minimized, $\hat{c}(g)=\mathrm{argmin}\;\int dc\; (c-\hat{c})^2P(c|g)$ \cite{spikes}.

\begin{figure}
\includegraphics[width =  \linewidth]{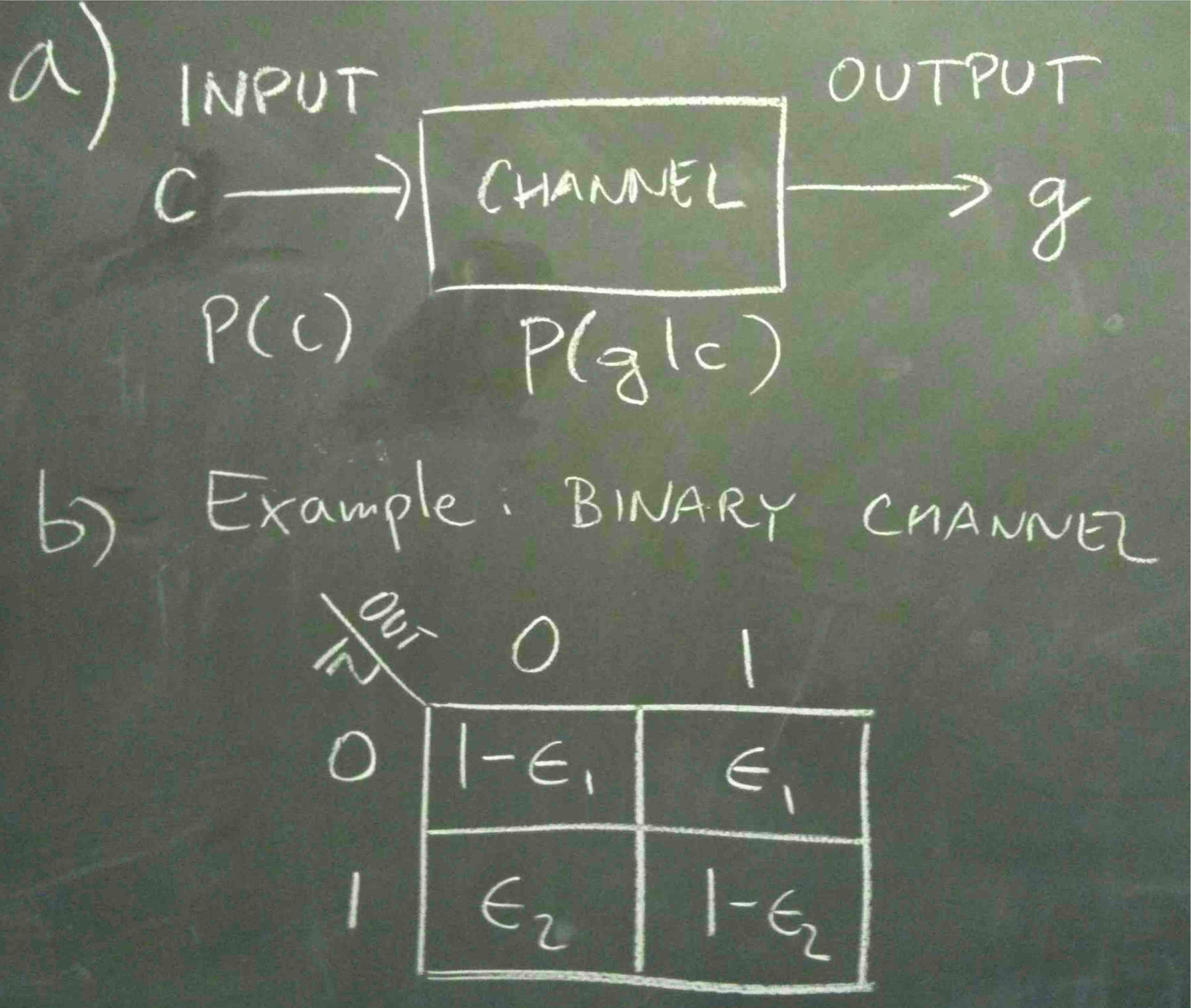}
\caption{{\bf a)} The schematic diagram of the channel receiving inputs $c$ and generating outputs $g$ according to a probabilistic map $P(g|c)$. The inputs are drawn from $P(c)$; the distribution of outputs is then fully determined, $P(g)=\sum_c P(g|c)P(c)$. {\bf b)} A simple example of the binary channel. The input and output values can be $0$ or $1$. If 0 is sent as an input, with probability $1-\epsilon_1$ the channel transmits the 0 without a mistake, but with probability $\epsilon_1$ the channel corrupts 0 into a 1. Similarly, $\epsilon_2$ gives the probability of the channel mistakingly transferring a $1$ into a $0$. }
\label{f-chanel}
\end{figure}

We return to an in-depth discussion of probabilistic encoding, decoding, and information transmission in Section~\ref{lec4}. To proceed, we note that there is a useful simplification at hand in the case when the symbols $c$ and $g$ are continuous: suppose that given $c$, the responses $g$ are well clustered around some \emph{mean response}, $\bar{g}(c)$, but there is also some ``spread'' around the mean, characterized by the variance $\sigma_g^2(c)$. The connection between these two quantities  and the fully probabilistic picture is as follows:
\begin{eqnarray}
\bar{g}(c) &=& \int dg\;g P(g|c)	\label{condmean},\\
\sigma_g^2(c)&=&\int dg\;(g-\bar{g})^2 P(g|c). \label{condvar}
\end{eqnarray}
These two functions are known as conditional mean and conditional variance, and they can easily be extracted from the distribution $P(g|c)$, if it is known. A noise-free deterministic limit is recovered as $\sigma_g^2(c)\rightarrow 0$, in which case $P(g|c)$ tends to a Dirac-delta distribution, $P(g|c)=\delta(g-\bar{g}(c))$.

Unfortunately, the full conditional distribution of responses given the inputs, $P(g|c)$, is usually only available in theoretical calculations or simulations, since in reality we rarely have enough data to sample it. In the case of gene regulation, sampling would involve changing the input concentration of TF, $c$, and for each input concentration, measuring the full distribution of expression levels $g$. More often than not we only have enough samples to measure a few moments of the conditional output distribution, perhaps the conditional mean and conditional variance. Given these measurements and $P(g|c)$ that is experimentally inaccessible directly by sampling, we can try making the approximation
\begin{equation}
P(g|c)\approx \mathcal{G}(g; \bar{g}(c), \sigma_g^2(c)),
\end{equation}
that is, we \emph{assume} that $P(g|c)$ is a Gaussian, with some input-dependent mean and variance.

In the presented setting, the mean input/output response and the noise in the response cleanly separate: one is given by the conditional mean, and the other by conditional variance. The noise can be thought of as \emph{the fluctuations in the output variable while the input is held fixed}. Recall that we are discussing all information processing systems in equilibrium, that is, when the dynamics in $g$ has reached steady state (and all variation in $g$ at given $c$ is due to noise).

\begin{figure}
\includegraphics[width =  \linewidth]{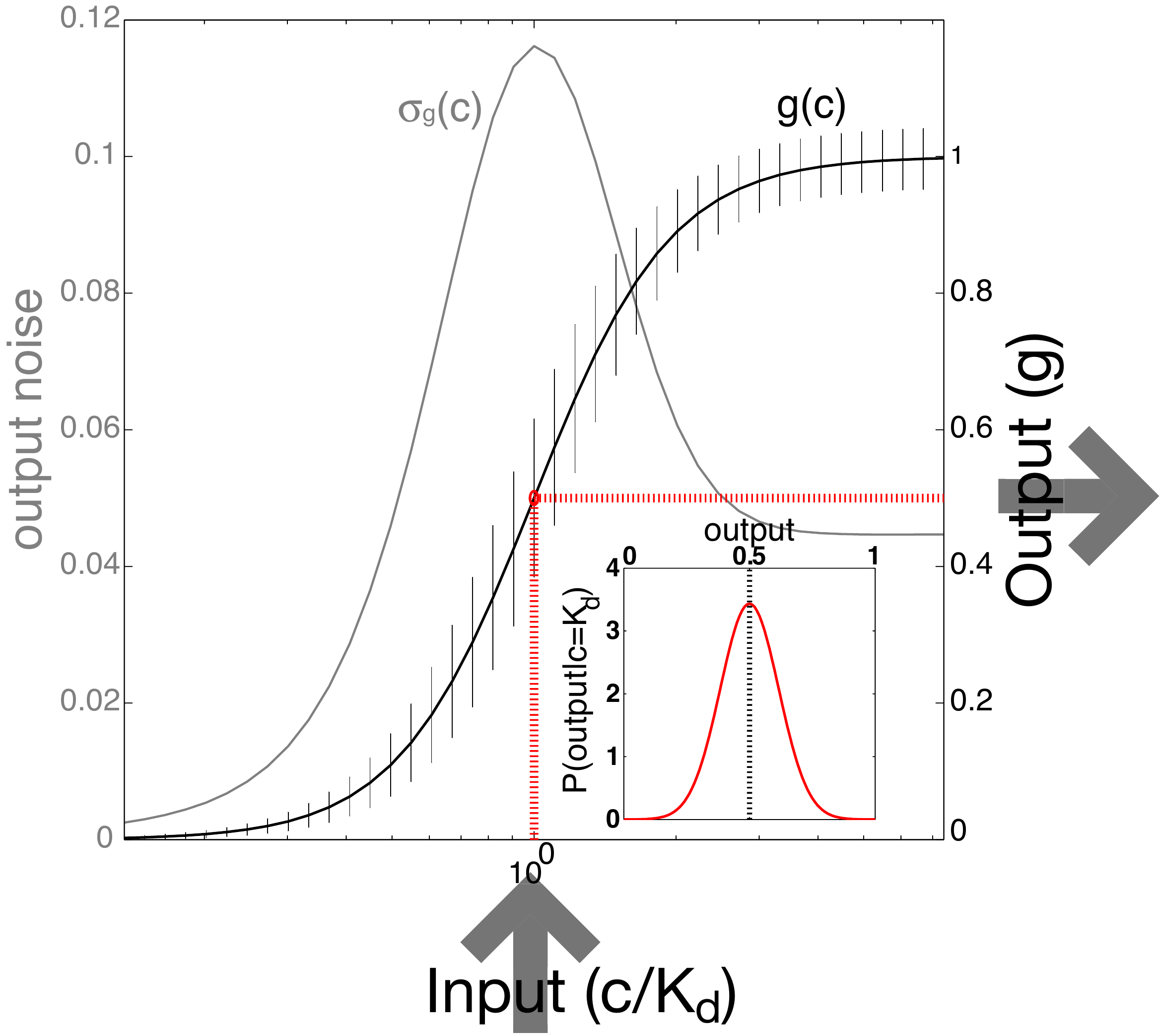}
\caption{The input/output relation, showing the mean response as well as noise. The input $c$ on the horizontal axis is mapped into the output $g$ on the right vertical axis through the input/output relation (thick black line); this summarizes the mean response, $\bar{g}(c)$. For each $c$,  there is a distribution of possible values of $g$, shown in the inset for $c/K_d=1$. The width of that distribution, $\sigma_g(c)$, is the measure of input-dependent noise, and is plotted as errorbars on the input/output relation. The errorbars themselves can depend on the input, as shown by the gray line and the corresponding scale on the left vertical axis. }
\label{f-schemeio}
\end{figure}

With the conditional mean and variance in hand, we can now create a detailed characterization of a noisy input/output regulatory element by means of two functions, as shown in Fig.~\ref{f-schemeio}.

\subsection{Sources of noise}
What factors contribute to the noise  in the response? 

Let us start by briefly considering a purely physical system first.
In the times of analog modems and noisy telephone lines, the modems' information rates increased  but started to saturate at about 30kbit/s. This is close to the theoretical limit predicted by Shannon's information theory (that we introduce in Section \ref{lec4}), given the level of noise and available bandwidth in the transmission lines. 
Such noise in electronic devices is well understood. Two fundamental sources of noise in electronic equipment that contribute are the \emph{Johnson noise} and the \emph{shot noise}. Johnson noise is due to random thermal fluctuations that jiggle the charges and thus induce random fluctuations in voltage (or current flowing through the resistance). 

Shot noise is due to quantal nature of charge carriers. ``Current'' is a macroscopic (average) quantity, and is a result of an integer number $n$ of elementary charges $q$, flowing through the cable in a time interval $T$. We can write the current as the total charge flowing during time $T$, i.e. $I=n q/T$. If we repeated the measurement of duration $T$ several times and were able to count individual charges, we would see that \emph{on average}, $\bar{n}$ charges flow, but this number has a Poisson fluctuations around the mean across our repeats of the experiment. A characteristic of a Poisson random process\footnote{Suppose that discrete point events (each of infinitely short duration) occur independently, and that on average we observe $\bar{n}$ such events in a time window $T$. The probability of observing $n$ events in each instance of the time window $T$ is then Poisson distributed, i.e. 
\begin{equation}
P(n|\bar{n})=\frac{1}{n!}e^{-\bar{n}}\bar{n}^n.
\end{equation}
The mean number $\bar{n}$, or alternatively the rate $r=\bar{n}/T$, are sufficient statistics for Poisson processes.
} is that the mean is equal to the variance, that is
\begin{equation}
\sigma_n^2=\bar{n}.
\end{equation}
We can then compute the observed fluctuations in the current: $\sigma_I^2=\sigma_n^2 q^2/T^2=I q / T$. In an experiment in which a constant current $I$ is flowing, and we measure for time $T$, $\sigma_I^2$ is the variance in the current that we would observe across the repeats of the experiment simply due to the fact that the charge is composed of elementary units $q$ which are not infinitely divisible.

Sources of noise that are directly analogous to the Johnson noise and the shot noise also act on the molecular level and can be observed in a wide variety of settings in biology. Before continuing, consider two additional interesting examples.

{\bf In human vision,} the photoreceptors in our retinas are very efficient at responding to small photon fluxes at low ambient light levels, when the retina is dark adapted. In an interesting set of experiments, various groups have delivered flashes of light of mean intensity $I$ to human observers that needed to report whether they saw the flash or not. This procedure allowed the experimenters to trace out the psychometric response curve, with mean intensity $I$ on $x$ axis, and the probability of detection on $y$ axis, summarizing the limits to human vision. This curve is not a step function, but rather has a smooth transition region, and for a while it was hypothesized that this ``fuzziness'' in transition might be due to the processing downstream from the retina: after all, the signals need to travel into the brain, where we take a conscious decision that might be corrupted by noise, because, for example, our attention during the experiment faltered, or for many other possible reasons.

Retinal processes that underlie vision are able to detect to single photons: a photon hitting retinal (the pigment in  rhodopsin molecules) can cause a chain of chemical reactions with high gain that ends up delivering a pulse at the photoreceptor output, and that pulse subsequently gets reported to the central brain in the spike trains of retinal ganglion cells. The fuzziness in the transition region that was observed in the experiments was not due to imperfect circuitry of the retina or the brain; rather, when delivering light pulses with intensity $I$, the experimenters were really delivering $\bar{N}\pm \sqrt{\bar{N}}$ photons (variance due to Poisson shot noise) in a given interval of time\footnote{The number of photons emitted is related to the wavelength of the light and the duration of the pulse. The number of photons incident on the photoreceptors in addition depends on the geometry of the eye and the light source, and possible scattering and absorption of light in the intervening medium.}. The smooth rise in the psychometric curve was due to this spread in the number of photons actually delivered -- humans really respond reliably to as few as 6 photons, and at those low light levels, the fractional fluctuation in the number of emitted photons at each pulse is significant (i.e. $\sim 1/\sqrt{\bar{N}}$). Remember that the inability to deliver a precise number $N$ of photons is not due to experimenters' lack of attention to detail: the emission of photons is a quantum probabilistic process and the Poisson shot noise is a basic physical limit that cannot be circumvented by a classical light source.

In addition to shot noise, we can also find the analogs of thermal (Johnson) noise in early vision. Due to thermal fluctuations, there is a small chance that the molecule of retinal will undergo a spontaneous conformational transition exactly mimicking the one that would normally have been caused by an impinging photon. There is no way for the downstream neural circuitry to distinguish whether such an event was a ``real,'' photon-induced transition, or a thermal fluctuation. Both in our eyes and in CCD cameras, this so-called ``dark current'' acts as a constant background hash causing false positives even if the detection circuit were otherwise perfect \cite{billphotons}.

{\bf In bacterial chemotaxis,} bacteria like \emph{Escherichia coli} implement a well-studied strategy of swims and tumbles, that is, periods of swimming in straight lines when the flagella coherently bundle in a cork-screw-like propeller, and periods when flagella turn incoherently, causing the bacterium to randomize its direction. The bacteria also like to swim towards sources of molecules that they find useful (``chemoattractants''), and they achieve this by  modulating the frequency of random tumbles: as long as the concentration of chemoattractants that the bacterium senses is increasing with time, the tumbling is repressed (since the swimming is in the correct direction); if the concentration is decreasing, tumbling is enhanced. 

Often, these chemoattracting chemicals will be present at very low concentrations, and one can ask the question ``how well can bacteria, even in principle, tell along which direction in space the concentration is increasing''? If the chemoattractant $c$ were at high enough concentration, one could imagine the bacterium having a detector for $c$ in its front and its back, which would allow it to measure the gradient of $c$ by simply measuring the front-to-back difference. Operating at very low $c$ concentrations, however, we need to stop thinking about continuous and infinitely precise concentration fields and start thinking about single ligand molecules. As this hypothetical bacterium ``sniffs'' with its front detector, it might measure, in time $T$, $\bar{N}_{\rm front}$ molecules on average at the front, and $\bar{N}_{\rm back}$ molecules in the back. But the molecules are discrete entities, so each of these measurements will fluctuate around the mean by $\sqrt{N_{\rm front}}$ and $\sqrt{N_{\rm back}}$ from measurement to measurement. Taking a difference between the front and the back counts, in the presence of this noise (and with no temporal averaging), is a very bad strategy for inferring the direction of the chemoattractant source. If $N_{\rm front}-N_{\rm back}\ll \sqrt{N_{\rm front}}$, then the estimate of the gradient will be completely swamped by noise. 

While this direct  strategy for estimating the gradient is implemented by certain  eukaryotic cells (which are much bigger and therefore able to measure the concentration differences better), bacteria -- to deal with the noise problem  -- implement a very different strategy of integrating the change of measured chemoattractant concentration over time. The trick is to realize that $dc(\vec{x}(t))/dt = \partial c/\partial t + \nabla c \cdot \vec{v}$; therefore, in a temporally constant gradient, the time derivative of the concentration measured by the bacterium at its position, $\vec{x}(t)$, swimming with velocity $\vec{v}(t)=\dot{\vec{x}}(t)$, is related to the spatial gradient of the concentration. It is also clear that with this strategy, bacteria can be experimentally ``tricked'' into believing that they move in a spatial gradient of the concentration, where really the concentration is spatially uniform but variable in time; this has been the basis of many beautiful chemotactic experiments by H. Berg and coworkers \cite{hcberg}.
%

In summary, in biology many important processes depend on events that occur between very small numbers of molecules. This can either be the detection of a single photon by the photopigment in the retina, or the detection of chemoattractant molecules by the swimming bacterium, or perhaps the binding of the transcription factor to its binding site. After all, in the last example one can think of  the binding site trying to detect or ``measure'' the TF concentration, making this example analogous to detecting chemoattractants or photons; it doesn't really matter that in one case the ligands are internal to the cell and in the other the ligands are external. On general grounds, we expect that in all instances where small numbers of molecules are involved in control processes, noise might be an issue.

\subsection{Impact of noise in biological systems}
When we study  problems in mechanics,  we begin by clearly delineating  what is our system of study and what is the environment. The same distinction is useful when we talk  about noise. For example, our system of study might be a single genetic regulatory element, denoted by $c\rightarrow g$, and its environment is the cellular environment of the particular single cell in which this system is embedded. This distinction is central because in an experiment, we will often observe many \emph{instances} of the system, and in each instance measure the input $c$ and output $g$ -- this is how many noise experiments using single-cell microscopy or FACS (fluorescence activated cell sorting) are performed. 

Once we measure the response $g$ given the input $c$ and the fluctuation in the response $\sigma_g^2(c)$, we need to ask what sources of variability contribute to that measured fluctuation. One source is the \emph{intrinsic noise} in the regulatory element $c\rightarrow g$ itself, contributed by the inherently stochastic molecular processes that we discussed above and will soon return to. Another source of noise, however, called \emph{extrinsic noise}, arises because the cellular environment changes from cell to cell; that is, one cannot guarantee that the system of study is always exposed to the same conditions. For example, the RNA polymerase concentration might fluctuate from cell to cell, and even if the process $c\rightarrow g$ would in itself contribute no intrinsic noise, the fact that this is transcriptional regulation process that involves RNAP and RNAP varies in the environment (from cell to cell), will induce some variance in measured $g$ across the population of cells. There is nothing fundamentally different about the intrinsic and extrinsic noise; the distinction is useful solely when a system of study can be  enclosed into a conceptual box and separated from the environment, and the noise in the system can be separated from the noise in its environment by clever experimental techniques\footnote{This is similar to internal and external forces acting on the system in mechanics. There is nothing physically different between external and internal forces. However, we need to account for them properly and not mix them up when writing down the dynamical equations for the system of interest.} \cite{swain02}.

We introduced the abstraction of  input / output devices in the previous lecture. Correspondingly, it makes sense to divide the intrinsic  noise into the input and output contributions. In a regulatory process $c\rightarrow g$, where TF $c$ controls gene $g$, the total noise in $g$, $\sigma_g^2(c)$, can arise from two kinds of sources. The first source, the \emph{output noise}, deals with the generation of output. In our case this is the transcription of mRNA molecules and their translation into protein molecules $g$. This source of noise would be present even if there were no transcriptional regulation whatsoever, i.e. if the gene were constitutively expressed. By making more and more mRNA and protein molecules, the relative impact of the output noise can be reduced. 

The second source of noise is called the \emph{input noise}, and this arises because the concentration $c$ at the binding site location itself is fluctuating (we will discuss why soon). This source of noise is important for two reasons: {\bf (i)} it gets mapped through the nonlinear input/output relation $\bar{g}(c)$ to give rise to the total measured noise in $g$; {\bf (ii)}  the input noise cannot simply be reduced by clever design at the output end. This is familiar to anyone who has dealt with electronics -- the noise in the input to the amplifier is the noise that no  amount of gain can reduce away.

When studying biological systems, we are faced with additional sources of fluctuation that are sometimes also referred to as noise, but contribute in addition to the fundamental sources of variability (such as thermal or shot noise sources discussed  above). The fundamental sources of noise, or physical limits to precision in biology, are the lower bound on the noise that biology cannot avoid. But clearly biological processes can be more noisy than the lower bound set by physics\footnote{An interesting choice of systems for study are therefore those systems which are believed to be under strong evolutionary pressure to reduce noise as far as possible, perhaps down to the physical limits to precision. It is thought that dark vision is one of such systems, because it gives selective advantage to both predators or pray that can see after sunset, or even in starlight. Perceptual studies that show human visual sensitivity approaching the limit set by the Poisson statistics of incoming neurons support this hypothesis.}.

One such additional source is \emph{experimental noise} that corrupts our measurements, independently of the actual noise in the biological system. Often, we don't have any theoretical understanding of what form this noise has, unlike in physics where the responses of the measurement devices can precisely be characterized. For example, microarray assays are a very popular high-throughput way of measuring the activity levels of various genes. However, there is no clear understanding of how the real level of mRNA in some physical units maps into the log light intensity ratios usually reported [in other words, what is $P($log light ratio$|$mRNA level$)$]. Therefore, inference from the measurement  to the underlying quantity of interest (which you can think of as decoding  the raw experimental data into the the true mRNA levels) is often done using some ad hoc procedure; in Section \ref{lec5} we discuss how information-theoretic inference can circumvent precisely this problem of unknown experimental noise.

Finally, how important really is noise to biology? The rough answer is that this depends on the system, and in particular to whether ultimately we can trace cellular (or neural) decisions to single microscopic rare events. If that is the case, the expectation is that the noise will play a large role. For example, the $\lambda$ switch in the phage controls the fate of the virus, i.e. whether it will stay lysogenic or turn lytic. The bistable switch is controlled by a small number of transcription factor repressor molecules, called cro and cI, that compete for the same binding sites \cite{ptashne}. In this case, the fate of the cell is tied to a single molecular decision, and therefore the stochasticity is important \cite{McAdams1997}. The same can be said for rates of ``spontaneous switching,'' where thermal noise is able to flip a genetic switch. This is a rare, yet potentially important event, and considerable theoretic effort goes into computing the frequencies of such rare events.

Overall, in both genetic regulation and neural systems, the noise limits the amount of information that the network can transmit. Nevertheless, the noise can often be treated as a small fluctuation riding on top of the signal. In protein-protein signaling networks (such as the two-component systems in bacteria), the intrinsic noise is thought to be small, because the reactions involve hundreds to thousands of signaling molecules. On the other hand, these molecules are proteins, transcribed from the genes and regulated by transcription factors, so the extrinsic noise can be large due to the slow (compared to the timescales of signaling reactions) random fluctuation in the total numbers of signaling proteins. There is evidence that biology tries to choose network wirings that make signaling networks \emph{robust} with respect to this extrinsic source of slow fluctuations \cite{barkaileibler}. 
\subsection{Derivation of noise for simple gene regulation}
Let's return to the simple gene regulation scenario of Fig.~\ref{f-sscheme}. We will sketch how the noise can be derived in this model using the Langevin approximation, and give a back-of-the envelope estimate for the terms that we do not compute here. The reader is invited to view the full derivation in Ref~\cite{plosone}.

We start with the dynamical equations:
\begin{eqnarray}
\frac{dn}{dt}&=& k_+c(1-n) -k_-n + \xi_n	\label{noise1}\\
\frac{dg}{dt}&=& Rn - \frac{1}{\tau}g + \xi_g, \label{noise2}
\end{eqnarray}
where again we take the binding site occupancy $n$ to be between 0 and 1, and the expression level of the output gene is $g$; $g$ is produced with rate $R$ when the binding site is occupied, and the proteins have a lifetime of $\tau$. We have already shown that the equilibrium solution of this system is $\bar{n} = k_+c/(k_+c + k_-)$ and $\bar{g} = (R\tau) \bar{n}$. Here we are interested in the fluctuations, $\sigma_g(c)$, around the steady state, that arise purely due to intrinsic noise sources: {\bf (i)} the fact that the binding site only has two binary states that switch on some characteristic timescale, {\bf (ii)} the fact that we make a finite number of discrete proteins at the output, and {\bf (iii)} the fact that the input concentration $c$ might itself fluctuate at the binding site location.

\begin{figure}
\includegraphics[width =  \linewidth]{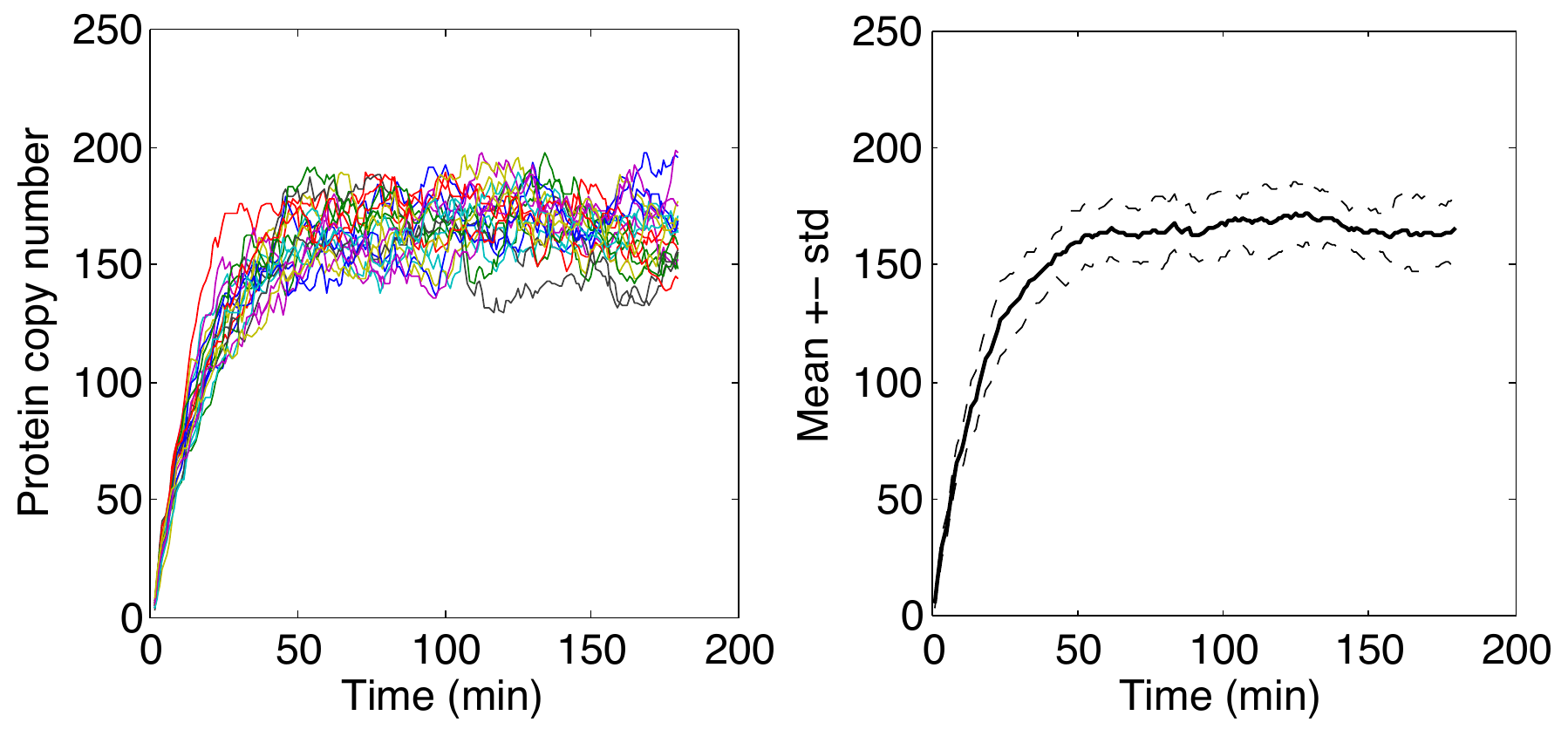}
\caption{A fully stochastic simulation of a simple model of gene expression using reactions specified in Eq~(\ref{cgillespie}). The simulation starts with $g(t=0)=0$ proteins; the steady state is reached after about 70 minutes. On the left, the trajectories of 20 simulation runs. On the right, the mean trajectory plotted in a solid line; mean $\pm$ 1-std plotted in dashed lines. The envelope measures the steady state level of noise due to (i) random promoter switching and (ii) the shot noise in producing the output molecules. }
\label{f-gillespie}
\end{figure}

One approach would be to simulate the system of Eqs~(\ref{noise1},\ref{noise2}) exactly using the Gillespie SSA algorithm. For a given and fixed level of input $c$, the results of 20 such simulation runs are shown in Fig.~\ref{f-gillespie}.

To compute this noise analytically instead of using the simulation, we have introduced random Langevin forces $\xi_n$, $\xi_g$. Consider the second equation, Eq~(\ref{noise2}). A single protein is produced anew, or is degraded, as an elementary step (since you don't make half a protein). In equilibrium, the production term $R\bar{n}$ balances the degradation term, $\bar{g}/\tau$. Now consider some time $T$ in which $RT\bar{n}=\bar{g}T/\tau \approx 1$, i.e. one molecule is produced or destroyed on average and with equal probability. While the expected change in the total number in equilibrium in time $T$ is zero, the variance is not: the variance is equal to $\frac{1}{2}\times$ (production of 1 molecule)$^2$ + $\frac{1}{2}\times$ (degradation of 1 molecule)$^2$ = 1. In general, the variance will be $T(R\bar{n} + \bar{g}/\tau)$ if we measure for time $T$. If you are familiar with random walks in 1D, this sounds very familiar: the mean displacement is 0 (because leftwards and rightwards steps are equally likely), but the variance in displacement from the origin grows with time $T$. 

Statistical physics tells us that in order to reproduce this variance in a dynamical system, we have to insert Langevin forces with the following prescription:
\begin{eqnarray}
\langle \xi_g(t)\rangle &=& 0	\nonumber\\
\langle \xi_g(t) \xi_g(t')\rangle &=& (R\bar{n} + \bar{g}/\tau) \delta(t-t'). \label{lngforce}
\end{eqnarray}
The mean random force is zero, it is uncorrelated in time, and it has an amplitude such that the random kicks have variance equal to the leftward and rightward step size; this will recover our intuition about 1D random walks. Similarly, $\langle\xi_n(t)\xi_n(t')\rangle=(k_+c(1-\bar{n})+k_-\bar{n})\delta(t-t')$.

To proceed, we first linearize Eqs~(\ref{noise1},\ref{noise2}) around the equilibrium, by writing $n(t)=\bar{n}+\delta n(t)$, $g(t)=\bar{g}+\delta g(t)$. Then we introduce Fourier transforms:
\begin{eqnarray}
\delta n(t)& =& \int \frac{d\omega}{2\pi} \delta \tilde{n}(\omega) e^{-i\omega t}\\
\delta g(t)&=& \int \frac{d\omega}{2\pi}\delta \tilde{g}(\omega) e^{-i\omega t}.
\end{eqnarray}
Fourier transforms of $\xi_n$ and $\xi_g$ are simply $\tilde{\xi}_n=2k_-\bar{n}$ and $\tilde{\xi}_g=2R\bar{n}$, respectively (because the Fourier transform of a delta-function is 1, and we have also used the fact that in equilibrium, the two terms that contribute to each Langevin force are equal). 

With this in mind, the system of equations in the Fourier space (denoted by tildes) now reads:
\begin{eqnarray}
-i\omega \delta \tilde{n}&=& -\frac{1}{\tau_c}\delta \tilde{n} + \tilde{\xi}_n \label{deriv1}\\
-i\omega\delta\tilde{g}&=&R\delta\tilde{n}- \frac{1}{\tau}\delta\tilde{g} + \tilde{\xi}_g, \label{deriv1a}
\end{eqnarray}
where $\tau_c^{-1}=(k_+c+k_-)$. 

We ultimately want to compute $\sigma_g^2(c)$. The total variance is composed from fluctuations at each frequency $\omega$, integrated over frequencies\footnote{Because the noise process is stationary (time-translation invariant), the noise covariance $\langle \delta g(t)\delta g(t')\rangle=C_g(|t-t'|)$ will depend on the difference in time only, and going into the Fourier basis will diagonalize the covariance matrix. The total noise variance is the integral over these independent Fourier components, and that is equal by Parseval's theorem to the total noise variance obtained by doing the corresponding integral in the time domain. }:
\begin{equation}
\sigma_g^2=\int \frac{d\omega}{2\pi}\langle \delta\tilde{g}(\omega) \delta\tilde{g}^*(\omega)\rangle = \int\frac{d\omega}{2\pi}S_g(\omega),
\end{equation}
where $S_g(\omega)$ is called the \emph{noise power spectral density} of $g$, and the asterisk denotes complex conjugate.
We see that we need to solve for $\delta\tilde{g}$ first from Eqs.~(\ref{deriv1},\ref{deriv1a}):
\begin{eqnarray}
\delta\tilde{g} = \frac{R\tilde{\xi}_n}{(-i\omega + \tau_c^{-1})(-i\omega+\tau^{-1})} + \frac{\tilde{\xi}_g}{-i\omega+\tau^{-1}}.
\end{eqnarray}
Next, we compute $\langle \delta\tilde{g}(\omega) \delta\tilde{g}^*(\omega)\rangle$. Recalling the definitions of $\langle \tilde{\xi}\tilde{\xi}^*\rangle$ [Eq~(\ref{lngforce})], we find that 
\begin{eqnarray}
S_g(\omega) &=& \frac{R^2(2k_-\bar{n})}{(\omega^2+\tau_c^{-2})(\omega^2+\tau^{-2})} + \frac{2R\bar{n}}{\omega^2+\tau^{-2}}.
\end{eqnarray}
The binding and unbinding of the promoter is usually much faster than the protein decay time, $\tau_c\ll \tau$. Using this and the fact that $\int_{-\infty}^{\infty}dx (x^2 + 1)^{-1}=\pi$, we finally find
\begin{equation}
\sigma_g^2(c) = \bar{g}(c) + \frac{(R\tau)^2}{k_-\tau}\bar{n}(1-\bar{n})^2. \label{rawnoise}
\end{equation}
If we normalize the expression level $g$ such that it ranges between 0 (no induction) to 1 (full induction) by defining $\hat{g}=\bar{g}/(R\tau)$, then the noise in $\hat{g}$ is 
\begin{equation}
\sigma_{\hat{g}}^2(c) = \frac{1}{R\tau}\hat{g} + \frac{1}{k_-\tau}\hat{g}(1-\hat{g})^2. \label{fnoise1}
\end{equation}
Our result is lacking at least one important contribution to the total  noise. The formal derivation of this term is involved \cite{simapaper, plosone}, so we will estimate it here up to a prefactor. In our derivation we have not taken into account that the molecules of transcription factor are brought to the binding site by diffusion. The diffusive arrival of molecules into a small volume around the binding site is a random process as well: it will induce some noise in occupancy of the binding site, and thus in the expression level $\hat{g}$. This is the contribution we are going to estimate.

Suppose that the binding site is fully contained in a physical box of side $a$. When the average TF concentration in the nucleus is fixed at $\bar{c}$, the average number of molecules in the box is $\bar{N}=a^3\bar{c}$. This, however, is only the mean number; if we were to actually sample many times the number of molecules  in the box, we would find that our counts are distributed in a Poisson fashion, with a variance equal to the mean: $\sigma_N^2=\bar{N}$. This is just the familiar shot noise in a new,  molecular disguise!

How can one reduce the fluctuations $\sigma_N^2$? As always, one can make more independent measurements, and average the noise away. With $M$ independent measurements, the effective noise should decrease, $\sigma_{N,\mathrm{eff}}^2=\sigma_N^2/M$. Suppose the binding site measures for a time $\tau$ (the protein lifetime, the longest time in the system). How many independent measurements  were made in the best possible case? It takes $t_0=a^2/D$ time for the molecules to diffuse out of the box of size $a$ and be replaced with new molecules; if we take snapshots and count the molecules at intervals faster than $t_0$, we are not making independent measurements. Therefore $M=\tau/t_0=\tau D/a^2$. Plugging this into the expression for effective noise, we find $\sigma_{N,\mathrm{eff}}^2=a^3\bar{c}\times a^2/(D\tau)$. Since $\bar{N}=a^3\bar{c}$, it follows that $\sigma_N^2=a^6\sigma_c^2$, and finally:
\begin{equation}
\sigma_{c,\mathrm{eff}}^2=\frac{\bar{c}}{Da\tau}. \label{cnoise}
\end{equation}
Equation (\ref{cnoise}) is a fundamental result: any detector  of linear size $a$ measuring concentration $c$, to which ligands are transported by diffusion with coefficient $D$, and making measurements for time $\tau$, will suffer from the error in measurement in concentration, given by $\sigma_c$. This contribution to the noise is called \emph{diffusive noise}, and it is a special form of input noise.

\begin{figure}
\includegraphics[width =  \linewidth]{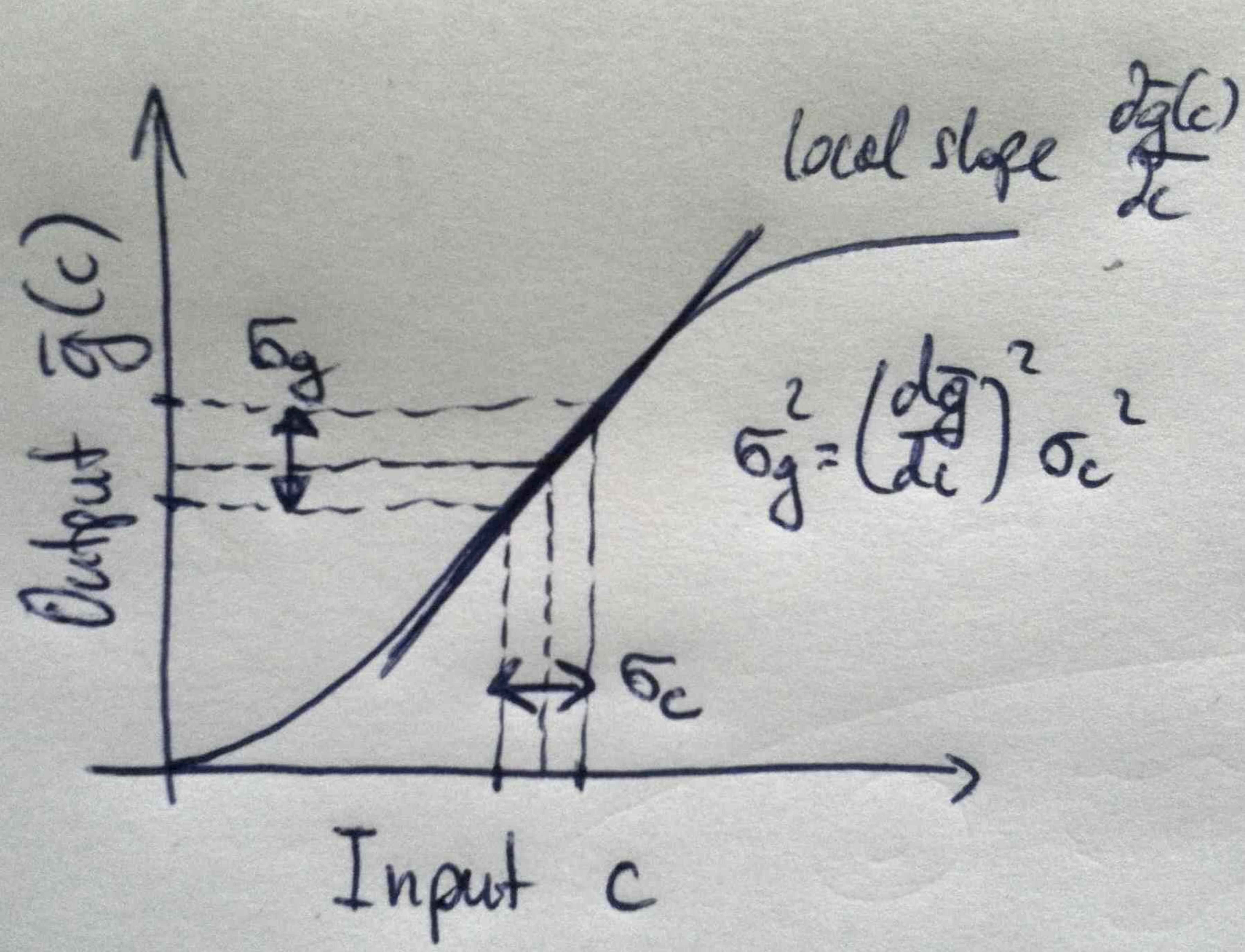}
\caption{Propagating the noise in the input $\sigma_c$, through the mean input/output relation, $\bar{g}(c)$, into the effective noise in the output, $\sigma_g$. The variances are related by the square of the local slope of the input/output curve, $d\bar{g}/dc$.}
\label{f-noiseprop}
\end{figure}

To assess how this error maps into the error in the gene expression $g$, note that any error at the input can be propagated to the output through the input/output relation, $\bar{g}(c)$ [see Fig~\ref{f-noiseprop}]:
\begin{equation}
\sigma_g^2=\left(\frac{d\bar{g}}{dc}\right)^2\sigma_c^2. 
\end{equation}
Adding the diffusive noise to previously computed terms in Eq~(\ref{fnoise1}), we find:
\begin{equation}
\sigma_{\hat{g}}^2(c) = \frac{1}{R\tau}\hat{g} + \frac{1}{k_-\tau}\hat{g}(1-\hat{g})^2 + \frac{\hat{g}^2(1-\hat{g})^2}{Da\bar{c}\tau}. \label{noisefinal}
\end{equation}

Let us stop here with the derivation, interpret the terms and summarize what we have learned so far. We tried to compute various contributions to the noise in the expression of gene $g$, in a simple regulatory element where the TF $c$ regulates $g$. In any real organism, such a small regulatory element will be embedded into the regulatory network, and $c$ will experience fluctuations on its own that will be transmitted into fluctuations in $g$, the so-called \emph{transmitted} noise, in addition to intrinsic noise calculated here \cite{pedraza}.

On top of intrinsic and transmitted noise sources, the output will also fluctuate due to the extrinsic noise because the cellular environment of the regulatory network is not stable. But even without these complications, we can identify at least three contributions intrinsic to the $c\rightarrow g$ regulatory process:

{\bf Output noise.} This is the first term in Eq~(\ref{noisefinal}), where the variance $\sigma_{\hat{g}}^2\propto \hat{g}$. Funamentally, this is a form of shot noise that arises because we produce a finite number of discrete output molecules. In the simple setting discussed here, the proportionality factor really is 1 [when $g$ is measured in counts, as in Eq~(\ref{rawnoise})], and this is a true Poisson noise where variance is equal to the mean. If we treated the system more realistically, with separate transcription and translation steps, the proportionality constant could be different from 1; a more careful derivation shows that then, $\sigma_{\hat{g}}^2=(1+b)/(R\tau)\hat{g}$, where $b$ is the burst size, or the number of proteins produced per single mRNA transcript, on average \cite{plosone}. This is easy to understand: the ``rare'' event is the transcription of a mRNA molecule, and that has true Poisson noise statistics, but for each single mRNA the system produces $b$ proteins, and the variance is thus multiplied by $b$.

{\bf Input promoter switching noise.} This is the second term in Eq~(\ref{noisefinal}). The source of this noise is binomial switching of the promoter, as it can only be in an induced ($n=1$) or empty ($n=0$) states. If we interpret $\bar{n}$ as the probability of being occupied, then the variance must be binomial $\bar{n}(1-\bar{n})$. Fluctuations between empty and full states of occupancy happen with the timescale $\tau_c$ [see Eq~(\ref{deriv1})], and the system averages for time $\tau$, so $\tau/\tau_c$ independent measurements are made, reducing the binomial variance to $\bar{n}(1-\bar{n})\tau_c/\tau$. Since $\tau_ck_-=(1-\bar{n})$ and $\bar{n}=\hat{g}$, we recover the switching term, $\hat{g}(1-\hat{g})^2/(k_-\tau)$. 

This term depends on the microscopic way the promoter is put together, hence the dependence on the kinetic parameter $k_-$. Regardless of these details, however, every promoter that has an ``on'' and ``off'' state will experience fluctuations similar in form to these derived here. In our example, $k_-$ is the rate of TF unbinding from the binding site and this is usually assumed to be very fast compared to the protein lifetime (in other words, the occupancy of the promoter is equilibrated on the timescale of protein production). In other scenarios that effectively induce gene switching, however, this assumption of fast equilibration might not be true. In particular, attention has lately been paid to DNA packing and regulation via making the genes (in)accessible to transcription using chromatin modification. The packing / unpacking mechanisms are thought to occur with slow rates, and such switching term might be an important contribution to the total noise in gene expression \cite{rajplos}.

{\bf Input diffusion noise.} The last term in Eq~(\ref{noisefinal}), as discussed, captures the intuition that even with the fixed average concentration $\bar{c}$ in the nucleus (that is, even if $c$ did not undergo any fluctuation relating to its own production, degradation and regulation), there would still be \emph{local} fluctuations at its TF binding site location, causing noise in $g$. This contribution is important when $c$ is present at low concentrations. As an exercise, one can consider the approximate relevance of this term in case of prokaryotic transcriptional regulation, where $D\sim 1\e{\mu m}^2/\e{s}$, the size of the binding site $a\sim 3\e{nm}$, the relevant TF concentrations are in nanomolar range, and the integration times in minutes. It has been shown that this kind of noise also represents a physical limit in the sense that it is independent of the molecular machinery at the promoter, as long as the predominant TF transport mechanism is free diffusion.

What we presented here theoretically was a simple example, but even so we'll see in the next section that the correspondence with the experiment is unexpectedly good. What is important for the lecture, however, can be summarized in the following observations: 
{\bf (i)} Not only can we make models for \emph{mean} input/output relations, but we can compute the noise itself, as a function of the input. Noise behavior is connected to the kinetic rates of molecular events, which are inaccessible in any  equilibrium measurement of mean input/output behavior. Therefore, if noise is experimentally accessible, it provides a powerful complementary source of information about transcriptional regulation. 
{\bf (ii)} There are fundamental (physical) sources of noise which biology cannot avoid by any ``clever'' choice of regulatory apparatus; thus the precision of every regulatory process must be limited. These sources all fundamentally trace back to the finite, discrete and stochastic nature of molecular events. In theory, the corresponding noise terms thus have simple, universal forms, and we can hope to measure them in the experiment.  
{\bf (iii)} There are sources of noise in addition to the fundamental, intrinsic ones, including extrinsic, experimental, etc. The hallmark of a good experiment is the ability to separate these sources by clever experimental design and/or analysis. We'll show how extrinsic and intrinsic noise sources can be separated in the examples below; in Section \ref{lec5} we show how to deal with unknown experimental noise.
\subsection{Experiments}

With the advent of quantitative microscopy, the use of protein-GFP fusions and FACS measurements,  noise, precision and reproducibility in gene regulation have become central themes in biophysics and molecular biology. A milestone has certainly been the \emph{two-color experiment} of Elowitz and coworkers \cite{elowitztwo}, allowing the separation of intrinsic and extrinsic noise sources. 

The basic idea of the two-color experiment is simple. To study a genetic regulatory mechanism $c\rightarrow g$, one engineers a bacterium to have two (almost) identical genes differing only in the color protein fusions; say $g_1=g-CFP$ and $g_2=g-YFP$. Both of these genes are regulated by the same transcription factor, $c$, and have the same promoters. The bacteria grow under the microscope, and for each bacterium, it is possible to collect the joint measurements of $(g_1,g_2)$ at a given fixed induction level (related to $c$ or the concentration of its inducer). 

\begin{figure}
\includegraphics[width =  \linewidth]{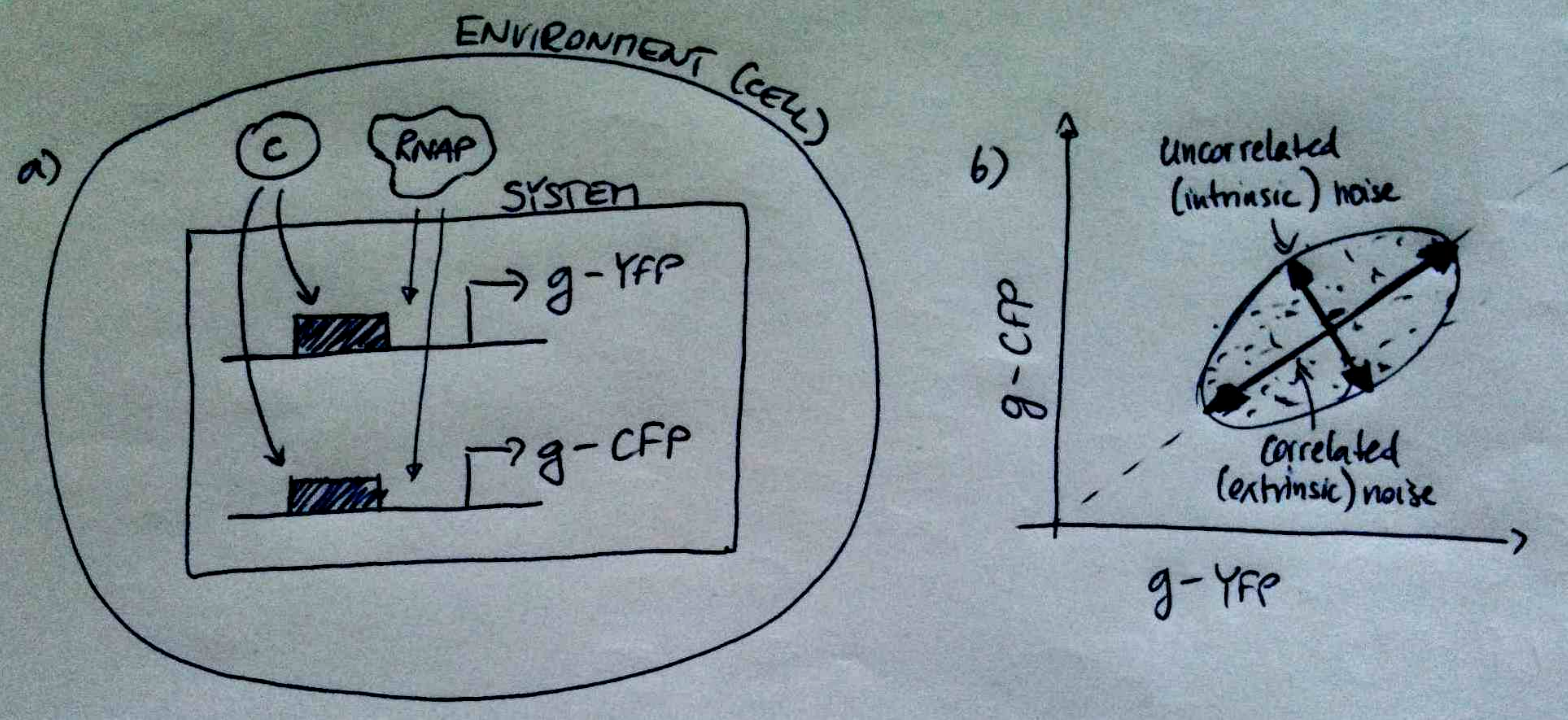}
\caption{The schematic diagram of the two color experiment. {\bf a)} The system of interest (in a square box) is a gene $g$ regulated by the transcription factor $c$. Two identical copies of the gene and its promoter are inserted; one gene is fused to yellow, while the other to the cyan fluorescent protein. The system is embedded into the cellular environment, and both copies of the gene share the same pool of transcription factors, RNA polymerase molecules etc. At a fixed level of $c$, the joint readouts in the yellow and cyan channels are taken and scatter-plotted against each other, as in {\bf b)}. Some of the variance is correlated, because the two copies in our system share the same cellular environment and are coherently impacted by the fluctuations in $c$, RNAP etc. This is the extrinsic noise component. Orthogonal to that is the uncorrelated (or intrinsic) noise component, due to the fluctuations that happen separately and independently in each of the two identical gene copies in our system.}
\label{f-2color}
\end{figure}

The basic realization of the experiment is as follows: scatter-plotting the values of $(g_1,g_2)$ collected from a cell, one can split the total variance in this cloud of points into the variance along the ``correlated'' axis (along the equality line), and the ``perpendicular'' axis, as in Fig.~\ref{f-2color}. These two orthogonal contributions are then identified with the extrinsic and intrinsic noise strengths, respecticely. The fluctuations in $g_1$ and $g_2$ can be correlated only because they don't happen in the system $c\rightarrow g$ itself, but in its intracellular environment, which affects both copies of $g_1$ and $g_2$ equally, in a correlated fashion. The intrinsic noise, however, is due to the terms we previously discussed in relation to Eq~(\ref{noisefinal}), and would be present even if the cellular environments were perfectly stable and reproducible. The intrinsic noise is uncorrelated, because random events in the $c\rightarrow g$ regulatory system happen independently in each of the two-color replicas in the cell. This ``two-color'' trick has enabled a whole set of experiments in which contributions from various steps in more complicated regulatory schemes, e.g. cascades $c\rightarrow g_1 \rightarrow g_2$, have  been teased apart \cite{pedraza, hooshangi}. Again, note that ``intrinsic'' and ``extrinsic'' are a matter of where one chooses to draw the boundary of  the system and  which part of the regulatory element gets replicated to apply the two-color paradigm.

A number of studies have since focused on the noise in prokaryotic gene expression. The findings indicate that the dominant sources of noise are the output (intrinsic) noise of making mRNA transcripts with relatively short correlation time (in minutes), and the extrinsic noise with long correlation time (of the order of a cell cycle) \cite{rosenfeld}. The noise in units of the mean, $\sigma_g/\bar{g}$, is very roughly of the order of $\sim 20\%$. A similar set of results was obtained in a high-throughput essay in yeast, where the noise was found to scale with the mean with a large prefactor, consistent with bursty expression \cite{noiseyeast}. 
An earlier study in the GAL promoter in yeast has, however, found a significant contribution of a noise that looks like the input switching noise, and has explained it by the cycling of the promotor through its microscopic states \cite{cantor}.

Later work in higher organisms has revealed an ever more important role of the intrinsic input noise, both the switching and the diffusive components. In mammalian cells, it has been possible to observe the number of mRNA molecules using the FISH method, and it was found that whole genes (or even sets of collocalized genes on the DNA) stochastically switch on and off; this might be a consequence of chromatin remodeling mechanisms \cite{rajplos}.

\begin{figure}
\includegraphics[width =  \linewidth]{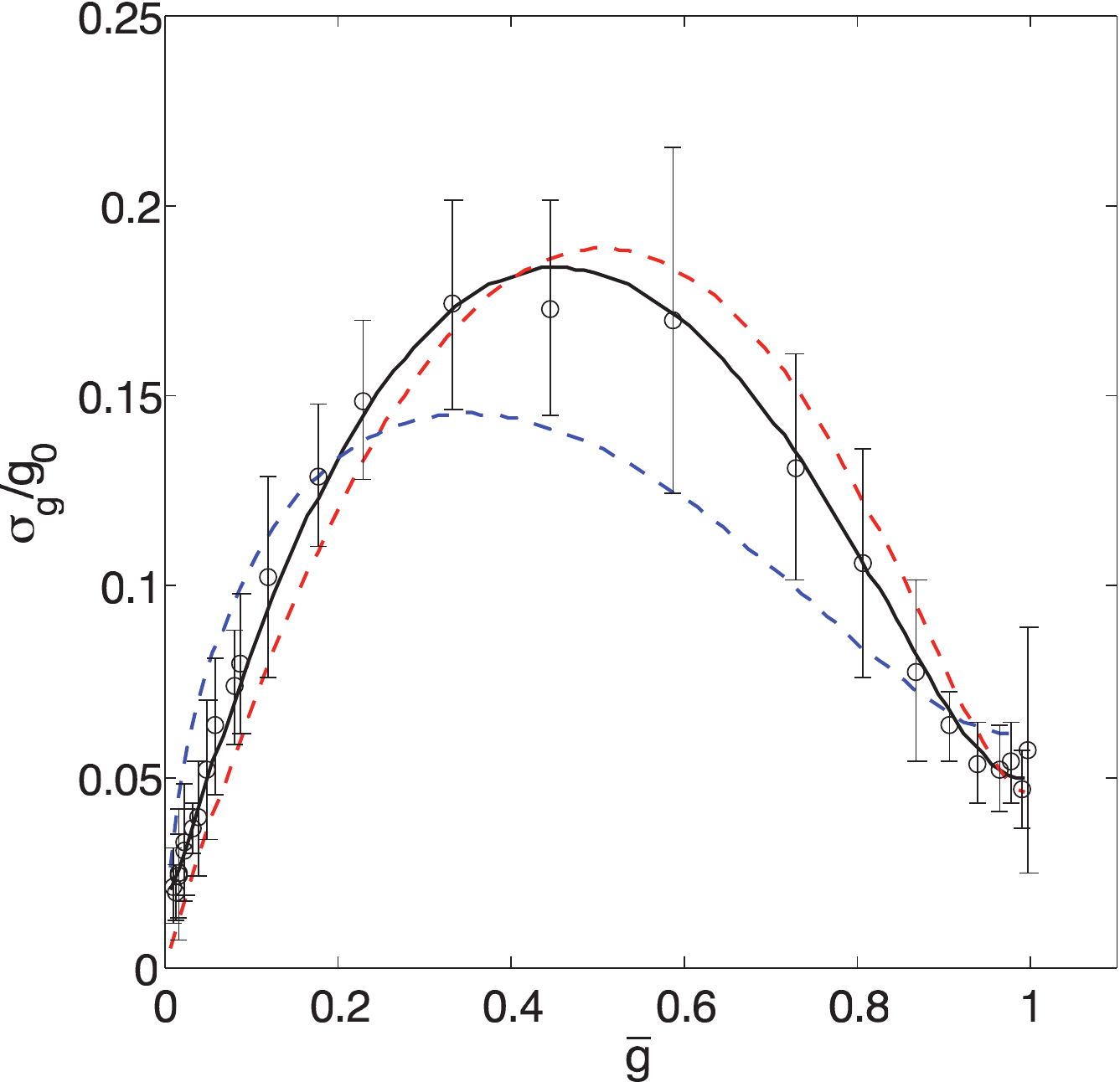}
\caption{The noise in the expression level of hunchback, $\sigma_g$, plotted as a function of mean hunchback induction $\bar{g}$, normalized to span the range from 0 to 1. Experimental data is shown as circles, with error bars denoting the std across 9 embryos. The solid black line is a fit of the model with output and input diffusive noise terms, assuming known Hill coefficient of $5$ for the hunchback regulation by bicoid. In red dashed line, the same fit using the  Hill coefficient of infinity (step like response). In blue, the fit using the input switching noise and the output noise contributions. The black two-parameter fit describes the data very well.}
\label{f-noise5}
\end{figure}

Our work in gene regulation during early \emph{Drosophila} morphogenesis has tried to quantitatively connect the model of Eq~(\ref{noisefinal}) with the experimental data \cite{thomas,plosone}. During development, each nucleus in the embryo of the fruit fly experiences a spatially varying concentration of the input transcription factor, bicoid, $c$, and activates the expression of hunchback, $g$; both of these quantities can simultaneously be measured using immunostaining and microscopy methods. Many nuclei experience the same level of input, $c$, in the same embryo. We can thus ask, for each input $c$, about the mean input/output relation, $\bar{g}(c)$, and also for the noise $\sigma_g(c)$, and these two measurements can be combined into $\sigma_g(\bar{g})$, the noise in hunchback expression as a function of mean induction, shown in Fig.~\ref{f-noise5}.

All three fits of the theoretical models to the data shown in Fig.~\ref{f-noise5} are two parameter fits. We fit essentially the same model as that of Eq~(\ref{noisefinal}); the only exception is that the noise is derived for a cooperative promoter with cooperativity of $5$, which can be read off from the Hb/Bcd input/output relation, see Fig.~\ref{f-noise2}. One parameter is always the magnitude of the output noise [the prefactor to the output noise term of Eq(\ref{noisefinal})]. The second parameter is the magnitude of the input diffusive noise (for solid black and red lines), or the magnitude of switching noise (blue line). The solid black fit describes the data excellently, and the following conclusions can be drawn from this study:
{\bf (i)} Qualitatively, input noise sources produce a peak in the noise at intermediate expressions when one plots $\sigma_{\hat{g}}$ vs $\hat{g}$. The input noise goes to 0 for zero or full induction. In contrast, the output noise increases monotonically with the induction. Therefore, whenever the experiment claims a peak in the noise, this is likely due to the non-negligible contribution of input noise to the total.
{\bf (ii)} The previous intuition also enables us to interpret the two fitted parameters. The strength of the output noise is proportional to the noise magnitude at full induction, $\bar{g}=1$, while the fitted strength of the input noise is proportional to the size of the peak at intermediate induction.
{\bf (iii)} We find that the fitted values of parameters are consistent with plausible values for transcriptional regulation: i.e., the magnitude of output noise is related to the number of  molecules of hunchback per nucleus (estimated from the fit to be $\sim 700-4000$), and the magnitude of the input noise is consistent with the measured bicoid diffusion constant and integration time.

\subsection{Relation to neuroscience}
Processes that underlie the limits to temporal precision of single neurons are physically analogous to the noise sources discussed in gene regulation; indeed, the study of temporal jitter in neural spiking predates the study of noise in gene regulation by several decades. 

A neuron generates the ionic currents by opening and closing of ion channels. These proteins have voltage-dependent probabilities, $p(V)$, of being open or closed. This nonlinear dependence, combined with the cable equation for propagating voltage disturbances in a medium with some capacitance, gives rise to self-excitability and generation of action potentials.

By making an analogy to the binomial ``switching noise'' in gene regulation, we see that for each ion channel, there will be an associated binomial variance in its activity, proportional to $p(V)(1-p(V))$\footnote{In principle, there could also be an associated ``shot noise'' variance in the number of ions that flow through these channels.}. This variance is reduced because there are many channels in each neuron that open and close independently (here we have population noise averaging, similar to temporal averaging in gene expression), but it is not reduced to zero. Using patch-clamp techniques it has been possible to electrically isolate a small patch of the membrane containing a very small number of channels, and to observe the quantal pulses of current flowing through single channels. This randomness in opening and closing of ion channels is one of the reasons why the neurons do not respond with identical spike trains to presentations of exactly repeated inputs.

\section{Introduction to information theory}
\label{lec4}
Up to this point we have stressed the role of noise in biological networks and mentioned several time that noise limits the ability of the network to transmit information; in this lecture we will turn this intuition into a mathematical statement. 

Recall that in our introduction to noise, we started with a probabilistic description of an information transmission system: given some input $c$, the system maps will map it into the output $g$ using a probabilistic mapping, $P(g|c)$. In case there were no noise, there would be no ambiguity, and $g=g(c)$ would be a one-to-one function.

Suppose that the inputs are drawn from some distribution $P(c)$ and fed into the system which responds with the appropriate $g$. Then, pairs of input/output symbols are distributed jointly according to
\begin{equation}
P(c,g)=P(g|c)P(c)	\label{joint}
\end{equation}
In what follows, we will be concerned with finding ways to measure how strongly the inputs ($c$) and the outputs ($g$) are dependent on each other. It will turn out that the general measure of interdependency will be tightly related to the concept of \emph{information}.

\subsection{Entropy and mutual information}
Let's suppose that our information transmission ``black box'' would be a hoax, and instead of encoding $c$ into $g$ in some fashion, the system would simply return a random value for $g$ no matter the input $c$. Then $c$ and $g$ would be \emph{statistically independent}, and $P(c,g)=P(c)P(g)$; such a box could not be used to transmit any information. As long as this is not true, however, there will be some statistical relation between $c$ and $g$, and we want to find a measure that would quantify ``how much'' can I know, in principle, about the value of $c$ by receiving outputs $g$, given that there is some input/output relation $P(g|c)$ and some distribution of input symbols $P(c)$. 

\begin{figure}
\includegraphics[width =  \linewidth]{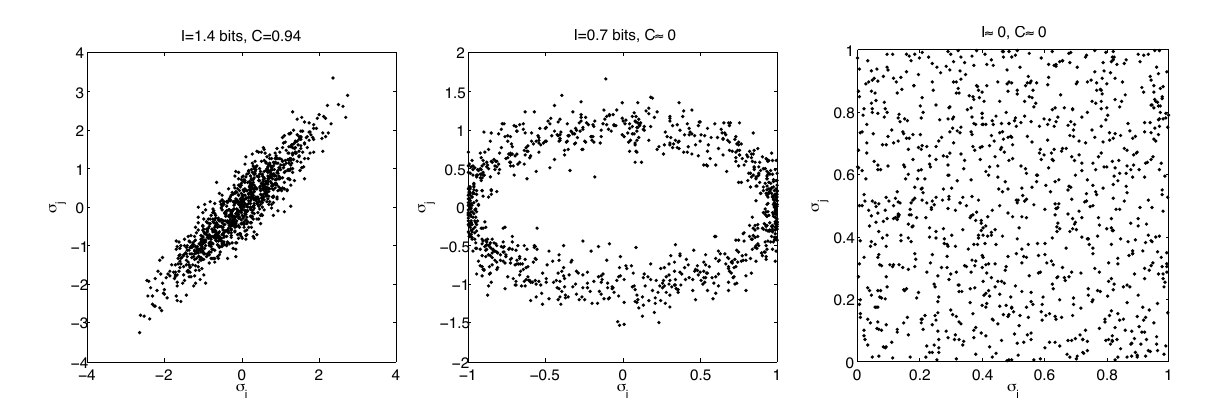}
\caption{Examples of two variables, drawn from three joint distributions. Shown are the scatterplots of example draws. On the left, the variables are linearly correlated, and the correlation is close to 1. In the middle, the variables are interdependent, but not in a linear sense. The correlation coefficient is 0, but measures of statistical dependence, such as \emph{mutual information}, give non-zero value. Note that we are looking for a general measure of interdependency: if we had a model that \emph{assumes} that $x$ and $y$ lie on a circle, we could fit that particular model or use a measure that makes the circular assumption. Instead, we would like to find a measure that detects the dependency without making any assumptions about the distribution from which the data has been drawn. On the right, the variables are statistically independent, and both linear correlation and mutual information give zero signal.   }
\label{scall}
\end{figure}

The first quantity that comes to mind as the interdependency measure between $c$ and $g$ is just the covariance:
\begin{eqnarray}
\mathrm{Cov}(c,g)=\int dc \int dg (c-\bar{c})(g-\bar{g}) P(c,g);
\end{eqnarray}
it is not hard, however, to construct cases in which the covariance is 0, yet $c$ and $g$ are statistically dependent. Covariance alone (or correlation coefficient) only tells us about whether $c$ and $g$ are \emph{linearly related}, but there are many possible nonlinear relationships that covariance does not detect; for example, see Fig~\ref{scall}.

Moreover, we would like our dependency measure to be very general (free of assumptions about the form of the probability distribution that generated the data) and definable for both continuous, as well as discrete outputs\footnote{Covariance can be problematic when used on discrete quantities.}.

We will claim, following Shannon, that there is a unique, assumption-free measure of interdependency, called the \emph{mutual information} between $c$ and $g$. Before we define it, however, we need to define another quantity, called the \emph{entropy} of a distribution $P(c)$:
\begin{equation}
S[P(c)]=-\int dc\; P(c) \log_2 P(c)
\end{equation}

To keep things simple, let's for now assume that the value of interest is discrete, that is, that $c$ can only take on the values $c_i$, $i=1,\dots,K$. In that case 
\begin{equation}
S[P(c)]=-\sum_{i=1}^K P(c_i) \log_2 P(c_i).
\end{equation}
If the data were continuous, we would have discretized it prior to computing the entropy (we discuss the discretization later); the measure that we are after, the mutual information, will turn out to be independent of discretization.

Entropy can be defined for any distribution. It is always positive, measured in bits, and always takes a value between two limits: $0\leq S[P(c)] \leq \log_2 K$ (in discrete case). The entropy is zero when the distribution has its whole weight of 1 concentrated at a single $c_i$. The entropy is maximal when $P(c_i)=1/K$, i.e. $P$ is a uniform distribution. The entropy is a unique measure of uncertainty about the value of $c$: the uncertainty is 0 when the distribution is peaked at a single value, and maximal when all $c_i$ are equally likely.

In one of the fundamental works of the 20th century science, Shannon has argued that this quantity, the entropy, is connected to the amount of \emph{information} that needs to be supplied to specify a particular value of $c$. Suppose $c$ can take on 8 different values $c_i=1,\dots 8$, and $P(c_i)=1/8$, that is, the distribution is uniform, and by our definition, has an entropy of $S=\log_2 8=3$ bits. I draw a particular $c_i$ from the distribution and don't share the value with you. You are allowed to ask a series of yes/no questions about the value I  chose and your task is to determine my choice in as few questions as possible. What is \emph{the minimum number of questions that you need to ask, on average}? This turns out to be the same as the entropy, and in case when $P(c_i)$ is uniform, one of the optimal strategies is bisection: ``Is the chosen $c_i$ larger than 4?" If yes, the next question could be ``Is $c_i$ larger or equal to 7?'' Otherwise, you could ask ``Is the chosen value less or equal to 2?'' and so on, until the correct $c_i$ is identified. Bisection is optimal because with a single binary question, it partitions the set of possible values into two equally likely subsets. To show that the number of questions is equal to the entropy even in the case of non-uniform $P(c_i)$ one needs some extra work, but the intuition remains the same.

Note that the entropy is the minimum number of questions, \emph{on average} -- across many repeats of the game. In single instances of the game you might be lucky and hit the correct number by simply asking ``Is the correct value $c_i$ equal to 7?''. However, this strategy will only terminate in $1/8$ of the games with a single question, and on average, it is worse than the strategy of bisection. 

In case of continuous variables, no number of questions can pin down the exact value of a real number, because, mathematically, real numbers  have  infinite precision. In practice this is not a concern, because the precision of physical data is always limited by noise or measurement error. We can therefore always discretize a continuous measurement into bins with the size of the error bar. Formally, all information theoretic quantities can be properly defined also for continuous variables, but in the interest of clarity we will skip these generalizations here.

To summarize: a ``bit'' is thus amount information contained in a single binary response to an optimally posed question -- in other words, a bit is the maximum amount of information that a binary question can convey. For each distribution $P(c)$ of some quantity $c$ we can define a measure called its entropy, a positive number expressed in bits, that quantifies the uncertainty about the value of $c$, and is connected to the minimal number of yes/no questions that need to be asked to find out the value of $c$ on average.

Let's now return to the original problem, where we think about two variables, $c$ and $g$, jointly distributed as $P(c,g)$, and we would like to quantify how statistically dependent the two variables are. Let's start by computing $S[P(g)]$, and assume that our values are discretized (so we are dealing with $g_j$ and $c_i$, where $i$ and $j$ run over all possible discrete choices for $g$ and $c$, respectively):
\begin{equation}
S[P(g_j)]=-\sum_j P(g_j)\log_2P(g_j).
\end{equation}
According to what we have just learned, this is the uncertainty about the value of $g$. We can also compute the following:
\begin{equation}
S_{n}(c_i) = S[P(g_j|c_i)]=-\sum_j P(g_j|c_i) \log_2 P(g_j|c_i).
\end{equation}
This entropy still depends on $c$, because the sum is taken only over the possible values for $g_j$. The interpretation of this quantity is the \emph{uncertainty about the value of $g$ if we know the value of $c$ to be $c_i$}. If $c$ and $g$ are related in any statistical way whatsoever, we expect that by knowing $c$, our uncertainty about $g$ will be less, on average, than if we don't know $c$. In other words, $c$ will give us some information about $g$, if there is any statistical interdependency. In equations, let's define mutual information $I$ as:
\begin{equation}
I(c;g) = S[P(g)]- \sum_i P(c_i) S_{n}(c_i).  \label{mut1}
\end{equation}
The crucial insight is that without any knowledge of $c$, $g$ will have the uncertainty $S[P(g)]$; with the knowledge of $c$, the average uncertainty about $g$ is smaller, $\sum_i P(c_i) S_{n}(c_i)$. The difference between these two quantities is the mutual information, and Shannon has shown that this is a unique assumption-free measure of any statistical dependency that does not make any assumption about the form of the joint distribution $P(c,g)$. 

Mutual information is a number in bits, and can be shown to be always positive. Algebraic manipulation of Eq~(\ref{mut1}) quickly produces a more compact expression:
\begin{equation}
I(c;g) = \sum_{i,j}P(c_i,g_j)\log_2\frac{P(c_i,g_j)}{P(c_i)P(g_j)} \label{mut2}
\end{equation}
where $P(c_i)$ and $P(g_j)$ are marginal distributions of the joint $P(c_i,g_j)$. Note that the mutual information is a single number, not a function, although it is conventionally written with the arguments in parenthesis, $I(c;g)$, to denote that it is information between $c$ and $g$.

Mutual information has a number of very appealing properties:
\begin{itemize}
\item {\bf It can be defined for continuous or discrete quantities.} Mutual information is a functional of a probability distribution, and probability distributions are very generic objects. $c$ and $g$ could both be continuous, or any one or both can be discrete.
\item {\bf It is reparametrization invariant.} Mutual information betwen $c$ and $g$ is the same than mutual information between any one-to-one function of $c$, $f(c)$, and any one-to-one function of $g$, $h(g)$, that is $I(c;g)=I(f(c);h(g))$. In biological experiments, this is a great asset, as we will soon see: experiments often report, e.g. intensities or log-intensities on the microarray chips or in FACS sorting, and there is a lot of discussion about how this data should be normalized and transformed prior to any analysis. This is important because some statistical measures of correlation, like correlation coefficients, depend on it. Mutual information, in contrast, is invariant to such reparametrizations of the variables.
\item {\bf It is symmetric.} Mutual information tells us, in bits, how much I learn about $g$ if I know the value of $c$. As is evident from Eq~(\ref{mut2}), $I(c;g)$ is symmetric with respect to the change in $c$ and $g$, i.e. $I(c;g)=I(g;c)$. This means that I can equally well learn about $c$ by knowing the value of $g$.
\item {\bf It obeys data processing inequality.} Suppose that $g$ depends on $c$ and $k$ depends on $g$ (but not directly on  $c$), in some probabilistic fashion. In other words, one can imagine that there is a Markov process, $c\rightarrow g \rightarrow k$, where arrows denote a noisy mapping from one value to the next one: $c$ gives rise to $g$ and $g$ to $k$. Then $I(c;k)\leq I(c;g)$, that is, information necessarily either gets lost or stays the same at each noisy step in the transmission process, but it is never ``spontaneously'' created.
\end{itemize}

There is a number of powerful theorems relating to mutual information which we will not go into here, but the interested reader is referred to the classical text of Thomas and Cover for details. 

To summarize, Shannon has shown that there is a unique, positive, assumption-free measure of statistical interdependency of two variables $c$ and $g$ that is called the mutual information and which is given by Eq~\ref{mut2}. This measure is zero if and only if the two variables are statistically independent. It does not describe \emph{what} kind of interdependency there is between the two variables, rather, it only quantifies how much of interdependency there is, in bits.

Since this is a measure over probability distributions, it is extremely generic. Before focusing on one in-depth example in Section~\ref{lec5}, let's first enumerate a number of possible applications relevant to biology:
\begin{figure}
\includegraphics[width =  \linewidth]{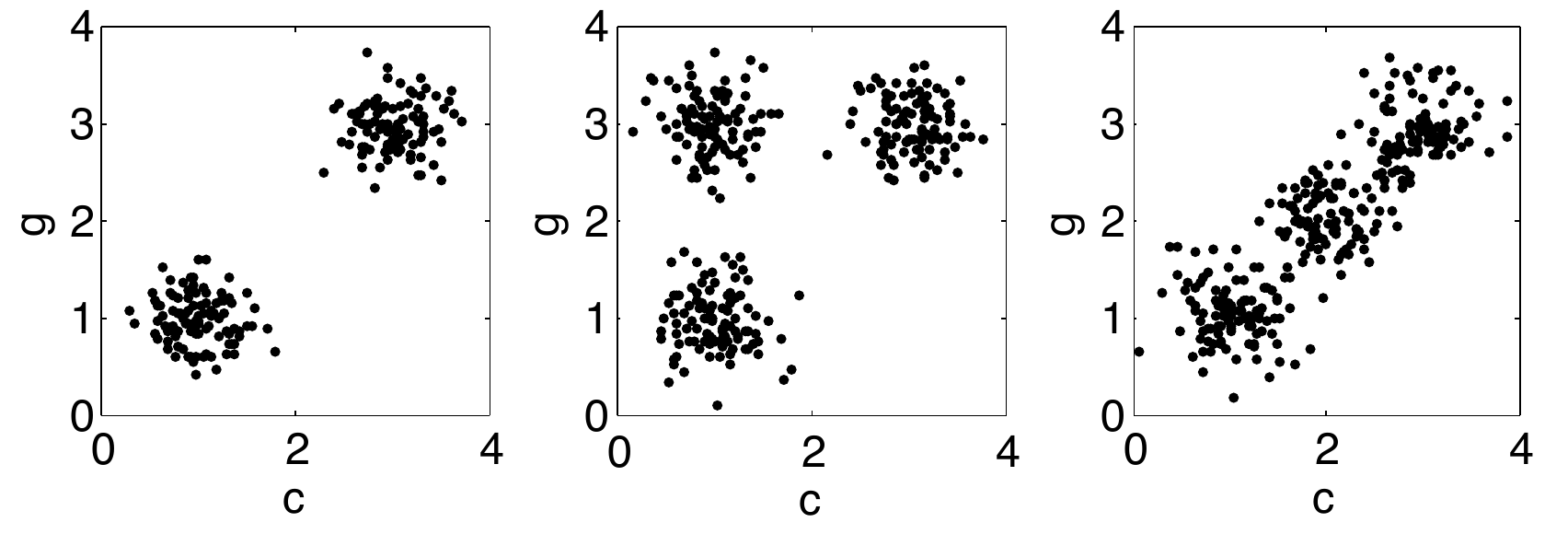}
\caption{Interpreting the mutual information values. On the left, the mutual information between $c$ and $g$ is 1 bit. This is because both $c$ and $g$ have two ``states'' (each corrupted by some noise), and the states are in a one-to-one relationship: knowing which state $c$ is in tells me about what state $g$ is in. In the middle, the connection between $c$ and $g$ is no longer trivial. When $c$ has low value, $g$ can either be low or high. The only information we have is that it is impossible for $c$ to be high when $g$ is low. This yields $\sim 0.25$ bits of information between $c$ and $g$. On the right, there are three possible states for $c$ that are distinguishable given the noise, and each maps into a unique state for $g$. The mutual information is $I\sim \log_2(3)$ bits. }
\label{f-miex}
\end{figure}

{\bf First,} $c$ and $g$ could be expression levels of different genes. The levels can be statistically correlated because the genes are regulated by a common transcription factor, because they regulate each other, or because their expression is modified coherently by some other cellular property (such as volume change). Mutual information can be used to measure the dependency between the two expression levels. In particular, in microarray experiments one measures the simultaneous expression of many (possibly all) genes, across a range of conditions, such as changes in nutrient concentrations, pH, temperature etc. Then, we can construct a similarity matrix of $K\times K$ mutual information values between all $K$ genes, computed across all experimental conditions. This matrix measures the degree to which the genes are coexpressed. Commonly, correlation coefficients of log intensity values are used for this purpose, but it has been shown that mutual information can discover nonlinear dependencies, such as that in Fig.~\ref{f-miex}, which the correlation measure will miss.

{\bf Second,} in these lecture notes we have set up an example where the transcription factor $c$ regulates the expression of gene $g$. Computing the mutual information between $c$ and $g$ then can answer, in a mathematically precise way, an interesting question: ``Are genetic regulatory elements binary switches (i.e. do they have the capacity of 1 bit) or are they able to convey more than 1 bit of information, and if so, how much?'' We explore this in Section \ref{lec7}.

{\bf Third,} one could ask about how strongly  any DNA sequence, for instance that of a TF binding site, influences the expression level of a reporter gene. This is an interesting problem that is looking for a relation between a short segment of DNA sequence of length $L$, $\vec{s}$, and the expression level $g$. Using the usual measures of dependency we might face practical difficulties because the sequences live in a space of $\vec{s}\in\{A,C,T,G\}^L$, whereas the expression levels are real-valued measured quantities, $g\in \Re$. Since mutual information is defined on probability distributions and $P(\vec{s},g)$ is a well-defined object, finding $I(\vec{s};g)$ is possible. We explore this in Section \ref{lec5}. 

{\bf Fourth,} we could ask about the information between the presence or absence of a given gene, or set of genes, and the phenotype of the organism \cite{slonim06}.

{\bf Fifth,} in visual neuroscience, neurons in the retina are sensitive to spatio-temporal correlations of light; when exposed to the preferred stimulus, the neuron might spike. How much information does a spike carry about the neural input? Again, answering this question requires us to find statistical dependency between the stimulus and a point object in time, such as a spike, and $I(\mathrm{spike};\mathrm{stimulus})$ is a natural way of quantifying this dependency.

{\bf Sixth,} one could use information as a measure over discrete objects, such as sequences. For instance, how much information does a single base-pair carry about whether the region of the genome in which the base pair is located is a coding vs noncoding region? How much do pairs, or consecutive triplets of base-pairs carry about the coding vs non-coding region? Another possible example would be to compute the mutual information between the sequence snippets of related organisms, $I(\vec{s}_1,\vec{s}_2)$, as a measure of evolutionary distance. 
\subsection{Estimation techniques: an example}
To be concrete, let's focus on a real dataset. The data was reported in Ref.~\cite{sachs+al_05}, where the authors studied a MAP kinase cascade signaling network in human immune system cells. To this end, they designed probes against specific phoshorylated (or otherwise activated) forms of 11 signaling proteins in the network; each probe was tagged with a fluorophore of a different color. Many cells were fixed and stained, and ran through the FACS machine to obtain  simultaneous readouts of the activation levels of 11 proteins, for each cell. The measurements were done in 9 conditions $\mathcal{C}_k$, $k=1,\dots,9$, and in each condition a sample of around $\sim 600$ cells was recorded. Conditions differed in the chemical environment that the cells were exposed to: in some conditions, naturally occurring ligands were presented, while in the others, artificial blockers or activators, specific for some of the signaling proteins in the network, were applied.

\begin{figure}
\includegraphics[width =  \linewidth]{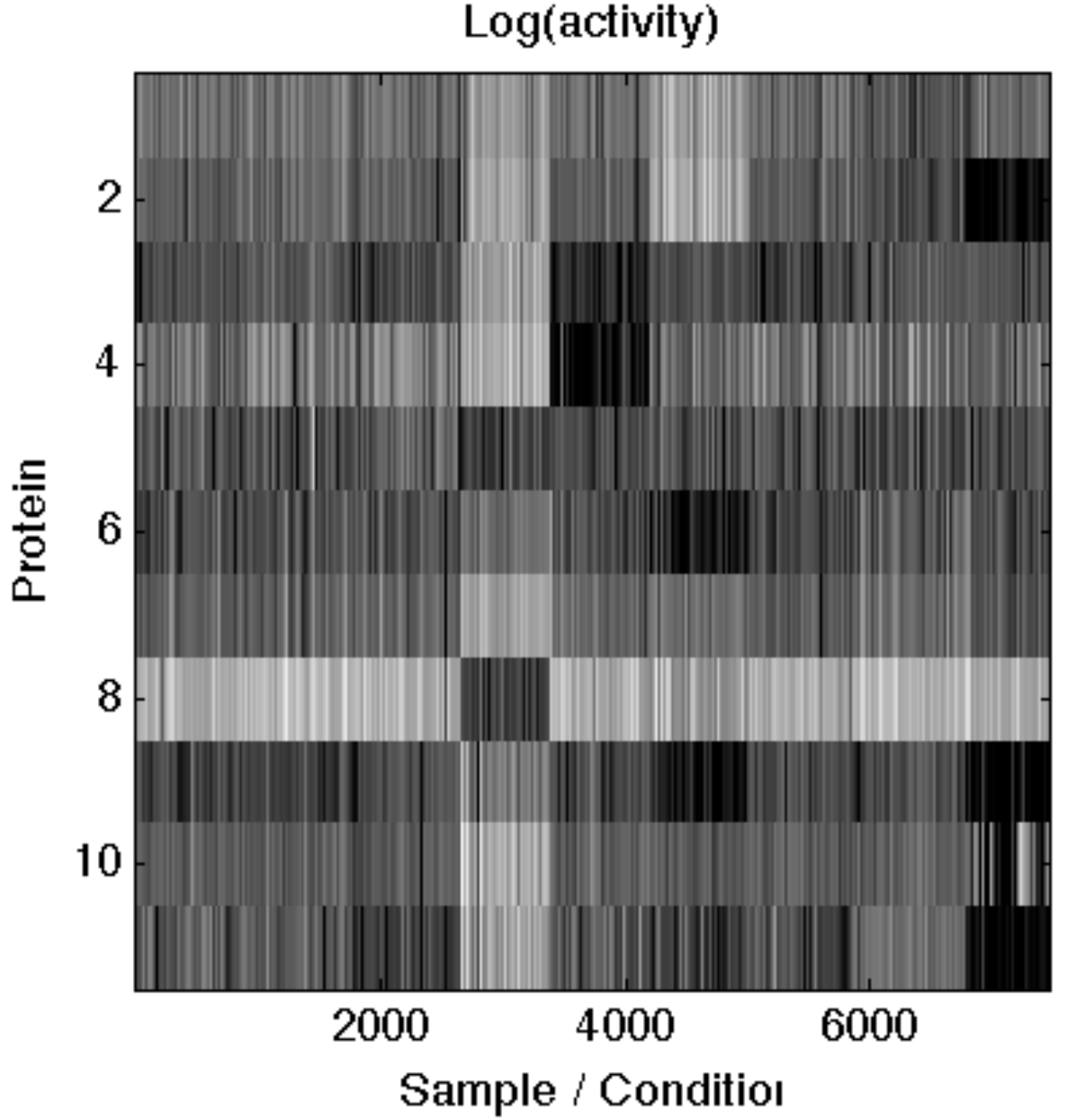}
\caption{Raw data of Ref~\cite{sachs+al_05}. The activation level (grayscale) of 11 signaling proteins (vertical axis) was recorded in $\sim 7000$ single cells. Samples are collected into consecutive blocks, for each of the 9 conditions (horizontal axis). }
\label{f-expt}
\end{figure}

For our purposes, the experiment provides us with a dataset  of $\sim 7000$ samples, where each sample is a simultaneous recording of 11 activation levels, $g_i$, $i=1,\dots,11$; see Fig.~\ref{f-expt}.

\begin{figure}
\includegraphics[width =  \linewidth]{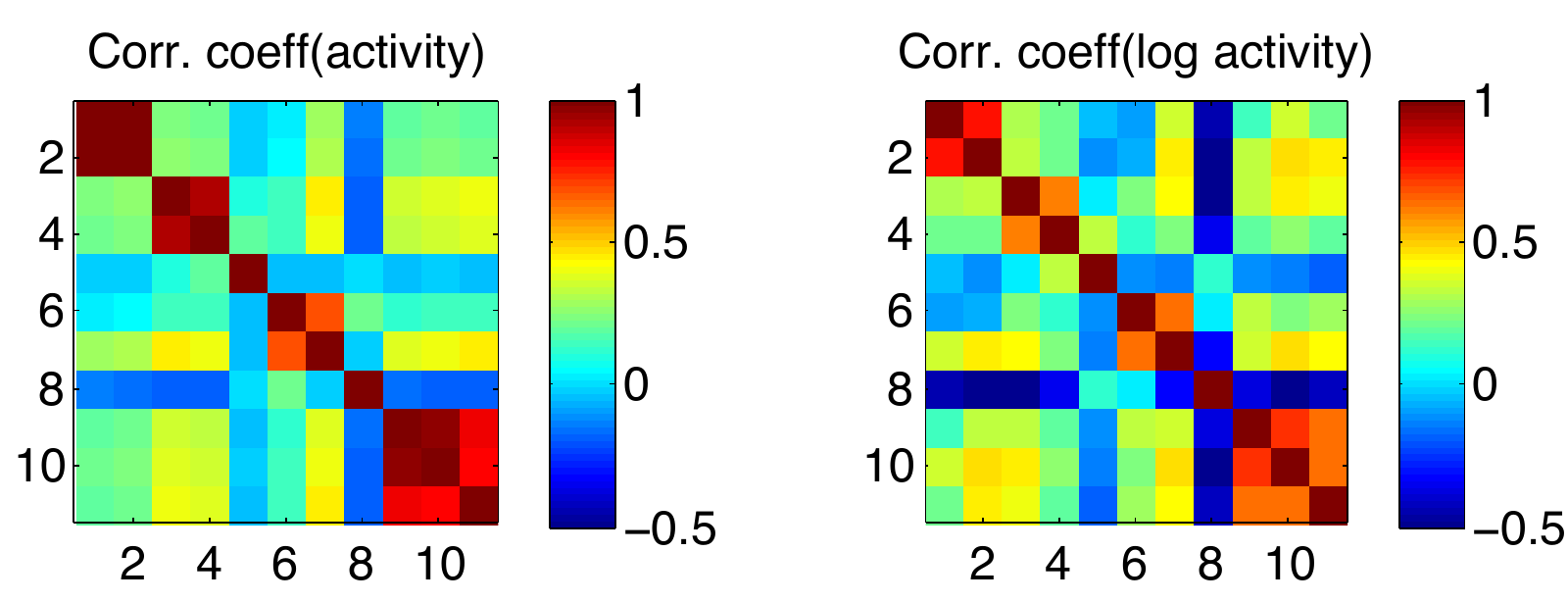}
\caption{The $11\times 11$ matrix of correlation coefficients between raw values reported in the experiment (left) and the log-transformed values (right). Note that the two matrices are significantly different. We also see that the activity of each network element is (significantly, not shown) correlated with most of the other elements. }
\label{f-pn-cc}
\end{figure}

\begin{figure}
\includegraphics[width =  \linewidth]{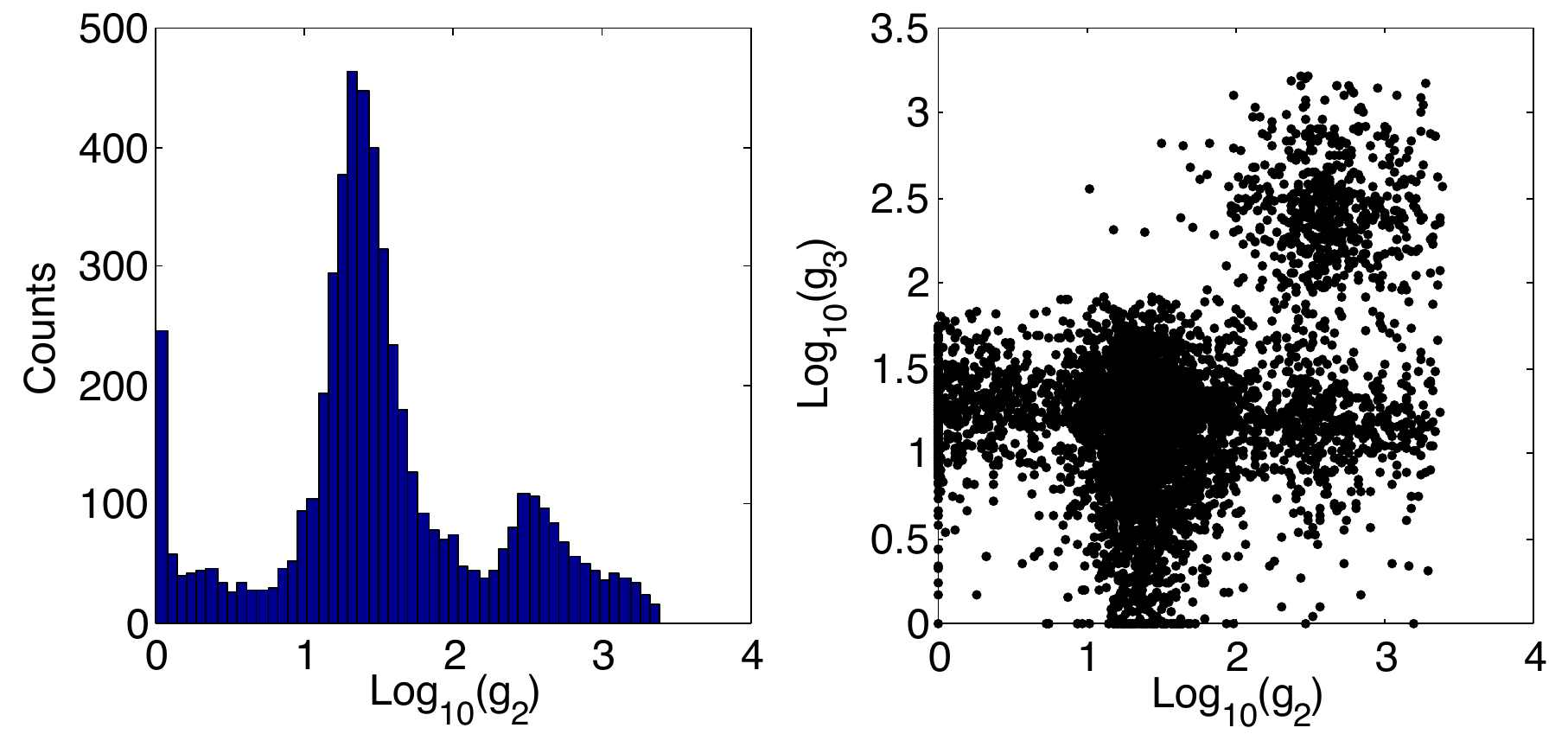}
\caption{The non-Gaussian nature of the data. On the left, the histogram of the log-activity level of signaling protein $g_2$ across all 9 conditions. We see 3 distinct ``activation states'' as peaks in the distribution. On the right, a scatter plot, across all conditions, of the simultaneously recorded levels of $g_2$ and $g_3$, showing a very nonlinear dependence.}
\label{f-pn-scatter}
\end{figure}

Let us first quantify how correlated are pairs of elements in this signaling network, across all conditions presented; this is similar to the procedure that would be used in microarray experiments. Figure~\ref{f-pn-cc} shows the pairwise correlation coefficients using two different normalization schemes.
One immediate problem that we face is that the correlation values strongly depend on the normalization of our data. More problems, however, are revealed when we look at the histograms and scatter-plots of the activities -- Fig.~\ref{f-pn-scatter} shows that the histograms $P(g_i)$ have very nontrivial, multi-peaked structure, and the scatterplots of pairs of activity levels reveal very non-linear dependence. 

Our first task will be to measure the pairwise dependencies using the mutual information measure of Eq~(\ref{mut2}). To compute the mutual information, the equation instructs us discretize both $c$ and $g$ into $b$ bins, histogram $P(c,g)$ from the data ($P(c,g)$ will be a $b\times b$ matrix), and from there  compute $I(c;g)$ in a straightforward way -- we will refer to such direct  calculation with binned and sampled data the \emph{naive} estimator with $N$ samples and $b$ bins, and denote it by $\hat{I}_{N,b}(c;g)$. Two problems, however, are in our way:  {\bf (i)} We need to discretize (bin) the data and there is a question of where to draw the discretization boundaries to gain the best statistical power; {\bf (ii)} It can be shown that mutual information is a sensitive statistic to compute; in particular, it gives a biased estimate of the true value when evaluated on probability distributions sampled from any finite dataset.  This happens because the empirical probability, $\hat{P}(c,g)$, that is obtained by sampling, has bins with small (or even zero) counts.

One of the most straightforward ways to correct for this bias is the so-called \emph{direct} estimation method, which addresses both issues (i) and (ii) simultaneously. The crucial realization was to derive the small-sample effect on the naive estimator:
\begin{equation}
\hat{I}_{N,b}(c;g) = I_{\infty,b}(c;g) + \frac{1}{N}A(b)+\dots. \label{estI}
\end{equation}
This equation says that if we hold the number of bins $b$ fixed and change the number of samples, the naive estimator and the true value (which we would get if the number of samples were infinite), differ by a bias that scales as $1/N$. We can use the equation as a prescription for getting rid of the bias: if our true dataset is of size $N_{\rm tot}$, we can subsample the data at fractions of the total size, for example at $N=0.7N_{\rm tot}, 0.8N_{\rm tot}, 0.9N_{\rm tot}, 0.95N_{\rm tot}$, many times, and compute an average naive estimator at each fraction of $N_{\rm tot}$. With these estimators in hand, we can use linear extrapolation in $1/N$ from Eq~(\ref{estI}) to obtain an unbiased estimate of information with $b$ bins, $I_{\infty,b}(c;g)$. What remains to be done, then, is to choose a correct number of bins for discretization. With too small a number, we will lose the structure in the joint distribution -- e.g. if one only discretizes into 2 levels, for instance, fine scale details in $P(g_1,g_2)$ of Fig.~\ref{f-pn-scatter} might be lost. If one discretizes into too many bins, however, then the linear correction term in Eq~(\ref{estI}) will no longer suffice to counter the sampling problems, and our estimates will be wrong.  For a reference on direct estimation technique, consult Ref.~\cite{minoam}.

\begin{figure}
\includegraphics[width =  \linewidth]{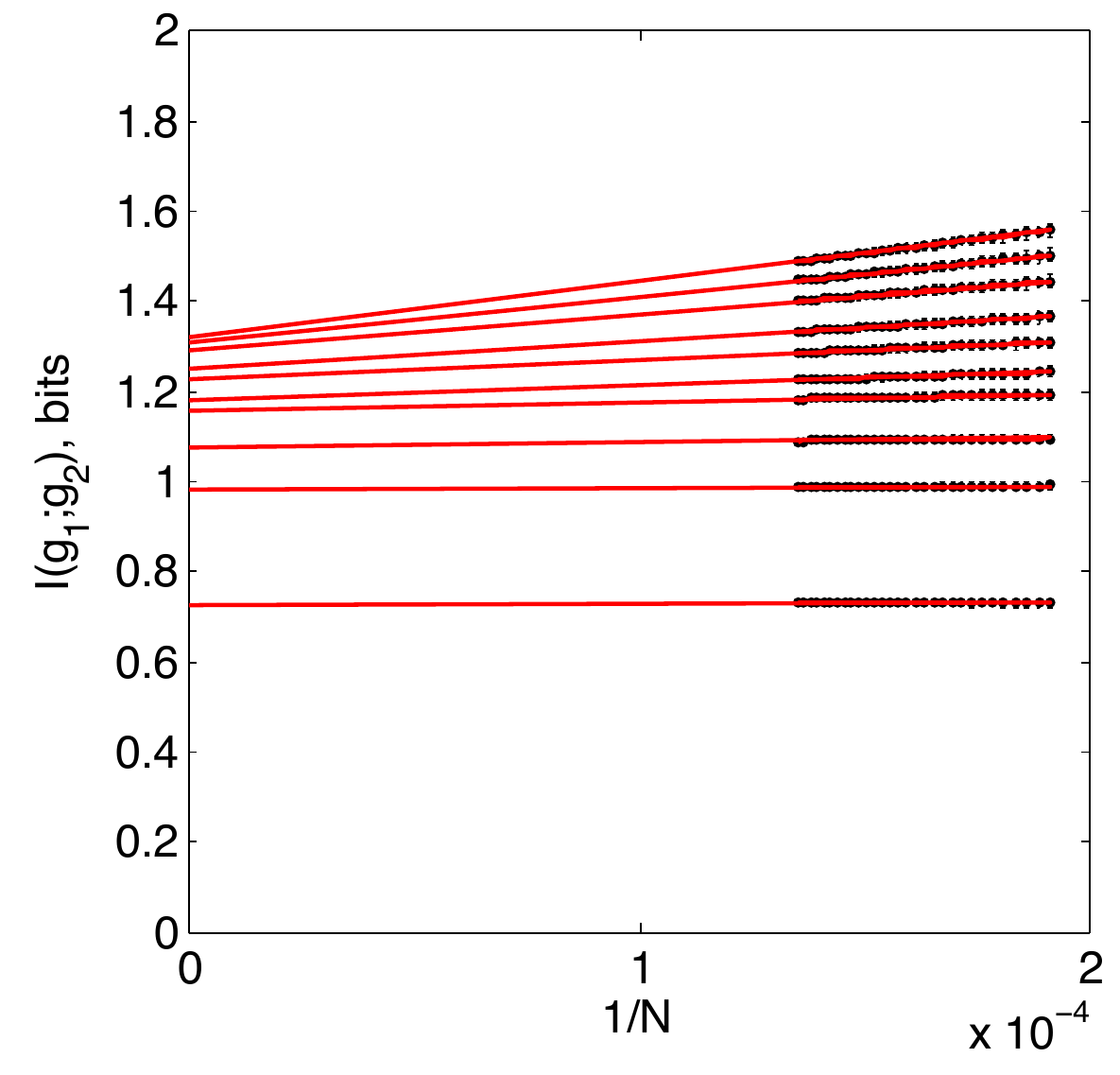}
\caption{The direct mutual information estimation procedure. This is an example where we compute the information between the activity level of signaling proteins $g_1$ and $g_2$. At each number of bins, $b=5,10,\dots, 50$ (different lines), one subsamples repeatedly the whole dataset of 5400 samples into $70\%, 71\%, \dots, 99\%$ of the samples, 500 times. For each fraction of the data, 500 naive estimators $\hat{I}_{N, b}$ are thus computed; their mean and std are plotted as black points with error bars. For each fixed number of bins $b$, naive estimates at different fractions of the data $N$ fall onto a line when plotted against $1/N$. The linear extrapolations are shown in red, and the extrapolation is from the finite dataset into $1/N\rightarrow 0$, i.e. to the limit of infinite data, which is the intercept of the red line with the vertical axis. As the number of bins is increased, we get higher and higher information estimates, until they start converging to the same value of $I(g_1;g_2)\sim 1.33$ bits, at $b=50$ bins.  }
\label{f-estMI}
\end{figure}

Figure~\ref{f-estMI} documents the \emph{direct} estimation procedure. We see that as the number of bins $b$ grows, we capture more and more information in the data, and the slope of the extrapolation line [and thus finite-size corrections of Eq~(\ref{estI})] are increasing in size. If we used $b\gg 50$ bins, the extrapolation would break down and we would lose control over information estimation.

If the data were inherently discrete, then the technique is more straightforward: one only does the $1/N$ extrapolation to correct for the sample size at the given number of bins that correspond to the number of discrete levels in the data.

\begin{figure}
\includegraphics[width =  \linewidth]{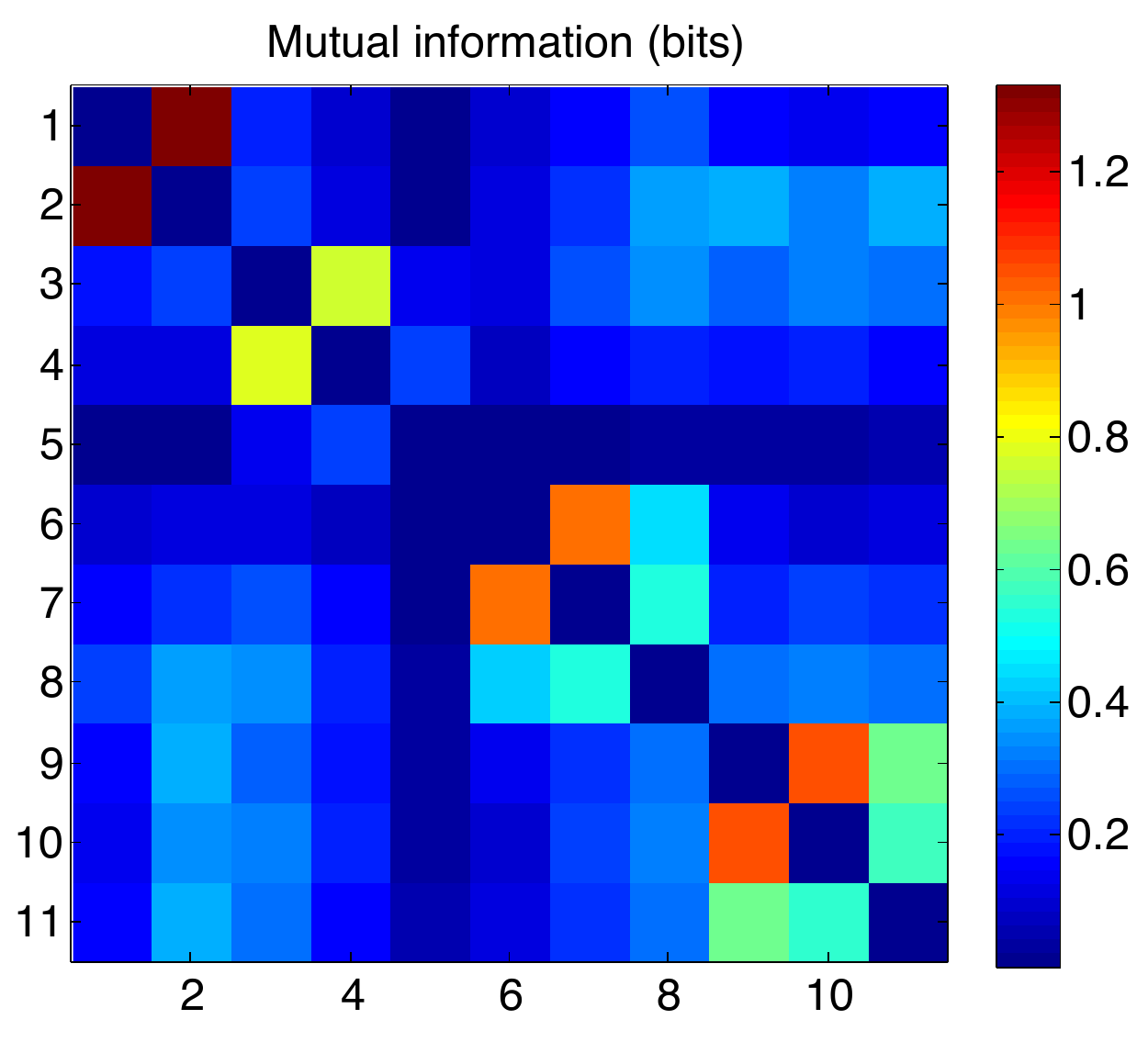}
\caption{A matrix of pairwise mutual information values $I(g_i;g_j)$, in bits (color), between pairs of activation levels of signaling proteins in the MAP cascade. The direct estimation procedure was used, with the maximal number of bins $b=50$, and extrapolation in Eq.~(\ref{estI}) was used to correct for finite sample size effects.}
\label{f-minolan}
\end{figure}

Finally, we can apply the estimation procedure to the MAP kinase signaling network dataset to extract the $11\times 11$ pairwise mutual-information matrix, summarizing the statistical dependencies between all pairs of activities $g_i$, $g_j$, see Fig.~\ref{f-minolan}. Interestingly, for example, the mutual information analysis reveals that the pair $(g_6,g_8)$ has about 0.4 bits of mutual information, yet the pairwise correlation coefficient is only $0.019$, signaling that most of the statistical dependency is not linear.

There are many other ways to estimate mutual information differing in how they handle the small-sample bias, which is the biggest technical difficulty with estimations of this sort; the naive estimators are very quick, but resamplings necessary to handle the bias lengthen the computational time considerably. Nevertheless, mutual information has been used to compute pairwise similarity matrices between all pairs of genes in high throughput experiments \cite{iclust}, and overall, this statistic is gaining in popularity  in life sciences. 
\subsection{Generalizations of mutual information}
\begin{figure}
\includegraphics[width =  \linewidth]{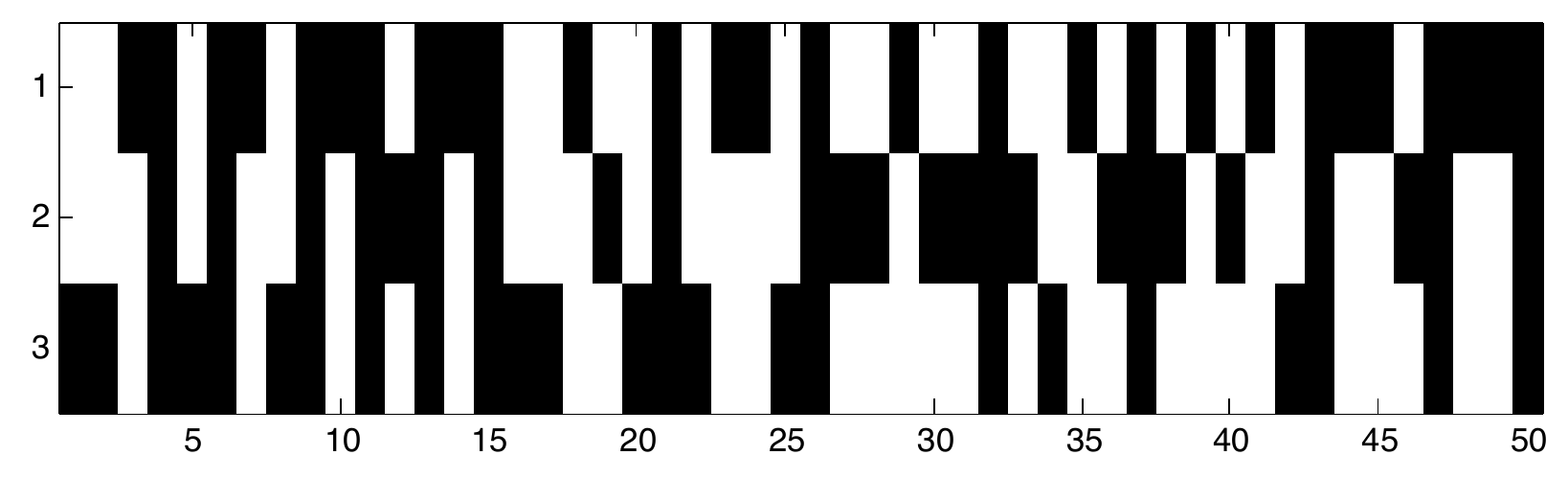}
\caption{A synthetic dataset where 50 samples for $g_1$ and $g_2$ (top two rows) are drawn independently with probability 0.5 for being 0 or 1, and $g_3=\mathrm{XOR}(g_1,g_2)$. If one looks at any pair $(g_i,g_j)$, there is no pairwise dependence, because all 4 combinations $00,01,10,11$ happen equally often. However, when one looks at the triplet $(g_1,g_2,g_3)$, there is a clear deterministic dependence of $g_3$ on the $g_1,g_2$.}
\label{f-xor}
\end{figure}

Before concluding this lecture, let us discuss two generalizations of mutual information. The first generalization extends the measure to include higher-than-pairwise structure. Suppose that we have three interacting elements, $g_1$, $g_2$, and $g_3$. There are 3 pairwise mutual informations that one can compute, $I(g_1;g_2)$, $I(g_1;g_3)$ and $I(g_2;g_3)$, describing pairwise statistical dependence. However, there might be statistical dependencies between three elements of the network that no pairwise measure can detect. The simplest example can be constructed if $g_1$, $g_2$, $g_3$ are binary variables, related by $g_3=\mathrm{XOR}(g_1,g_2)$, where the binary function XOR returns 1 exactly when $g_1$ and $g_2$ are different, and 0 otherwise; see Fig~\ref{f-xor}.

In order to detect this higher-order dependence, we need an information-theoretic measure that generalizes mutual information. This measure is called the multi-information, and is defined as follows:
\begin{equation}
I(\{g_i\})=\sum_{\{g_i\}}P(\{g_i\})\log_2\frac{P(\{g_i\})}{\prod_{i=1}^M P_i(g_i)}, \label{multii}
\end{equation}
where the numerator contains the joint distribution over $M$ elements, $g_1,\dots,g_M$, and the denominator contains the product of marginal distributions. Multi-information captures all statistical structure  between pairs, triplets, up to the complete statistical correlation between all $M$ elements. While powerful, this quantity is usually very hard to estimate because one would need to sample the full joint distribution over $M$ elements. We can, however, restrict ourselves to triplets of elements. For the synthetic XOR example we would find that the multi-information is 1 bit (because if $g_1$, $g_2$, $g_3$ were completely random, uncorrelated binary variables, they would have 3 bits of entropy; however, only $g_1$ and $g_2$ are randomly and independently drawn, while $g_3$ is a deterministic function of both, so the true entropy of the joint distribution is only 2 bits; this reduction of 1 bit is the multi-information). In case of the MAP network data, the full multi-information would require the 11 dimensional joint distribution, but we could restrict ourselves to triplets $(g_i,g_j,g_k)$ and ask about how much interdependency there is in such combinations.

Up to now, we have been looking for statistical structure among $\{g_i\}$ irrespective of the conditions, $\mathcal{C}$. We could also ask about the mutual information between the activation level of protein $i$ and condition $\mathcal{C}$, that is, $I(g_i;\mathcal{C})$, using the estimation techniques already developed. When  we consider such condition (or stimulus) dependence, there is also a new set of statistical questions that we can ask. For two signaling proteins, $g_i$ and $g_j$, we can compute  separately $I(g_i;\mathcal{C})$ and  $I(g_j;\mathcal{C})$, respectively. But we could also ask about  $I(\{g_i,g_j\};\mathcal{C})$: how much information does a pair $(g_i,g_j)$ of two activation levels together tell us about the condition $\mathcal{C}$. In principle, this can be more or less than the sum $I(g_i;\mathcal{C})+I(g_j;\mathcal{C})$. We define the quantity $R$ as:
\begin{equation}
R =  I(g_i;\mathcal{C})+I(g_j;\mathcal{C}) - I(\{g_i,g_j\};\mathcal{C}) ;
\end{equation}
when $R$ is negative, the pair of activity levels together is more informative about the condition than both levels considered separately; $(g_i,g_j)$ are said to be synergistic. Alternatively, when $R$ is positive, $g_i$ and $g_j$ are not providing independent information about the condition -- they are redundant.

Synergy and redundancy are extensively used in neuroscience to ask how groups of neurons \emph{together} encode the stimulus (stimulus in neuroscience is analogous to conditions $\mathcal{C}$ in our example), as compared to how single neurons on their own encode the stimulus. In our MAP kinase example, we find that pairs of activity levels provide redundant information about the condition. Further details can be found in Ref \cite{phd}.
\subsection{Use and interpretation}
Information theoretic measures, such as mutual information, multi-information, redundancy, synergy (and others that we don't discuss here, such as Kullback-Leibler divergence, Jensen-Shannon divergence etc) provide a very powerful, assumption-free framework for discovering statistical dependencies in the data. There exist systematic approaches that discover correlations, both linear and non-linear, between pairs of elements, triplets, quadruplets etc \cite{schneidman+al_03}; we are usually limited by the availability of data and run into sampling problems for such higher-order dependencies, but in principle they could be computed.
%
%
\section{Using information theory for inference: Transcription factor -- DNA interactions}
\label{lec5}
In our toy example of transcriptional regulation we have been assuming that transcription factors bind  to a well-defined ``binding site'' somewhere on the DNA. But what distinguishes the particular binding site -- a sequence of 10-20 nucleotides -- from all other possible short sequences in the genome? How does the TF molecule \emph{find} the correct binding site?

There is a substantial amount of existing work addressing each of the two questions which are stated more precisely below:

{\bf The ``specificity problem''} arises because the correct statistical mechanics problem for TF binding is not only that of a single (specific) site being occupied or empty, as schematized in Fig.~\ref{f-sscheme}. Instead of a single site, there is really a large number $M$ of sites on the genome, and they have a distribution of binding energies $\rho(E)$: most likely, the specific site is one of the best binders (having the most negative $E$), but there might also be some non-functional sites with low energies as well as a huge number (millions for a prokaryote, or billions for an eukaryote) of spurious non-specific sites that the TF molecule could bind weakly. Our partition function should reflect this: $Z=\sum_{i=0}^M e^{-\beta(E_i- \mu)}$, where the sum is taken across \emph{all} the sites in the genome (including the specific site that we denote as having the energy $E_0$). The real question is as follows: how does the cell make sure that the TF spends most of the time occupying the (functional) binding site, and not sitting wastefully on a large number of non-functional traps? Clearly, the binding energy to the specific site must be much stronger than to the non-specific sites, $E_{0}\ll E_i,i\neq 0$. But the sites only differ in their sequence of nucleotides $\vec{s}$, so there must exist an ``energy function'' $E(\vec{s})$ such that $E(\vec{s}_0)\ll E(\vec{s}_i), i\neq 0$. What are the energy functions for real transcription factors?

{\bf The ``search problem''} arises when we realize that even if \emph{equilibrium} occupancies were to work out and the specific site is occupied with larger probability than non-specific sites, there remains the question of the speed of equilibration. To equilibrate, TF molecules in the nucleus they must sample various sites on the genome and must therefore physically move from site to site on the DNA. Because the translocation of TFs is driven by random diffusion, this puts a computable upper bound on how quickly the sites can be sampled and how quickly the system can equilibrate. A lot of excitement was generated when it was observed that some transcription factors can find their sites faster than predicted given the 3D diffusion limit; more complex modes of TF translocation were proposed, including sliding and hopping of transcription factor molecules along the 1D contour of the DNA. It seems that a combination of 1D and 3D diffusion can reconcile the measured rapid TF search times with the theoretical expectations \cite{tfsliding}.

Here we will focus on the first problem of specificity in TF-DNA interactions. In particular, we will discuss how information theory can be used to infer the energy function, $E(\vec{s})$, for a given transcription factor whose binding has been probed in various high-throughput assays.

Why is the problem of DNA-TF interaction important? If we want to ultimately understand genetic regulatory networks, we need to know which transcription factors bind where in the genome -- in particular, which sets of genes they are regulating. The latter question can be answered if one knows the energy model for TF-DNA interaction and is in possession of the complete genome sequence. Whole genome sequences are available today, so the difficult part of this program is learning the model of TF-DNA interaction from various datasets, such as protein binding microarrays, chromatin immunoprecipitation assays, microarrays and high throughput sequencing techniques.

\subsection{Energy matrices}
We start by giving a brief overview of how the interaction between TF molecules and DNA has traditionally been described and inferred from data. We continue by pointing out the flaws in the traditional approach and show how it can generate biased models of TF-DNA interaction. We end by proposing a new information-theoretic inference method that can avoid  these problems if enough data is available.
\begin{figure}[!h]
  \begin{center}    
            \includegraphics[width=3in]{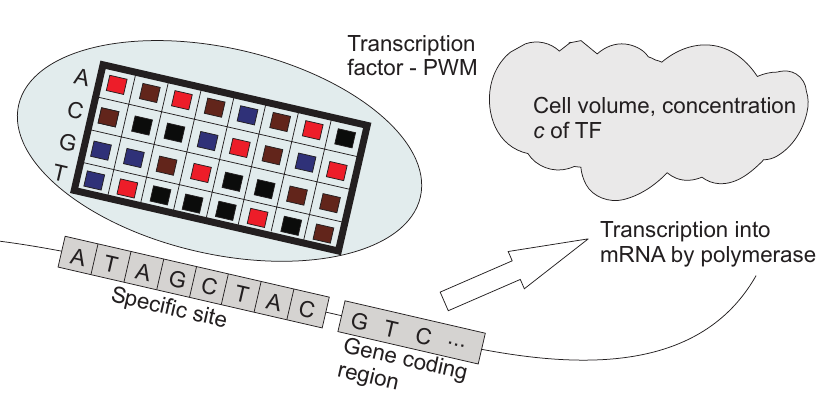}
                \caption{ A schematic representation of the interaction between a transcription factor molecule (ellipse) and a site on the DNA of length $L$ nucleotides and sequence $\vec{s}$. The energy of this interaction $E(\vec{s})$ is given by an \emph{energy matrix} $\epsilon_{ib}$ of dimension $L \times 4$, here represented as the matrix in the TF molecule. Each nucleotide $s(\alpha)=\{A,C,T,G\}$ in the sequence contributes independently to the total binding energy, which is  a linear function of sequence: $E(\vec{s})=\sum_{\alpha=1}^L \epsilon_{\alpha s(\alpha)} $. At each position $\alpha$ we look up the base at that position $b=s(\alpha)$ in the short sequence, then look up the corresponding entry in the energy matrix $\epsilon_{\alpha b}$, and add it to the total binding energy.} 
                     \label{f-pwm}
  \end{center}
\end{figure}

The simplest model of TF-DNA interaction is the so-called energy matrix model, shown in Fig.~\ref{f-pwm}. In this model, TF binds  short sequences $\vec{s}=\{s(1),s(2),\dots,s(L)\}$ of length $L$ on the DNA. Each base pair in a short sequence, $s(\alpha)=\{A,C,T,G\}$, contributes independently to the total binding energy. These energy contributions are parametrized by the energy matrix $\epsilon_{\alpha b}$ of dimension $L\times 4$, where each entry in the matrix at position $i$ specifies how much energy is contributed by a particular base $b=s(\alpha)$ at that position, as shown in Fig.~\ref{f-pwm}.

This independent energy matrix model is likely not literally true, but the number of free parameters $L\times 3$ can be small enough so that reasonable energy matrix models can be fit from available data\footnote{One can subtract a constant from each column of the matrix without affecting the binding because energies are significant only up to an overall additive constant factor. This is why the total number of free parameters is not $L\times 4$, but $L \times 3$ instead.}. Models that include higher-order contributions to the energy are more realistic, but the number of parameters explodes. Moreover, the simple model has had a number of successes in predicting TF binding sites, so we adopt it here.

\subsection{Connection between energy and position weight matrices}
When no high-throughput experiments were available, the primary data that could serve as input for inference of energy matrices were lists of known and experimentally verified binding sites. Frequently, these lists were incomplete, both because it was time-consuming to probe many candidates, and because the experimentalists preferred to set stringent criteria for a ``true'' site and thus avoid the controversial issue of possible weak, but functional, sites.

Suppose a list of $M$ known binding sites, $\{\vec{s}_i\}$, $i=1,\dots,M$, is given. Then, there exists a concise summary of TF sequence preference known as the \emph{position weight matrix}, or PWM:
\begin{equation}
w_{\alpha b}=\frac{1}{M}\sum_{i=1}^M \delta(s_i(\alpha)=b),
\end{equation}
that is, the PWM is simply a frequency table of how often a particular base $b$ appears at position $\alpha=1,\dots,L$ in the set of known binding sites. In a seminal paper, Berg and von Hippel have shown that under certain conditions there exists a simple relation between the PWM of a given transcription factor, and its energy matrix \cite{bvh}:
\begin{eqnarray}
\epsilon_{\alpha b}=-\log w_{\alpha b}, \label{bvh}
\end{eqnarray} 
up to the arbitrary energy offset in each row of the energy matrix, and the overall unit of energy (scale factor). Since this paper has appeared, the connection between PWM and energy matrix has become the cornerstone of inferring energy matrices: first, a list of known (putative) binding sites is produced, a PWM is extracted from it and the energy matrix is constructed using the equality in Eq~(\ref{bvh}), and the list of putative binding sites is then usually refined in some iterative procedure that involves experimental data. Even when the experimental techniques progressed and the data was not restricted to short lists of verified binding sites, most inference procedures still relied on the Berg-von Hippel equality. Moreover, PWM slowly emerged as \emph{the} relevant object that characterizes TF-DNA interaction, together with the picture that one must look for \emph{statistical signals} in the promoter regions of the genes that signify sequences different from some \emph{null background} expectation for what a non-functional sequence should look like. This ``statistical view,'' which regards TFs as objects that look for ``patterns in the sequence,'' can be deceiving, if one forgets that TFs are not algorithms, but physical objects, and therefore must be described by physical quantities: energy of DNA-TF interaction is such a quantity, while a PWM is not.

What are the assumptions that must be true in order for Eq~(\ref{bvh}) to hold? {\bf (i)} The true binding sites are embedded into genomic background which is large and where bases are used independently at each site; {\bf (ii)} The true binding sites have sequences that are as random as possible (maximum entropy), with the only constraint that the \emph{average} binding energy of the sites in the known sites list is fixed, $\bar{E}$. In other words, this assumption states that the only distinguishing feature of functional sites is that their binding energy is approximately $\bar{E}$, presumably lower than the other nonfunctional sites in the genome; {\bf (iii)} The concentrations of TF are not such that some sites would be fully saturated; {\bf (iv)} The known sites list is complete, or at least an unbiased sample of the true binding sites.

What role do these assumptions play in combination with modern data sets when inferring TF energy matrices?
\begin{figure}[!h]
  \begin{center}    
            \includegraphics[width=3.5in]{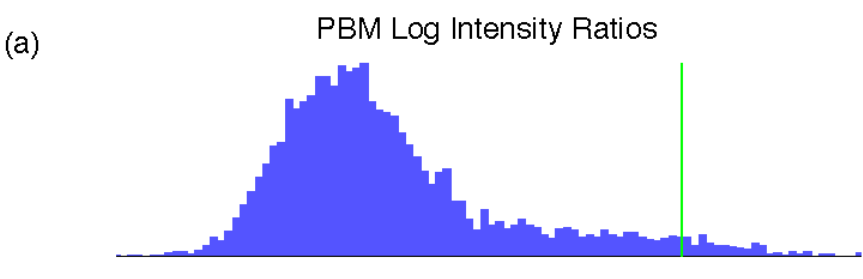}
                \caption{Experimental results from PBM (protein-binding-microarray) experiments of Mukharjee et al \cite{mukharjee}. In PBM assay, intergenic double-stranded DNA of yeast was spotted onto the array. Transcription factor of interest, Abf1p in this case, is introduced to the array and left to bind; there should be higher probability of binding somewhere in those intergenic regions  that contain the actual Abf1p binding sites. The TFs are fluorescently tagged. Each spot in the array, corresponding to one particular intergenic region in yeast, therefore provides a light intensity readout related to whether Abf1p is bound somewhere in that region or not. In this plot, the histogram of log light intensities for every spot in the PBM chip is shown, together with the experimentalist's stringent cutoff (green vertical line) that separates true positives (intergenic regions with TF bound) from the rest; the cutoff was chosen to minimize the amount of false positives. } 
                     \label{f-mukherjee}
  \end{center}
\end{figure}
Figure~\ref{f-mukherjee} shows data from one high-throughput experiment that can probe simultaneously the binding of a single transcription factor to all intergenic regions in yeast (see figure caption for details). Usually the analysis proceeds by thresholding the dataset to isolate sequences in which there are binding sites (true positives) and to minimize those sequences mingled in that do not have a binding site yet pass the threshold (false positives). Data above threshold is retained, while data below it is discarded. With the data above threshold in hand, some model is assumed that links the direct experimental readout (e.g. the light intensity in the PBM chip) with the putative binding of the transcription factor somewhere in the corresponding intergenic region. Often, an initial guess will be made for the energy matrix, which will then be scanned across the sequences passing the cut to identify possible `hits' in those regions, i.e. sites of length $L$ that score well with the assumed energy matrix. The predicted hits will give rise to predicted light intensity, which can be compared to the true intensity, and an update in the energy matrix guess can then be made. Most often, the iterative step will make use of Eq~(\ref{bvh}) to improve the guess of the energy matrix.

Our motivation for devising a new method for inferring TF-DNA interactions was based on the following observations about the existing approaches: 

{\bf (i) The assumptions underlying the Berg-von Hippel equation that links PWMs with energy matrices might not hold.} 

In particular, the genomic background is \emph{not} always simple, as has been amply demonstrated by failed attempts to identify transcription factor binding sites in \emph{Plasmodium falciparum}, the malaria parasite; this parasite has non-coding regions with a lot of statistical structure, including complicated repeats, and simple models of genomic background fail to capture these dependencies, leading to a failure in algorithms that exploit Eq~(\ref{bvh}) to search for binding sites \cite{fire}. Moreover the concentration of TFs in the experiment may be saturating for some sites, which is problematic for the original Berg-von Hippel formulation, but has been addressed by, e.g. Ref~\cite{senguptaetal}. 

{\bf (ii) Discarding most of the experimental data (e.g. measurements that lie below the threshold of light intensity in the PBM example) is wasteful.}

The experimental observation that the TF \emph{does not} bind in some intergenic region is in fact as informative as the observation that it does bind in certain other regions. In other words, these ``negative'' samples inform us about what the energy matrix cannot be (because with a wrong model, we could predict binding in those regions that indeed do not show any binding experimentally), and thus are informative about the model as well.

 {\bf (iii) We don't know the ``error model'' of the experiment.} For a principled, unbiased inference of any model from the data (including TF energy models), we would have to write down the likelihood of the observed data given the model, $P(\mathrm{data}|\mathrm{model})$. To do proper Bayesian inference, we would then maximize this likelihood with respect to the model parameters (e.g. the energy matrix)\footnote{More precisely, we could include a prior and maximize the posterior, which would be the really correct way of doing things, but this does not change the gist of our discussion.}. However, in most of the experiments, we have no idea about what form the error model -- the probability $P$ -- has. That is because experiments such as binding arrays, microarrays etc usually  involve many complicated biochemical and detection steps, such as hybridization, washing, sonication etc, and therefore there is no principled way of writing down the probability of raw experimental readout (i.e. light intensity) given that the TF is or is not bound. Without the knowledge of $P$ and the subsequent inability to do proper Bayesian inference, most researchers have resorted to ad hoc approaches, such as setting thresholds. Moreover, these ad hoc approaches usually contain many algorithmic details, involving data normalization and representation (e.g. does one analyze light intensity, or log light intensity), that can strongly influence the results. We would like to reexamine the premise that unbiased inference is impossible without the explicit knowledge of $P$.
 
 {\bf (iv) Analysis of the same TF using different methods often gives inconsistent results.} 

\begin{figure}[!h]
  \begin{center}    
            \includegraphics[width=3.5in]{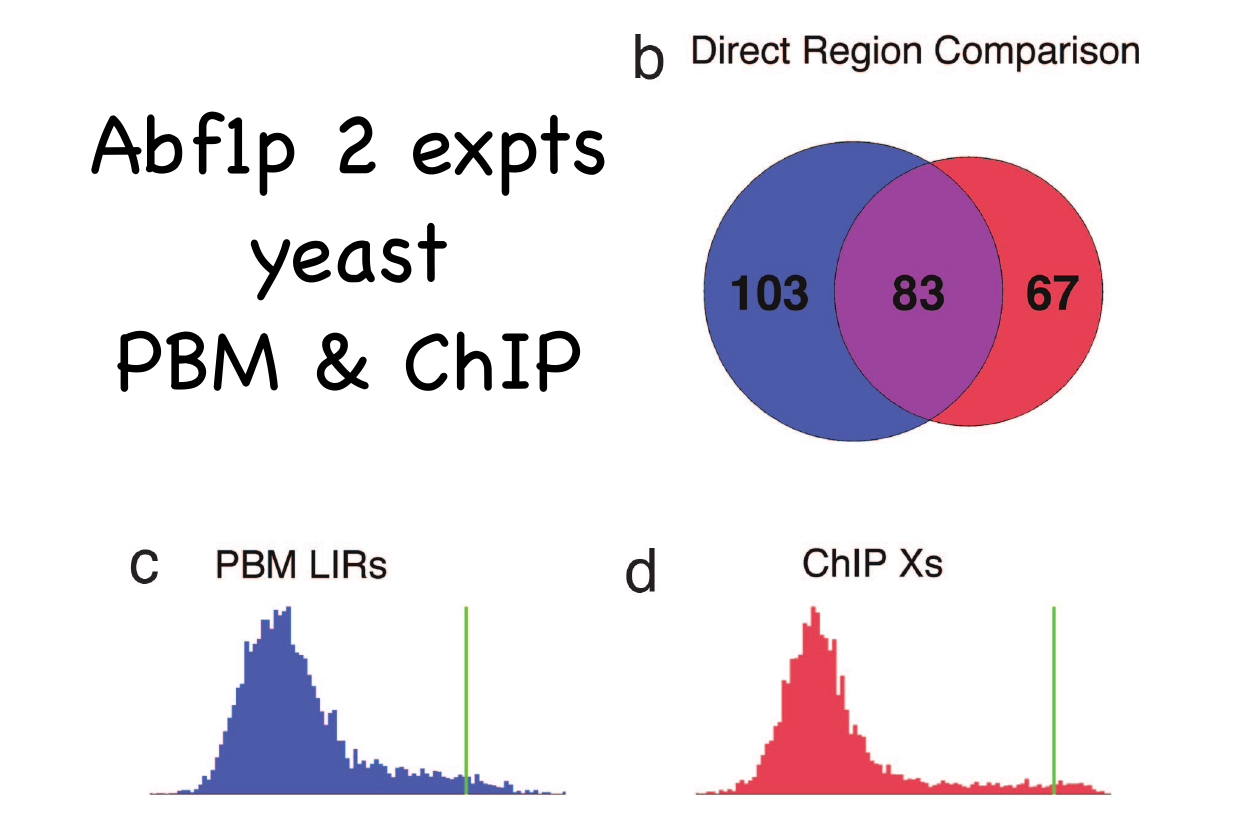}
                \caption{Comparison of experimental results about the binding of Abf1p in yeast using two assays, protein binding microarrays (PBM) from Ref.~\cite{mukharjee}, and chromatin immunoprecipitation (ChIP chips) from Ref.~\cite{lee}. In the bottom, the raw experimental results are shown with the thresholds determined by the experimentalists. The circles show the number of bound regions identified by the two experiments along with their intersection. Figure reproduced from Ref.~\cite{justin1}.} 
                     \label{f-pbmvschip}
  \end{center}
\end{figure}

As Fig.~\ref{f-pbmvschip} shows, despite best efforts of the experimentalists to select regions that are true positives, the intersection of regions declared bound by PBM and ChIP-chip experiments is rather low. While such inconsistent results are not rare in the field, it is not clear how to interpret the inconsistency: as the experimental imprecision, difference in experimental conditions (including the state of the yeast culture), biased inference of the bound regions, biological noise, etc\footnote{Generally, playing with the thresholds in order to maximize the intersection does not work well either.}.

{\bf (v) We would like to put proper error bars on any inferred model.}
\\

In a paper by Kinney et al \cite{justin1}, we argued that some of these problems are related: for instance, if we could compensate properly for our lack of knowledge about the error model, perhaps the inferred binding sites from different experiments could be made consistent; likewise, if we did not need to assume anything about the statistics of the background sequence, perhaps the existing experiments would reveal binding sites in \emph{Plasmodium} parasite. In the next section, we present an information-theoretic approach to modeling transcription factor -- DNA interactions that will address some of these concerns.

\subsection{Mutual-information based inference of TF energy matrices}
We start by representing data from a typical high-throughput experiment in a new form. Suppose that a high-throughput experiment probes sequence fragments $\vec{s}_i$. These fragments can be longer than the suspected binding site size, which we assume is of length $L$; for example, sequence fragments $\vec{s}_i$ could be intergenic regions in yeast, as in Ref.~\cite{justin1}. We don't know where in these sequences the binding sites are, if they are present at all, or how many binding sites there might be for the TF of interest.

The experiment provides us with a raw experimental readout that corresponds to each intergenic region. In protein-binding microarrays, this is the light intensity level that is correlated with the probability that a fluorescently tagged TF is bound in that intergenic region. Similarly, we can use the ChIP arrays. But the framework is more flexible: \emph{any quantity, either continuous or discrete, that is experimentally accessible and is thought to correlate with the binding of TF of interest, may be used; this in principle includes a combination of such measured quantities}. To be concrete, consider a set of microarray experiments where expression levels of genes are probed at many different conditions. These results are then used as input to clustering, and genes are grouped into co-regulated clusters. For example, there is a group of genes that is determined by clustering to be in group $\mathcal{G}$. We can now assign to every intergenic region a value 0, if the corresponding downstream gene is not in $\mathcal{G}$, and value 1, if the corresponding gene is in $\mathcal{G}$. Instead of PBM or ChIP experimental data, in this case we would use the result of the clustering partition as the variable (binary in this case) that correlates with the presence or absence of a TF binding site.

Before we proceed, we will bin (or discretize) the data into a number of discrete bins, if the data is continuous. This discretization is in principle arbitrary, and for best results one should make a tradeoff between increasing the number of bins (to increase the resolution of the method) and the sample size (so that the number of samples in each bin is large enough). In PBM experiments, for instance, the log light intensity levels are discretized into $50,100,\dots,200$ bins (to show that the results will not depend on this discretization). The bins group together log light intensities that are similar. We will use this discretized data to estimate mutual information; as shown in Section~\ref{lec4}, mutual information is insensitive to monotonic transformations of data, and this is exactly what we want -- our inference will be insensitive to whether we use log light intensity, raw light intensity, or any nonlinear function thereof. We have also mentioned that in the limit of large enough data set, the way we discretize does not influence the computed mutual information, so our method is robust with respect to this choice\footnote{In practice, we do need to worry about the small-sample effects, but in cases that will be discussed these can be addressed.}.

To summarize, the input for our analysis will be pairs of (sequence, data bin), where sequences may or may not contain binding sites for the TF of interest, and the data is any experimentally determined quantity that is though to have a statistical dependence with the presence or absence of TF binding sites. In particular, after the discretization, our analysis does not even require the raw data values any longer, just the \emph{data bin} into which the data was assigned on discretization. Let us  denote the input data $\mathcal{D}$ mathematically as $K$ pairs indexed by $i=1,\dots,K$ pairs: $\mathcal{D}=\{ (\vec{s}_i, z_i)\}$, where $\vec{s}_i$ are the sequences potentially containing the binding sites, and $z_i$ are the (discrete) bins associated with those sequences and summarizing the experimentally measured values that correlate with TF binding.

We now explain the core of the argument on which our inference is based. We assume that the experimental results $z_i$ are some unknown function of the sequence $s_i$, but the statistical dependence between the two can only be established through binding energy $E$ of TFs to sites in the sequence $s_i$. Suppose that there are many subsequences in $s_i$ of length $L$, where the TF \emph{could} bind. If I knew the correct energy matrix  $\epsilon$ for the TF, I could evaluate the energy of the TF to bind onto each site in $\vec{s}_i$. In the simplest case, I could declare that if any of these binding energies is below some threshold, the TF will bind in that intergenic region (at least once), and I would declare that intergenic region bound, $x_i=1$. If there is no such sequence of length $L$ in $\vec{s}_i$, then I would declare the intergenic region unbound, $x_i=0$\footnote{The details of how to deal with multiple bound sites in the same region, or whether to use a hard threshold to declare a site being bound or a soft threshold where ``being bound'' is a real number denoting probability, can be looked up in Ref.~\cite{justin1} and its supplement. Interestingly enough, if mutual information is used to infer the energy matrices $\epsilon$, the matrices will be largely independent of these assumptions. Here, concretely, a simple rule was used: apply $\epsilon$ to any site of length $L$ within the region $\vec{s}_i$. If the energy of binding is favorable, i.e. below threshold $\Theta$, declare that site bound, and declare the whole region bound, $x_i=1$. If none of the energies is below the threshold, the region is unbound, $x_i=0$. The matrix $\epsilon$ can only be determined up to a scale, i.e. $\epsilon_{ib}$ and $\lambda \epsilon_{ib}$, where $\lambda$ is a positive scalar, are both equally good guesses (in this setup it can be shown that the experimental data cannot predict the absolute scale of the matrix in physical units of, say, joules). We can get rid of this arbitrariness by fixing the threshold for binding, $\Theta=1$. }.

Formally, the argument we are making can be represented as follows:
\begin{equation}
\vec{s}_i\rightarrow x_i=f(E(\vec{s},\epsilon))\rightarrow z_i, \label{condind}
\end{equation}
that is, the sequence determines the energy of binding (that depends on the energy matrix $\epsilon$ which we would like to infer), and the energy determines whether the region is bound or not $x_i=0,1$; that \emph{alone} determines the experimental data bin $z_i$ reflecting the measurement. There is no other statistical dependence between the sequence and the experimental data, \emph{on average}, than through the binding energy\footnote{In each particular sequence, there might be other statistical dependencies, such as binding of other transcription factors, chromatin influences etc, but those unobserved influences will act as noise in our inference. Since we are not assuming any noise model, they should not bias the inferred energy matrices.}!

\begin{figure}[!h]
  \begin{center}    
            \includegraphics[width=2.5in]{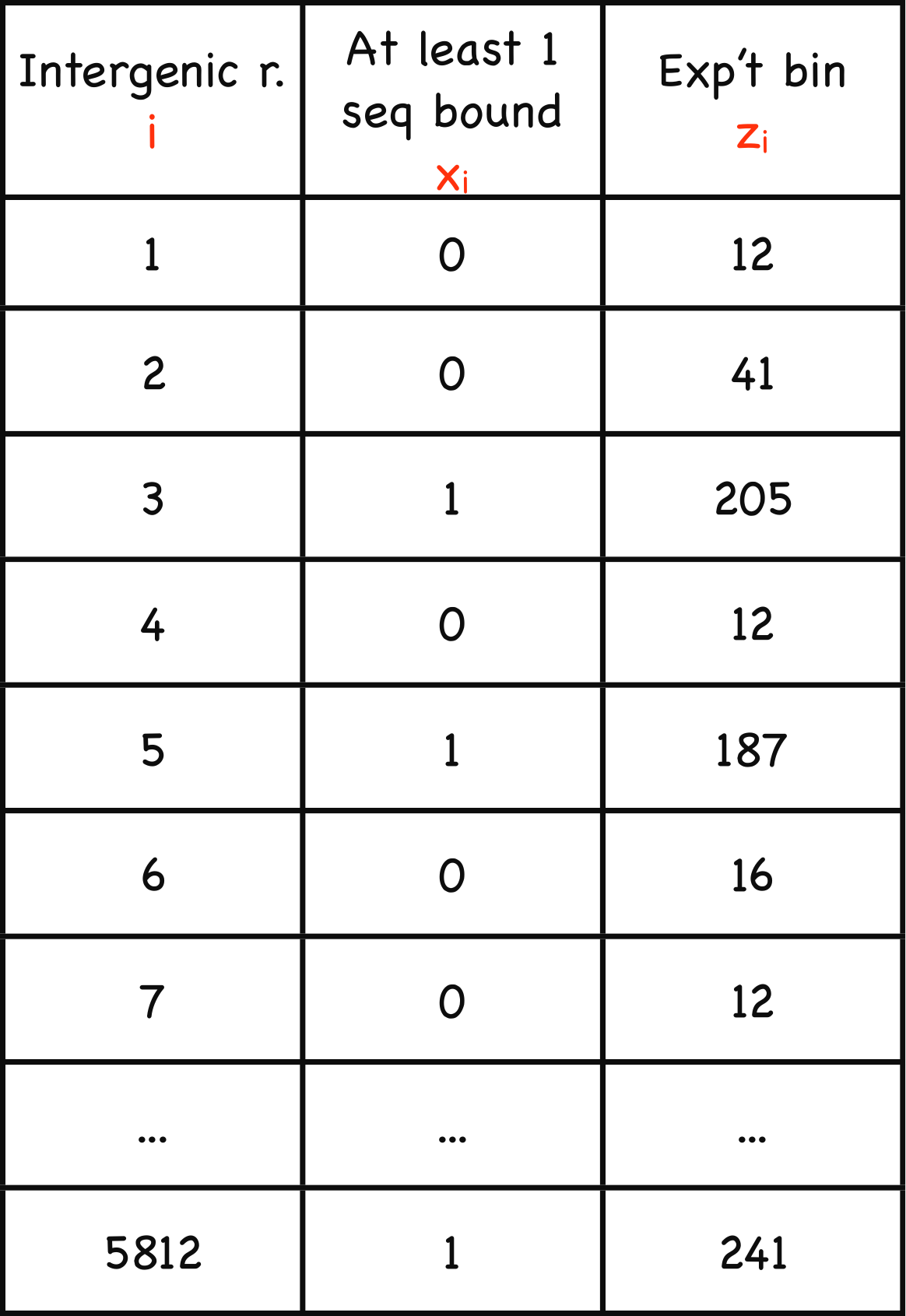}
                \caption{An example dataset for inferring the TF energy matrix.  Intergenic regions in yeast are ordered, $i=1,\dots,5812$. If we make a guess at the energy matrix $\epsilon$, we can decide whether each intergenic region is bound or not (see text), and assign $x_i=0,1$ to each region. In addition, each region has produced some experimental readout, that has been discretized and so each region is assigned to one of the bins $z_i=1,\dots,250$ (maximum number of bins is 250 in this example). The table illustrates  that there is some statistical dependence between the region being bound ($x_i=1$) and being assigned into a bin corresponding to e.g. higher light intensities on the PBM chip (higher corresponding bin number $z_i$). This table can be constructed if some energy matrix $\epsilon$ is assumed, because $x_i$ is a function of sequences $\vec{s}_i$ and $\epsilon$.} 
                     \label{f-mitable}
  \end{center}
\end{figure}

With these remarks in mind, a typical representation of a dataset might look as shown in Fig.~\ref{f-mitable}.

If the experimental result depends only on binding, and the binding depends on the sequence through binding energy alone as in Eq.~(\ref{condind}), we can formulate the following inference principle:
\begin{equation}
\epsilon^*=\mathrm{argmax}_\epsilon I\left(x_i(\epsilon, \vec{s_i}); z_i\right), \label{miinfer}
\end{equation}
where $\epsilon^*$ is the energy matrix that we are looking for, and argmax is returns that $\epsilon$ that maximizes the mutual information $I$. Let us now parse this equation in detail. 

We believe that the experimental results $\{z_i\}$ should be maximally dependent on whether the corresponding regions are bound or not, $\{x_i\}$. To characterize the full statistical dependency without making any assumption about the probability distribution $P(z_i|x_i)$, we compute the mutual information $I(x_i;z_i)$. Remember that this independence of any error model was one of our initial motivations. From a table like the one in Fig.~\ref{f-mitable}, $I(x_i;z_i)$ can be directly computed, by simply accumulating the joint probability distribution $P(x_i,z_i)$ across all $K$ pairs (sequence, data) [K=5812 in the example], and using Eq.~(\ref{mut2}) to compute the information\footnote{This can be corrected for sampling errors, as discussed in Section~\ref{lec4}}.

Information $I(x_i;z_i)$ can be computed for any choice of the energy matrix $\epsilon$. We should now search the space of all energy matrices ($3\times L$ parameters) for that matrix $\epsilon^*$ that will maximize this information. While nontrivial, this search can be implemented using Metropolis Monte Carlo (MMC) methods. This search does not only yield the best matrix $\epsilon^*$, but \emph{an ensemble} -- a solution set -- of matrices, all of which explain the data almost equally well and yield the same value for information $I$. We will not go into the details here beyond stressing that since we end up with a set of good solutions, we can compute how well constrained the energy matrix elements $\epsilon_{\alpha b}$ are and put rigorous error bars onto them.

\begin{figure}[!h]
  \begin{center}    
            \includegraphics[width=3.5in]{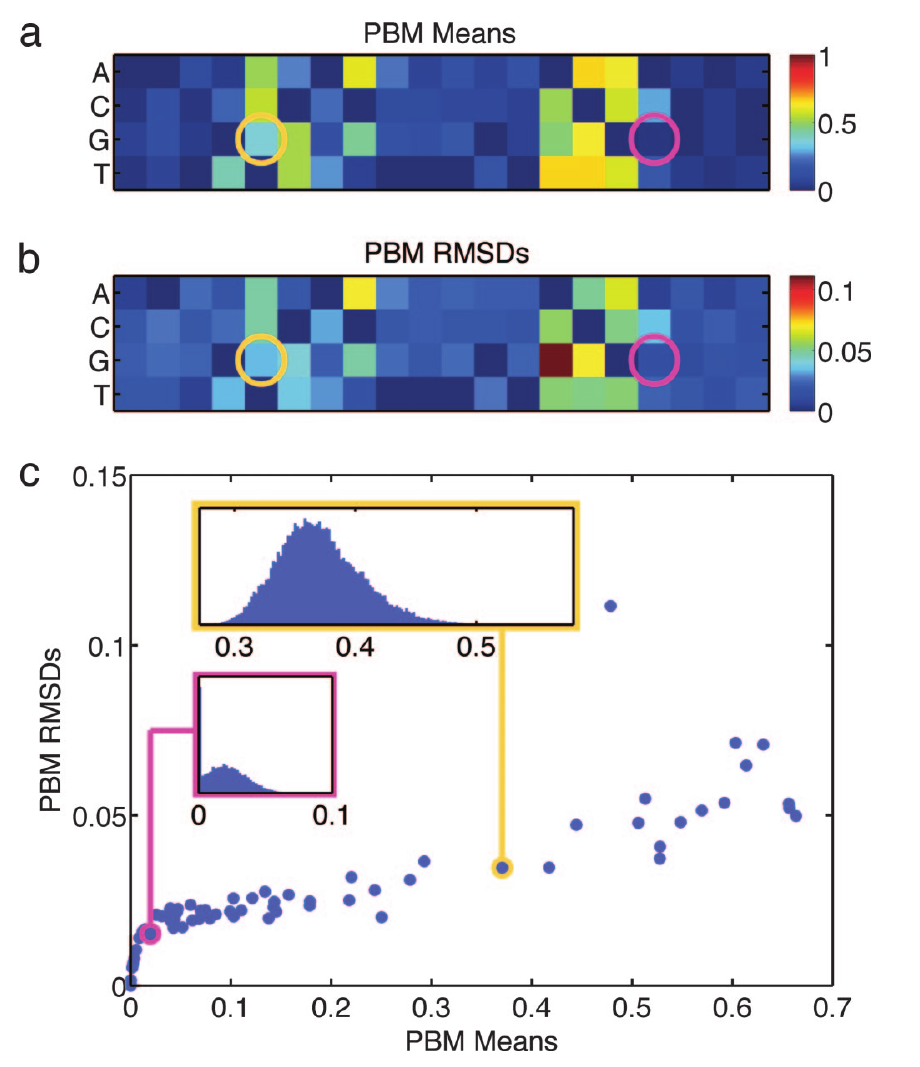}
                \caption{Results for Abf1p energy matrix in yeast inferred from PBM data from Ref.~\cite{mukharjee}; figure  reproduced from Ref.~\cite{justin1}. Panel a) shows the energy matrix $\epsilon_{\alpha b}$, with hotter colors indicating larger energy matrix contributions, and darker colors smaller contributions. Panel b) shows the error bars on each of the energy matrix elements, $\sigma_{\epsilon_{\alpha b}}$. Panel c) plots the matrix elements vs their errorbars (y axis), and also shows in the two example insets, the distribution of values for two particular energy matrix elements in the solution ensemble found by MMC optimization. Here we looked for an energy matrix of length $L=20$ basepairs. The search can be repeated for $L=14,16,18$ to show that the core elements remain unchanged, and the same energy matrix is found each time. The inference can also be performed by splitting the total dataset (5812 sites) into random halves, and showing that on each half consistent energy matrices are found, so that there is no overfitting [not shown, c.f. Ref.~\cite{justin1}].} 
                     \label{f-pbmmx1}
  \end{center}
\end{figure}

How do the results look like for the yeast Abf1p example inferred from PBM data? Figure~\ref{f-pbmmx1} is reproduced from Ref.~\cite{justin1}, and it shows the inferred energy matrix and the error bars for each of the matrix elements. We see that most of the energy matrix elements are constrained very well by the data; a pattern emerges where Abf1p makes contact to the DNA in two regions, with small (but still significant) energy contributions from the basepairs between the two regions. These results are broadly consistent with, but more precise than, previous energy models for Abf1p.
\begin{figure}[!h]
  \begin{center}    
            \includegraphics[width=3.3in]{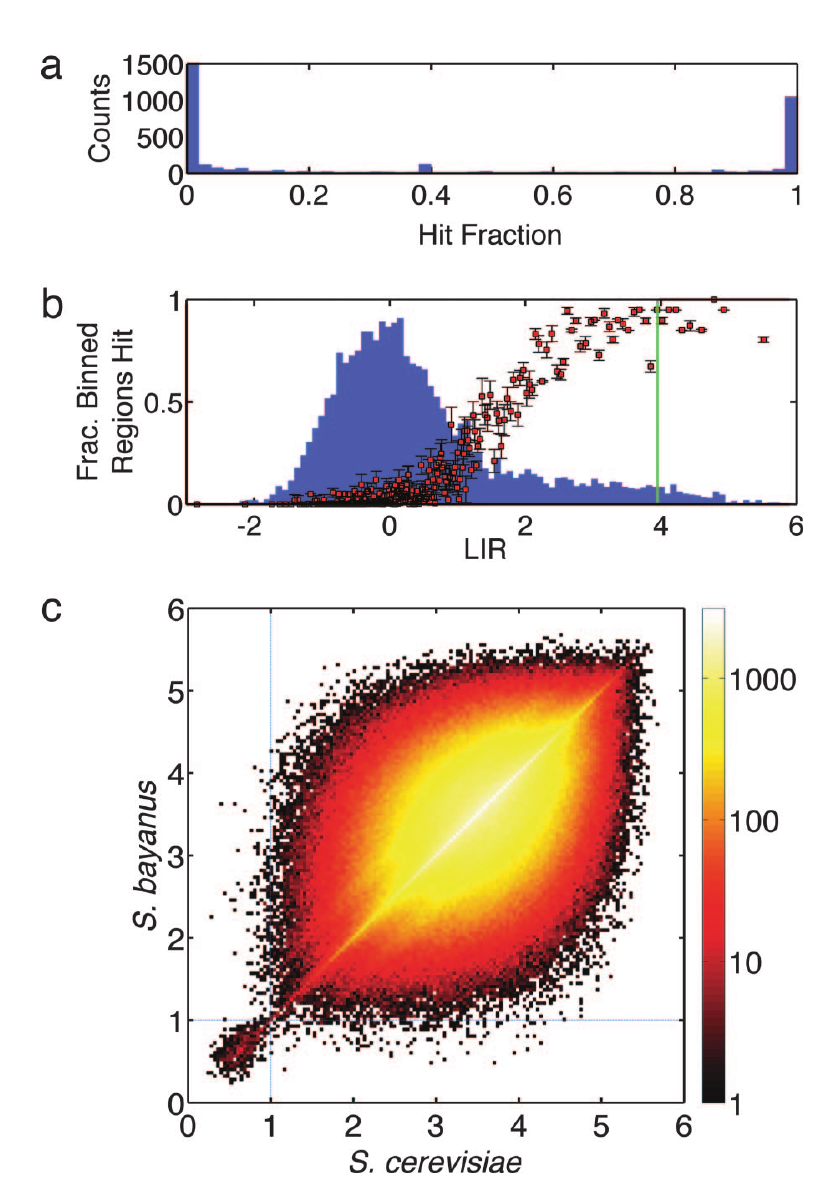}
                \caption{Further results for Abf1p energy matrix in yeast inferred from PBM data from Ref.~\cite{mukharjee}; figure  reproduced from Ref.~\cite{justin1}. Panel a) shows how $K=5812$ intergenic regions of yeast are categorized as bound $x_i=1$ or unbound $x_i=0$ by energy matrices in the solution set. Plotted is the histogram of the ``hit fraction'', $\langle x_i\rangle_{\epsilon}$, i.e. the average value of $x_i$ for each region across all solution energy matrices. The regions clearly separate into two classes: most of the regions are declared unbound (HF$\sim 0$) by all energy matrices (the histogram bar corresponding to HF$=0$ would extend beyond 1500 counts in the plot, but is cut for clarity), and a smaller set of regions that is declared bound (HF$\sim 1$), by most of the energy matrices. Panel b) shows the histogram of raw experimental data in blue, the experimentalists' threshold (green line), and the \emph{inferred} $P(x_i|z_i)$ (sigmoidally shaped scatterplot with errorbars). The sigmoidal shape has not been assumed, but is the result of our inference. Panel c) shows the energies assigned to sites of length $L$ in every intergenic region in \emph{S. cerevisiae} (the yeast species on which the analysis was run) vs the energies assigned to orthologous sites in a related yeast species, \emph{S bayanus}. Those sites that are declared bound (below threshold $\Theta=1$) in \emph{S cerevisiae} also have correspondingly low energies in the other yeast species, forming an island of points in the lower left corner of the plot. For sites above the threshold which are non-functional, the correlation between the energies in the two species is much weaker. } 
                     \label{f-jbk}
  \end{center}
\end{figure}

Figure~\ref{f-jbk} shows that the resulting energy matrices unambiguously split all intergenic regions into those that are bound and those that are not. More surprisingly, as part of our results (not as an assumption!), we also learn $P(x_i|z_i)$, the probability that the site is bound given the experimental data bin, $z_i$; this quantity is related to the error model $P(z_i|x_i)$, which we did not want to assume a priori. This curve has a sigmoidal shape, showing that no single hard threshold (as has traditionally been done) will perfectly separate regions that are bound from those that are not. This finding also has biological implication -- it is saying that there are both strong and weak binding sites, and some of the weakly bound sites are mixed into experimentally determined bins with regions that truly don't contain binding sites. Lastly, Fig.~\ref{f-jbk}c shows that our inferred binding energies are conserved across two yeast species despite significant difference in aligned sequences due to evolutionary distance. 

\begin{figure}[!h]
  \begin{center}    
            \includegraphics[width=3in]{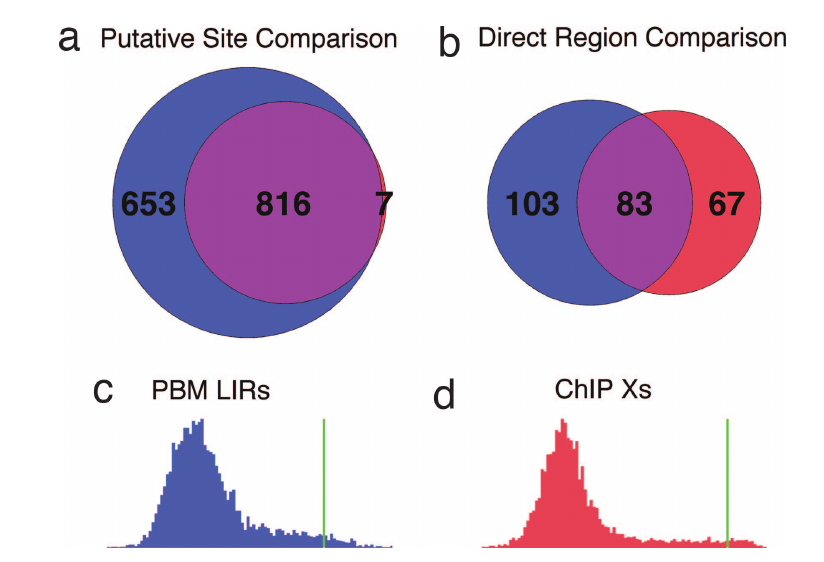}
                \caption{Lower row: raw data from PBM and ChIP experiments on Abf1p, along with experimental thresholds. An intersection of regions that pass the threshold in both experiments shows large inconsistencies, as shown in Panel b). Panel a) shows the intersection of binding sites inferred using the mutual-information method presented here. More sites are found, and the sites found in the ChIP experiment are almost a perfect subset of the PBM-identified sites. Figure reproduced from Ref.~\cite{justin1}.} 
                     \label{f-miexcomp}
  \end{center}
\end{figure}

We find many more binding sites than the number inferred by Ref.~\cite{mukharjee}, where they set a very strict experimental threshold to avoid false positives. Interestingly, we can run exactly the same analysis on the ChIP data by Lee et al \cite{lee}; we find that the inferred binding sites from two different experimental assays performed by two separate  experimental groups can be made consistent, unlike the discrepancy observed in direct region comparison, see Fig.~\ref{f-miexcomp}. We think that this finding illustrates that proper inference can lead to more complete and consistent identification of binding sites, including the weak ones.
In conclusion, let's comment on why mutual-information inference was able to provide us with good energy matrix models despite our inability to write down the likelihood (or error model) $P(\mathrm{data}|\mathrm{model})$. The answer lies in the observation that with enough data, one can simultaneously infer this error model along with the energy matrix. There is a formal way to show the connection between mutual information inference of Eq.~(\ref{miinfer}) and the Bayes maximum likelihood inference, and we point the reader to details of Ref.~\cite{justin1}. The key point is that one can formulate the problem as maximum likelihood inference using an unknown error model $P(z_i|x_i)$ and then average over all such error models with some prior; in that case one obtains an ``error model averaged'' log likelihood $\log P_{EMA}$ for the data, and this turns to be directly related to the mutual information of Eq~(\ref{miinfer}): $\log P_{EMA}=K I(z_i;x_i)+$ (irrelevant terms). Thus a mathematical connection can be established between mutual information and Bayesian inference.

To summarize, a new method for inferring TF energy models from a wide variety of experimental data has been proposed and shown to bring various existing experiments into concordance. High-throughput datasets provide enough data -- if \emph{all} data is indeed used for inferrence and not ignored by arbitrary thresholds -- to swamp the uncertainty introduced because of our ignorance of the real error model. It can be shown that this inference also works well when genomic background is complicated, as in Ref.~\cite{fire}; that reference also provides an online tool which uses the same framework to learn many TF energy models simultaneously across the genome. Mutual information inference procedure does not rely on the relation between the position weight matrix (PWM) and the energy matrix, and thus does not require the validity of assumptions underlying the Berg \& von Hippel argument \cite{bvh}.

\subsection{Probing combinatorial regulation}
We conclude this lecture by briefly reviewing the work of Kinney et al \cite{justin2}, that has combined the mutual-information inference with a new high-throughput deep-sequencing based approach, in order to {\bf (i)} precisely quantify the contribution of each basepair in the regulatory sequence to the function; and {\bf (ii)} build detailed biophysical models of combinatorial regulation.

In Ref.~\cite{justin2}, the authors decided to reexamine the regulatory region of Lac operon, where both the CRP transcriptional activator, and the RNAP polymerase bind to regulate the expression of lac genes; see Fig.~\ref{f-lac}.
\begin{figure}[!h]
  \begin{center}    
            \includegraphics[width=3in]{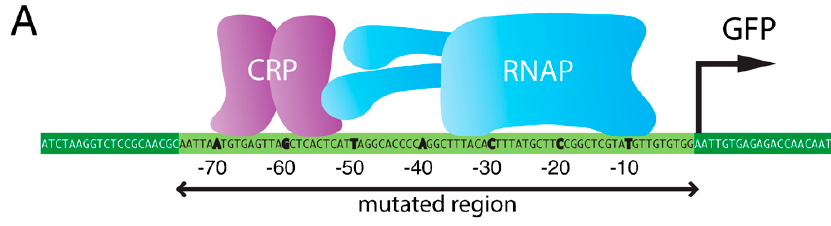}
                \caption{The regulatory sequence of the lac operon, with binding sites for the CRP transcription factor and the RNA polymerase, reproduced from Ref.~\cite{justin2}. To probe precisely how the sequence impacts function, a plasmid library was engineered in which mutagenized versions of the regulatory sequence (bright green) drove the expression of GFP. \emph{E coli} were transformed with the plasmids, the expression of the GFP was induced, and the fluorescence could be measured at the single cell level by FACS.  } 
                     \label{f-jbk2}
  \end{center}
\end{figure}

The key to the experiment was to create a large library of plasmids that differ only in that the regulatory sequence of interest has been mutated, see Fig.~\ref{f-jbk2}; this regulatory sequence controlled the expression level of GFP, which could easily be recorded in the experiment. The cells with different regulatory sequences $\vec{s}_i$ drove various levels of fluorescence, that the experimenters sorted into 9 bins, $z_i$. Despite being seemingly very different from the PBM / ChIP setups described in the previous section, here too one is working with a large number ($K\sim 5\cdot 10^4$ per experiment) of pairs of sequences and the experimental readouts, and an almost identical inference technique can be applied.

\begin{figure}[!h]
  \begin{center}    
            \includegraphics[width=3in]{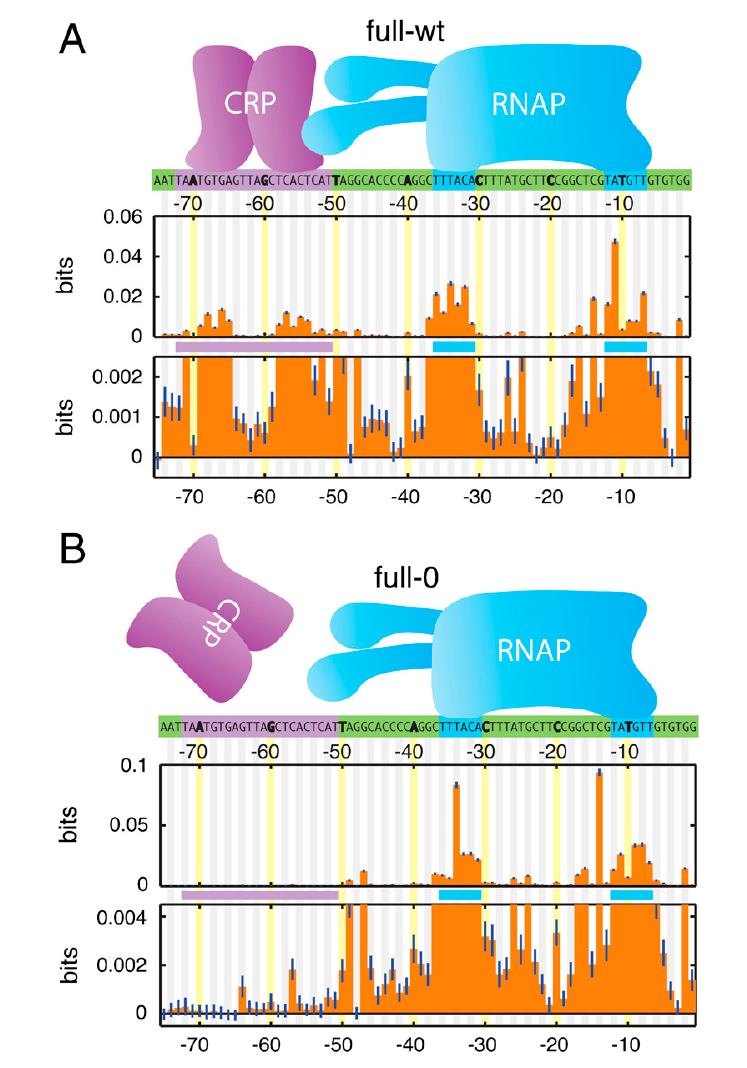}
                \caption{Information footprints of CRP and RNAP binding to DNA in the lac promoter, reproduced from Ref.~\cite{justin2}. The blue and violet regions on the sequence (green) denote where RNAP and CRP, respectively, are thought to contact the DNA. Panel a) shows the case where CRP is bound. Below the sequence, the information $I(b_\alpha;z)$ is shown, in bits, about the identity of a single nucleotide and the final GRP fluorescence, together with the error bars. In the third row we see the zoom-in of the small information values. The method yields precise results with very small error bars; the results are broadly consistent with what is known, but also show that some positions carry small, but significant information about promoter activity. Panel b) shows the same analysis performed when CRP is not bound. Within error bars, most of the positions that are informative about expression with CRP  bound are now 0, as expected.  } 
                     \label{f-jbk3}
  \end{center}
\end{figure}

First, one may ask directly about the mutual information $I(b_i;z)$, about the identity of the base pair $b_\alpha=\{A,C,T,G\}$ at position $\alpha=1,\dots,70$ in the mutated regulatory sequence, about the expression level $z$ of GFP, as measured by fluorescence activated cell sorting (FACS). This is a direct measure, without making any modeling assumptions, about how each base pair on its own affects expression; see Fig.~\ref{f-jbk3}.

As Fig.~\ref{f-jbk3} shows, the method yields extremely precise \emph{information footprints} that characterize the functional impact of every nucleotide on the promoter activity, using the mutual information measure. This method is clearly applicable to unknown regulatory regions if one wishes to quantitatively determine which nucleotides are functionally important.

\begin{figure}[!h]
  \begin{center}    
            \includegraphics[width=3in]{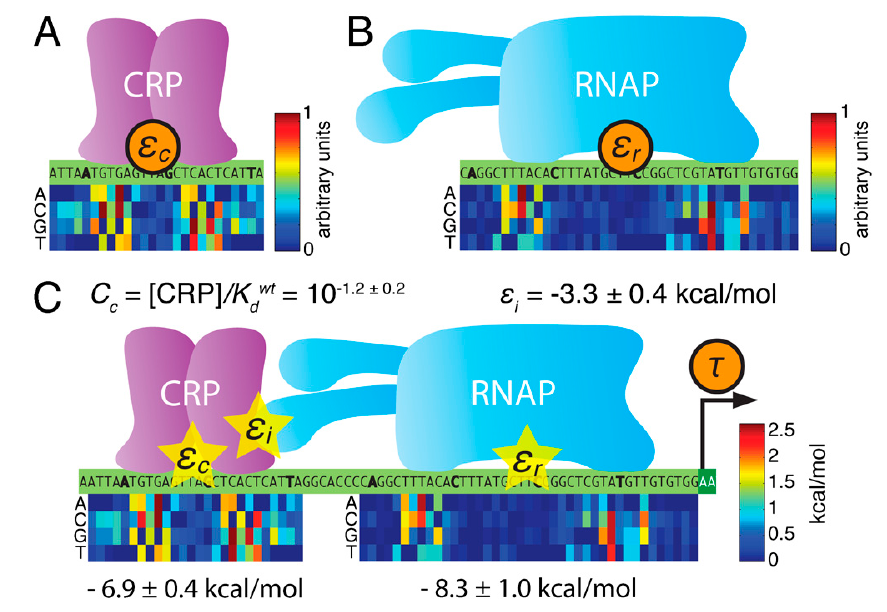}
                \caption{Energy matrix for CRP binding in panel a) and for RNAP in b), when  probed separately using mutual-information inference. Panel c) shows the joint inference of both energy matrices together with the energy of interaction, $\epsilon_i$. The matrices in c) are consistent with the separately learned matrices in a,b). Due to the presence of cooperative interaction $\epsilon_i$, it is possible to put absolute units on all energies.  Figure reproduced from Ref.~\cite{justin2}.} 
                     \label{f-jbk4}
  \end{center}
\end{figure}

Finally, it is possible to use the mutual-information inference to learn the energy matrices of the CRP and RNAP, $\epsilon_{CRP}$ and $\epsilon_{RNAP}$, and the possible energetic interaction between the two proteins. Here, the interaction even helps in constraining the absolute energy scale of the energy models\footnote{It can be shown mathematically that the interaction energy term breaks the arbitrariness about the scale of the energy matrices present when single TF binding is analyzed.}, so the binding energies can be expressed in physical units, as shown in Fig.~\ref{f-jbk4}. Using a convincing series of information-theoretic arguments, the authors can compare how much information the sequence gives about the promoter activity $I(\vec{s};z)$ (this is computable directly from the data without any model), with the fraction of that information that is captured by any particular model -- this can inform us how good any model is with respect to reality. They show that the thermodynamic model of combinatorial regulation, developed in line with the reasoning of Section~\ref{lec2}, does an excellent job of accounting for the data, and that the models inferred using mutual-information inference outperform the models constructed using the Berg \& von Hippel relation in Eq.~(\ref{bvh}). For many other interesting details and controls we refer the reader to the original reference \cite{justin2}.

Hopefully these examples present a strong case for the usefulness of information-theoretic measures and methods, and   demonstrate that inference  should develop in step with advances in experimental techniques. In particular, the examples highlight the difference between traditional physics experiments in which the instruments can be calibrated and understood well enough for us to obtain a handle on $P(\mathrm{data}|\mathrm{model})$, and quantitative biology, when such understanding is often lacking. If the latter case and when using biased or ad hoc inference methods, it is not clear that a larger dataset would actually lead to better models. On the other hand, principled methods can give a very detailed, quantitative and physical account of what is happening in the regulatory regions of the genome. The price for this performance is large required amount of data and computational time, but as the field progresses, those factors are becoming less and less important as practical constraints.
\subsection{Relation to neuroscience}
A very similar method has been devised to probe the behavior of sensory neurons. As we mentioned in Section~\ref{lec2}, sensory neurons are often described in terms of a linear-nonlinear (LN) model: a stimulus, e.g. a movie displayed to a retinal ganglion cell, is convolved first with a linear kernel $\mathcal{L}$ that parametrizes the preference of the neuron for some linear feature (dimension) in a normally highly-dimensional stimulus. The result of this convolution is then passed through a nonlinear function that yields the probability rate for generating a spike. As discussed, various methods have been developed to infer the linear filter $\mathcal{L}$ from traditional experimental setups, in which a large number of stimulus snippets $\vec{s}_i$ are presented, and the spike/no-spike $z_i$ is recorded. Note the analogy with gene regulation: we have stimulus of high dimensionality (sequence or a movie), experimental output (GFP level or spikes), we are looking for a linear function on the stimulus space (energy matrix $\epsilon$ or linear kernel $\mathcal{L}$), and we don't know the probabilistic model of how the output is generated given the product of stimulus with the kernel (or sequence with the energy matrix).

If we knew the nonlinearity that the neuron implements and its likelihood function for determining when it spikes, we could infer the linear kernel $\mathcal{L}$ by writing down the probability of data given the model, and do Bayesian inference of the model. But just as in the case of transcriptional regulation, these quantities too are often unknown. A method called \emph{maximally informative dimensions} that finds the linear kernel $\mathcal{L}$ by maximizing the mutual information between the stimulus projection and the spike trains has been developed and successfully applied in several sensory systems \cite{mid}.
\section{Reconstructing biological networks using the maximum entropy principle}
\label{lec6}
One of the most pressing questions in systems biology today deals with deciphering the structure of regulatory networks from data. The traditional way in which such networks were dissected was through genetic experiments, where painstaking experimentation and mutagenic studies helped deconstruct the networks one link at a time. With the advent of high-throughput experiments, such as microarrays, ChIP and protein binding arrays, as well as simultaneous stains and fluorescent protein fusions etc, the need arose for computational tools that would be able to infer networks from such datasets directly.

Apart from being high-throughput, some of these techniques have opened up another window: instead of looking at bunches of cells mixed together (like in microarrays), they are enabling experimenters to record \emph{simultaneous} expression or activation levels of the nodes of a single biological network, without pooling over many ``copies'' of such networks (e.g. extracted from different cells). In physics terms, this means that not only are the mean activation or expression levels indicative of the network activity, but so are the correlated fluctuations among the nodes. To make use of this fact, we are looking for physical models that include modeling of the (correlated) noise. The structure of fluctuations  can potentially tell us a lot about the wiring diagram of the system.

The simplest -- and still most widely used -- approaches for detecting network structure rely on studying correlations directly. In a typical microarray experiment, the cell cultures are exposed to various external perturbations and the mRNA levels for genes across perturbations are recorded. Then a pairwise correlation matrix is computed across the perturbations, yielding a $N \times N$ pairwise similarity matrix  between $N$ genes (generalizations are possible by measuring mutual information instead of correlation coefficients). Such a similarity matrix can then be used as an input to, for example, clustering. 

Clustering is one of the simplest and most scalable methods of understanding the collective behavior of a network. Consider the information matrix of Fig.~\ref{f-minolan} as a matrix of weights between the nodes of a graph: the graph has strongly connected components that correspond to clusters (blocks on the diagonal of the information matrix) and these blocks are weakly coupled to other blocks. One might even threshold the information matrix and draw binary links in the graph whenever the similarity measure exceeds the threshold value, and some researchers have indeed taken this approach.

Clustering turns out to be an extremely powerful approach for several reasons. Firstly, in gene regulation we know that out of the whole set of genes, the total number of genes that regulate other genes -- so called \emph{transcription factors} -- is on the order of a few percent. Although this, much smaller, group of genes with regulatory power could conspire combinatorially and still regulate every other gene in a complicated individual fashion, many genes need to be up- or down-regulated together, because they act as enzymes in connected reaction pathways or they need to be active in a specific tissue. This \emph {coregulation} is the basis for the success of the clustering approach: coregulated genes cluster and cluster members are assumed to be regulated in identical ways by their (one or few) transcription factor(s). 

Although clustering is clearly productive as a first step in understanding genetic regulatory networks, it is not a generative model of the network. It reorders the nodes so that the structure (hopefully) becomes apparent, but does not give any prescription about \emph{how} the activity of one gene influences the activity of the others --  the only input to the clustering procedure is the mutual information, and we explicitly stated that information measures dependency without revealing anything about underlying functional relationships. Moreover, as we will soon see, understanding that network elements $\sigma_i$ and $\sigma_j$ are correlated, which is the basis of clustering, tells us nothing about whether $\sigma_i$ is really directly influencing $\sigma_j$; in particular, in gene regulation, the genes are coregulated and are therefore coexpressed, and correlation does not imply causation or direct interaction. Despite being very practical, clustering leaves too many questions unanswered if we want to understand network behavior.
\subsection{Correlations vs interactions}
\begin{figure}[!h]
  \begin{center}    
            \includegraphics[width=2in]{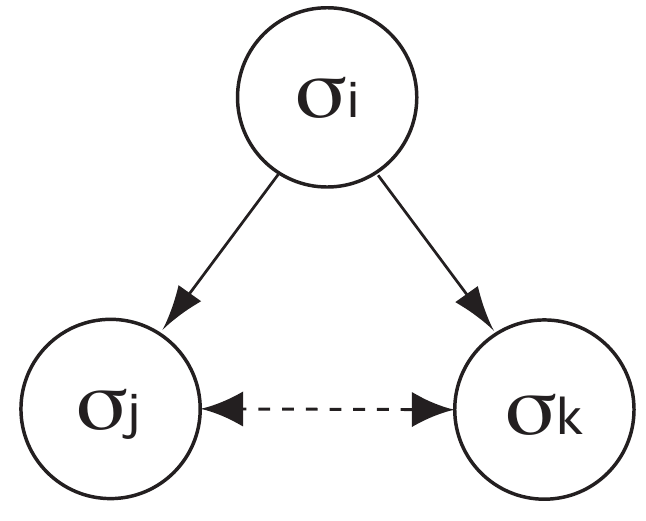}
                \caption[Difference between correlations and interactions.]{A small three-element section from a network of interacting nodes. Suppose that $\sigma_i$ modulates the activity of $\sigma_j$ and $\sigma_k$ through some microscopic mechanism (denoted by thick lines). We can expect to observe strong correlations between $\sigma_i$ and $\sigma_j$, and between $\sigma_i$ and $\sigma_k$ due to this direct influence. On the contrary, $\sigma_j$ and $\sigma_k$ are not directly coupled, but can still show significant correlation (dashed line) because of common control by $\sigma_i$. } 
                     \label{c2_i3}

  \end{center}
\end{figure}

Can we disentangle the mesh of correlations and separate the correlations caused by real underlying interactions from the correlations induced indirectly by other interactions, as is illustrated in Fig.~\ref{c2_i3}? 

To start, we recall a classic problem in statistical physics: we are given a lattice of Ising spins (binary variables), and some specification of exchange couplings (interactions) -- perhaps between nearest neighbor only -- and the exercise requires us to find the equilibrium correlation function between the spins, i.e.\ $\langle \sigma_i \sigma_{i+\Delta}\rangle$. In our case however, we will be dealing with network ``reverse engineering.'' The exchange interactions themselves will be unknown, yet we will observe a mesh of correlations. The problem will then be to compute the exchange interactions from the measured correlations, with the hope of finding a network defined by the interactions to be \emph{simpler} (for instance sparser) than the network of correlations.

Let us formulate the problem more precisely. The network consists of $N$ nodes with activities $\sigma_i$, $i=1,\dots,N$, which, we will for now assume, can take on only two values, $\sigma_i\in \left\{-1,1\right\}$. An experimental snapshot of the network activity is then described by a vector, $\vec{\sigma}=\{\sigma_1,\sigma_2,\dots,\sigma_N\}$. Our data consist of patterns $\mathcal{D}=\left\{\vec{\sigma}^1, \vec{\sigma}^2,\dots,\vec{\sigma}^T\right\}$, i.e.\ there are a total of $T$ simultaneous measurements of the activities at all nodes, while the network is in some stationary state. These samples can be thought of as ``instantaneous'' snapshots of the system or, in simulation, draws made during a Monte Carlo sampling run.  From the samples we can estimate the moments of $\{\sigma_i\}$ at successively increasing orders: first order moments are $N$ mean activity values, $\langle \sigma_i\rangle$; second moments are $N(N-1)/2$ correlations, $\langle \sigma_i \sigma_j\rangle$; and so on. Because the system is noisy, there will be fluctuations around the stationary state and not all $T$ patterns are going to be equal. We expect some patterns to be more likely than the others, and the full description of the system in equilibrium must be contained in a joint probability distribution, $p(\sigma_1, \sigma_2, \dots,\sigma_N)$. Getting a handle on this distribution is therefore our final goal, and as we will soon discover, computing successive approximations to it will give us  the desired interactions that underlie the observed correlations.

Except for a very small number of network nodes there is no hope of directly sampling the distribution from the data. Its size grows exponentially in $N$ and for a modest network of 10 binary nodes we would generally need to estimate $2^{10}\sim 1000$ parameters. To proceed, we clearly need a simplifying principle. 

\begin{figure}[!h]
  \begin{center}    

            \includegraphics[width=2in]{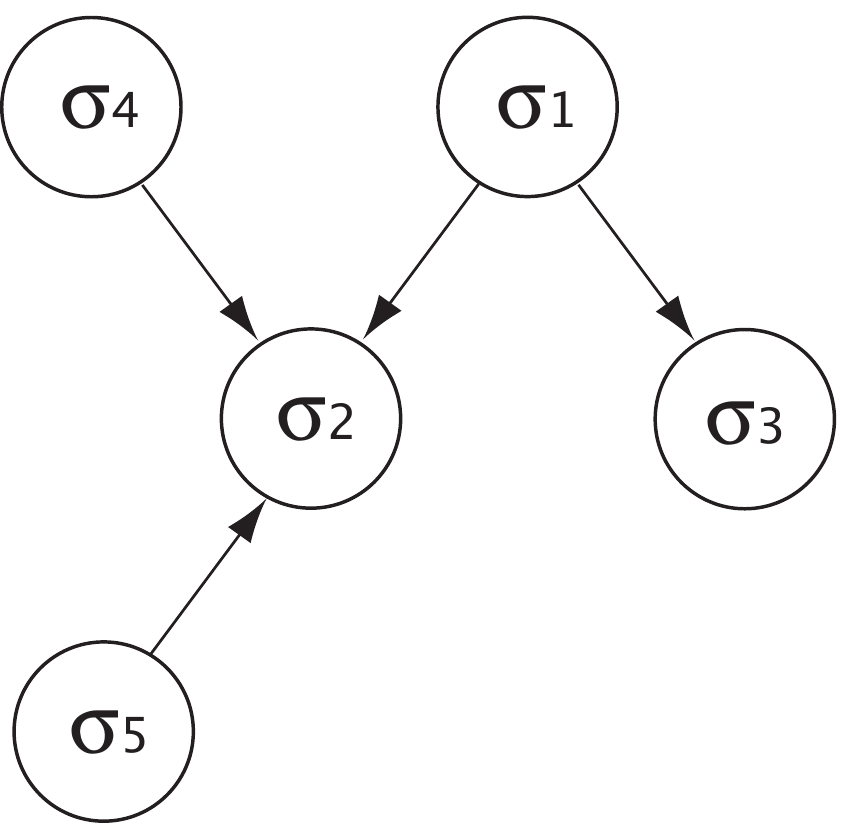}
              \caption[Graphical illustration of conditional independence.]{Bayesian factoring of the probability distribution over five nodes. The edges imply conditional dependence, and if two nodes are not joined by an edge, they are assumed to be conditionally independent. This example graph $\mathcal{G}$ implies that the joint probability distribution can be written as follows: $p(\sigma_1,\dots,\sigma_5)=p(\sigma_1)p(\sigma_4)p(\sigma_5)p(\sigma_3|\sigma_1)p(\sigma_2|\sigma_1,\sigma_4,\sigma_5)$. This form has $1+1+1+2+8=13$ free parameters in case of binary variables (remember that conditional probability distributions are normalized), and is much simpler than the completely arbitrary $p(\sigma_1,\dots,\sigma_5)$, that for 5 binary nodes would have $2^5-1$ free parameters. }. 
                     \label{c2_i4}

  \end{center}
\end{figure}

A commonly used procedure is called Bayesian network reconstruction \cite{friedman_03}, and it is a method from the more general class of graphical models. One starts by assuming a specific (initial) factorization of the joint probability distribution over all nodes and represents it as a graph $\mathcal{G}^0$, as in Fig \ref{c2_i4}. Remembering that the activities are discrete variables, all conditional distributions in the factorization can be represented as probability tables with unknown entries that need to be fit from the data. Such fitting procedure can be performed in many ways, and one can evaluate the likelihood of the fit $\mathcal{L}(\mathcal{G}^n)$\footnote{Bayesian network reconstruction is an iterative procedure, and at $n$-th step, we are considering graph $\mathcal{G}^n$, hence the index.}. Of course, we have no prior knowledge of what the correct graph factorization of the initial distribution is, therefore a procedure is devised that wanders in the space of possible graph topologies and tries a likelihood on each, producing a sequence $\left\{\mathcal{L}(\mathcal{G}^0),\dots,\mathcal{L}(\mathcal{G}^n),\mathcal{L}(\mathcal{G}^{n+1}),\dots\right\}$\footnote{This will usually be some sort of gradient descent or simulated annealing procedure.}. The complexity of each graph, e.g.\ the number of links, is penalized and combined together with the fit likelihood into a scoring function. The goal is to find the factorization of the probability distribution with the best score. Presumably, we will then have discovered a simple graph that fits the data well.

There have been successful network reconstructions using this approach \citep{sachs+al_05}. The key simplifying assumption that makes this approach feasible is that the graph of interactions is sparse, i.e.\ that there are many fewer real than potential interactions. Given such sparsity, the factorized probability distribution will have a far smaller number of unknown parameters than the full joint distribution, and there will be reasonable hope of fitting them from the data. The method allows interactions of arbitrary complexity (as many arrows converging on a single node as possible), but has some drawbacks. Firstly, there is an exploding number of graph topologies over $N$ elements, and no hope in exhaustively trying all of them; whatever algorithm one devises to explore the space of topologies, it can get stuck in local extrema of the scoring function. Secondly, due to computational constraints not all kinds of graphs can be explored -- usually one has to exclude loops and this is a big handicap for biological systems where feedback plays a very important role.  Finally, because we are looking for a tradeoff between the best likelihood fit and the simplicity of the model, we have to (arbitrarily) decide how to penalize complex topologies. It is not \emph{a priori} clear that one should simply minimize the number of links and disregard other features of the graph. In particular, we expect that for systems, in which collective effects are driven by the presence of weak interactions between lots of pairs, Bayesian method will perform poorly.

\subsection{Maximum-entropy models: general introduction}

Here we will try to take a radically different route to the solution, motivated by inverse problems in statistical mechanics. This methodology has been applied successfully to a diverse set of biological interacting networks, such as neurons, human immune system, and signaling networks \cite{eladn,gtkacikn,alext}; the full analysis of the dataset presented here can be found in Ref \cite{phd}.

We start with the realization that with a limited number of samples, $T$, we can successfully estimate several lowest-order moments of the sought-for joint distribution $p(\sigma_1,\dots,\sigma_N)$ that generated the data, for example, the means $\langle \sigma_i \rangle$ and covariances $C_{ij}=\langle \sigma_i\sigma_j\rangle - \langle \sigma_i\rangle\langle\sigma_j\rangle$, or, in general, a set of mean values of some of the statistics of the unknown distribution, also called its ``operators,'' $\langle \hat{O}_\mu(\vec{\sigma})\rangle$ (which can be arbitrary functions of $\{\sigma_i\}$). For any reasonable choice of the operators there is an infinite number of joint distributions over $N$ elements with the same mean operator values. Nevertheless, there is only one distribution that also has maximum entropy, i.e.\ there is one distribution that is \emph{as random as possible} but still satisfies those statistics that have been measured in an experiment.  This is the distribution that we would like to find, and the maximum entropy principle embodies the idea that any structure (or constraint) in the distribution has to be induced by the measurement (and not by explicit or hidden assumptions on our part).  In other words, we will approximate the true distribution $p(\sigma_1,\dots,\sigma_N)$ that we cannot measure in full (but can estimate some of its statistics $\hat{O}_\mu$), with the \emph{maximum entropy distribution} that is as random as possible but matches the true distribution in the values of all of the measured statistics.

Formally, we are looking for the extremum of the following functional:
\begin{eqnarray}
\mathcal{L}[p(\vec{\sigma})]&=&S[p(\vec{\sigma})] - \sum_\mu g_\mu \langle \hat{O}_\mu(\vec{\sigma}) \rangle - \Lambda \int d\vec{\sigma}\,p(\vec{\sigma})\nonumber\\
&=&-\int d\vec{\sigma}\,p(\vec{\sigma}) \log_2 p(\vec{\sigma}) -  \label{c2_mef}\\
&-&\sum_\mu g_\mu \int d\vec{\sigma}\,p(\vec{\sigma})  \hat{O}_\mu(\vec{\sigma})  - \Lambda \int d\vec{\sigma}\,p(\vec{\sigma}).\nonumber
\end{eqnarray}
The first term is the entropy of the distribution, and there are $\mu$ constraints enforced by their Lagrange multipliers $g_\mu$:
\begin{equation}
\langle \hat{O}_\mu(\vec{\sigma})\rangle_{p(\vec{\sigma})}=\langle \hat{O}_\mu\rangle_{\rm expt\,\,\mathcal{D}}, \label{c2_const}
\end{equation}
such that the average values of the operators over the sought-after distribution $p(\vec{\sigma})$ are equal to the averages over data patterns, $\mathcal{D}=\left\{\vec{\sigma}^1,\dots\vec{\sigma}^T\right\}$. The Lagrange multiplier $\Lambda$ enforces the normalization of the distribution. It is easy to take the variation in Eq (\ref{c2_mef}) and write the explicit form for the maximum entropy solution:
\begin{equation}
p(\vec{\sigma})=\frac{1}{Z}\exp{\left[\sum_\mu g_\mu \hat{O}_\mu(\vec{\sigma})\right]}. \label{c2_mem}
\end{equation}
We call Eq (\ref{c2_mem}) the maximum entropy distribution with constraints $\langle\hat{O}_\mu\rangle$.

Operators that constrain the distribution can be arbitrary, but we can gain further insight by restricting ourselves to the moments of increasing orders (the activity variables are still binary for simplicity). If one chooses $\hat{O}_\mu=\sigma_\mu$, then the mean values, $\langle\sigma_i\rangle$, are constrained, and the maximum entropy distribution is the factor distribution:
\begin{eqnarray}
p^{(1)}(\vec{\sigma})&=&\frac{1}{Z}e^{\sum_\mu g_\mu \sigma_\mu}=\prod_\mu \frac{1}{Z_\mu}e^{g_\mu\sigma_\mu}, \label{c2_memind} \\
Z_\mu&=&2\cosh(g_\mu).
\end{eqnarray}

This factor distribution is easy to compute, but it does not include any interactions -- each element behaves in an independent fashion (similar to the mean-field theories in physics). We could continue constraining the maximum entropy distribution with correlation functions of higher and higher orders. If we were to fix both mean values and two-point correlations, the resulting distribution, Eq (\ref{c2_mem}), would have an Ising form:
\begin{equation}
p^{(2)}(\sigma_1,\dots,\sigma_N)=\frac{1}{Z}e^{\sum_i h_i\sigma_i+\frac{1}{2}\sum_{ij}J_{ij}\sigma_i\sigma_j};
\end{equation}
this is the simplest distribution that contains pairwise interactions between the network elements.

Constraining the three-point correlations would induce a new term in the exponent of the form $\sum_{ijk}J_{ijk}\sigma_i\sigma_j\sigma_k$. There is clearly a ``ladder,'' where higher and higher order constraints are imposed on the distribution, and as a result, better and better maximum entropy approximations are constructed. Let us call, then, $p^{(k)}(\vec{\sigma})$ a maximum entropy distribution consistent with correlations of order $k$ and smaller, in line with our notation for the factor distribution, $p^{(1)}(\vec{\sigma})$. In an $N$-body system, the highest order of correlation is $N$, and $p^{(N)}(\vec{\sigma})$ must therefore be the exact joint distribution -- at this order our approximation \emph{is} the exact solution, with entropy equal to $S[p(\vec{\sigma})]$.  In Ref \cite{schneidman+al_03}  it has been shown that this sequence of ever better maximum entropy approximations defines a unique decomposition of multi-information of Eq~(\ref{multii}):
\begin{eqnarray}
I[p(\vec{\sigma})]&=&\sum_{k=2}^{N} I^{(k)}	\\
I^{(k)}&=&S[p^{(k-1)}(\vec{\sigma})]-S[p^{(k)}(\vec{\sigma})]. \label{c2_multiicon}
\end{eqnarray}
In words, the \emph{connected information of order k}, $I^{(k)}$, is the difference of the entropies of the maximum entropy distribution consistent with correlations of order $k-1$ and one higher order. For example, connected information of the second order is the reduction of the entropy due to pairwise interactions; one creates the best factor (independent) model for the data and the best pairwise (two-body Ising) model for the data, and compares their entropies to see how much of the total structure in the joint distribution has been explained by purely pairwise terms.
\subsection{Maximum-entropy models: pairwise interactions}
How do we use this framework to model real networks? Once we collected the measured correlations, we would postulate the maximum entropy model of Eq (\ref{c2_mem}) and solve the equations that determine all couplings, Eq (\ref{c2_const}); mathematically, we need to find $\{h_i,J_{ij}\}$ that solve the following equations:
\begin{eqnarray}
p(\sigma_1,\dots,\sigma_N)&=&\frac{1}{Z}e^{\sum_i h_i\sigma_i + \frac{1}{2}\sum_{i\neq j}J_{ij}\sigma_i\sigma_j} \label{max2}\\
\frac{\partial Z}{\partial h_i}&=&\langle \sigma_i\rangle_{\mathcal{D}}\\
\frac{\partial Z}{\partial J_{ij}}&=&\langle \sigma_i\sigma_j\rangle_{\mathcal{D}},
\end{eqnarray}
where the expectation values on the right-hand side are measured in the dataset $\mathcal{D}$.

This procedure yields two important results: {\bf (i)} since we have a generative model of the data, i.e. the probability distribution, we can calculate and predict any expectation value (especially of those statistics that were not used as constraints), and compare it to experiment; {\bf (ii)} we can examine the couplings $\{J_{ij}\}$, conjugate to the constrained operators, and interpret these as \emph{interactions} that cause and explain the observed correlations. 


As is done in Bayesian network reconstruction, once we have computed the couplings, we can draw a graphical model of the network with a link for each nonzero coupling $J_{ij}$ connecting the elements $\sigma_i$ and $\sigma_j$\footnote{Therefore, for instance, the graphical decomposition of the probability distribution plotted in Fig \ref{c2_i4} would correspond to the distribution $p=\exp (-H)/Z$, where $H=\sum_i h_i\sigma_i + J_{13}\sigma_1\sigma_3+J_{145}\sigma_1\sigma_4\sigma_5$ in the maximum entropy picture.}. These weighted links are undirected as there is generally no way of determining the ``direction'' of the interactions from an equilibrium model. 

Assumptions underlying maximum entropy reconstruction are quite different from its Bayesian relative: whereas in the latter case we assume sparse a network of (arbitrarily complex) interactions, we assume an arbitrarily dense network of simple (low order, e.g.\ pairwise or triplet) interactions in the former case. To explain all $N(N-1)/2$ pairwise correlations one needs the full matrix of $N(N-1)/2$ exchange couplings $J_{ij}$\footnote{For higher orders, there is similarly no restriction on the structure of, for example, three-point interactions, $J_{ijk}$.}, and therefore no discrete topology on the graph is assumed \emph{a priori}. There is hence no problem of searching and scoring the space of topologies, no exclusion of graphs that include loops, and reduced dependence on the implementation details of the algorithm. The drawback is the \emph{ab initio} exclusion of complex irreducible interactions between many nodes. Clearly, the real question to ask is about the approximation regime that is more suitable to biological systems, if  a general answer exists at all.

In practice, unfortunately, the maximum entropy network reconstruction is made difficult by two problems. One is technical -- solving coupling Eqs (\ref{c2_const}) is very hard. In essence, one needs to solve
\begin{equation}
\frac{\partial \log Z(\left\{g_\nu\right\})}{\partial g_\mu}=\langle \hat{O}_\mu\rangle_{\rm expt\,\, \mathcal{D}}, \label{ccsol}
\end{equation}
where $Z$ is the partition function of the maximum entropy distribution in Eq (\ref{c2_mem}). This set of equations is both nonlinear in couplings $g$ and requires the evaluation of the partition function, $Z(\left\{g_\nu\right\})$, or effectively a complete solution of the statistical mechanics problem. The other problem concerns the identification of the nodes that are observed in the experiment. First, one will usually  be able to take measurements of only a small subset of the nodes comprising the network and we need to be concerned about how the hidden nodes influence models of visible nodes. Second, even if all nodes were identified, there is an issue of ``coarse-graining.'' Is a node with two states really an elementary, physical object that only has two states (a protein with two phosphorylation states), or is it in itself a complex with many states, but for which a two-state model might (or not) be a valid approximation? We do not have time to systematically address these issues in the lecture notes, but do wish to point them out.

\begin{figure}[!h]
  \begin{center}    
     \includegraphics[width=\linewidth]{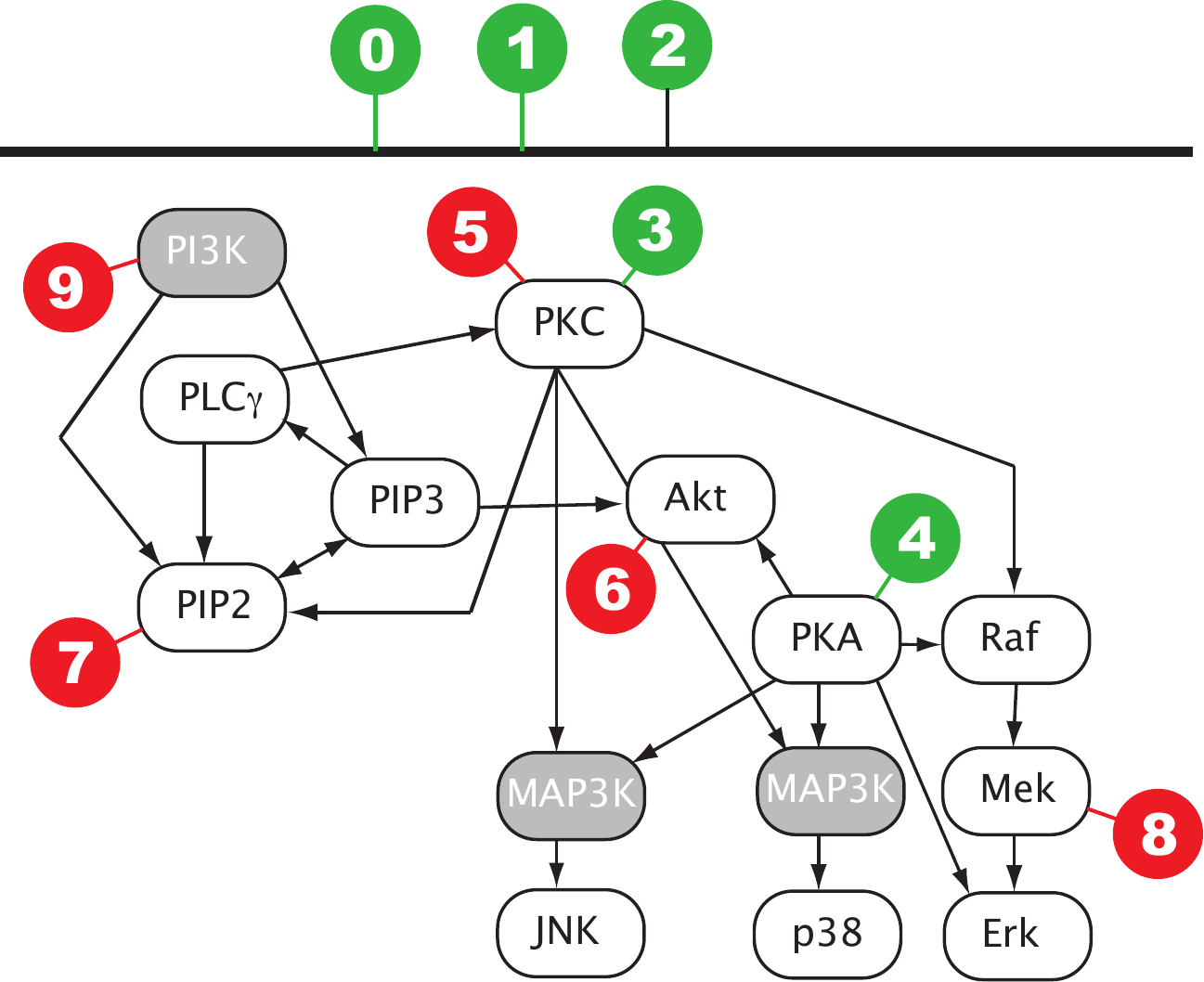}
       \caption[MAP signaling network in immune system cells.]{A diagram of MAP signaling network in human CD4 T-cells, reproduced and simplified from Ref \cite{sachs+al_05}. Phosphorylation level of 11 white nodes was observed; red and green numbers indicate points of intervention (i.e.\ the change of external conditions $\mathcal{C}$ in which the network operates). These chemical interventions change the state of the whole network by locking the activity of the nodes on which they act into activated (green) or deactivated (red) state. Chemicals 0, 1 and 2 represent naturally occurring stimulatory agents; 0 and 1 are present in all $\mathcal{C}$, while 2 is present in $\mathcal{C}=2$. The arrows represent experimentally verified chemical interactions; there are a number of known interactions through intermediaries that are known, but not plotted. Gray nodes were not observed in the experiment.}    \label{f_mapk}
  \end{center}
\end{figure}

\subsection{Reconstructing a biochemical network}

Here we present a maximum-entropy-based approach to biochemical network reconstruction following the steps outlined in previously. We tackle these questions on the set of 11 interacting proteins and phospholipids (jointly referred to as \emph{biomolecules} here) in a signaling network of human primary immune system T cells. We use data from  Ref \cite{sachs+al_05}, where approximately 600 single-cell measurements of the activity level of biomolecules have been made for each of the 9 available conditions $\mathcal{C}$ using flourescent cell cytometry; this dataset has already been presented in Section~\ref{lec5}. The network has been studied in detail and Fig.~\ref{f_mapk} shows the conventionally accepted interactions, placing the observed proteins into their biological context.

We will assume that, given a set of $N=11$ network nodes, their interactions can be well-described as occurring only between pairs or perhaps triplets, and not as combinatorial interactions involving quadruplets or larger groups.  We'll assess the validity of this assumption at the end of the lecture; detailed checks are presented in Ref \cite{phd}.

A typical experiment to which maximum-entropy network reconstruction can be applied will yield a large number of simultaneous observations of $N$ real-valued activation levels for each external stimulus. As a first step in the analysis, we discretize the data into two binary levels\footnote{Discretization can be regarded as a form of data compression; the original continuous data have some correlation structure, and as quantization maps the data into the discrete domain, we would like the structure to remain preserved. There are a lot of technical issues involved in discretization, especially when discretizing into more than 2 levels; for clarity we skip these problems here.}. We illustrate the maximum entropy reconstruction by focusing on each of the nine experimental conditions separately, and attempt to address the questions presented above. It is possible to formulate maximum entropy problem such that the network reconstruction takes advantage of all experimental conditions simultaneously; for details see Ref \cite{phd}.
\subsection{Analyzing a single condition}
The data collected for conditions 1 and 2 describe the activation levels of 11 biomolecules when the cells are exposed to their natural stimulatory signals. If we focus on each of the two conditions separately, we will be dealing with draws from two stationary distributions. We first discretize separately the data in each condition, and end up with 11 bit binary words that represent fluctuations around the steady state in that condition. Because the nodes are functionally connected, the fluctuations are not independent, and must reflect local couplings between nodes near the given steady state.  Can we learn something from the correlated fluctuations in the activities? 

Having quantized the data into two levels and calculated the correlations and mean values, we write down the form of maximum entropy distribution consistent with these operators; to be consistent with physics conventions, let's write $\tilde{\sigma}_i = 2\sigma_i-1$, such that $\tilde{\sigma}_1=\pm 1$ ($\tilde{\sigma}_i=-1$ denotes the ``off'' state and $\tilde{\sigma}_i=1$ denotes the ``on'' state):
\begin{equation}
p(\tilde{\sigma}_1,\dots,\tilde{\sigma}_N)=\frac{1}{Z}\exp\left\{\sum_i h_i\tilde{\sigma}_i+\frac{1}{2}\sum_{ij}J_{ij}\tilde{\sigma}_i\tilde{\sigma}_j\right\}.
\label{eq_ising}
\end{equation}
We proceed to calculate the \emph{interaction map} $J_{ij}$ and the biases $h_i$ that explain the measured observables [Eq \ref{ccsol}], by using a numerical nonlinear equation solver\footnote{There is a unique solution for the $\{h_i,J_{ij}\}$ and the problem is convex.}.
 \begin{figure} 
   \centering
    \includegraphics[width=\linewidth]{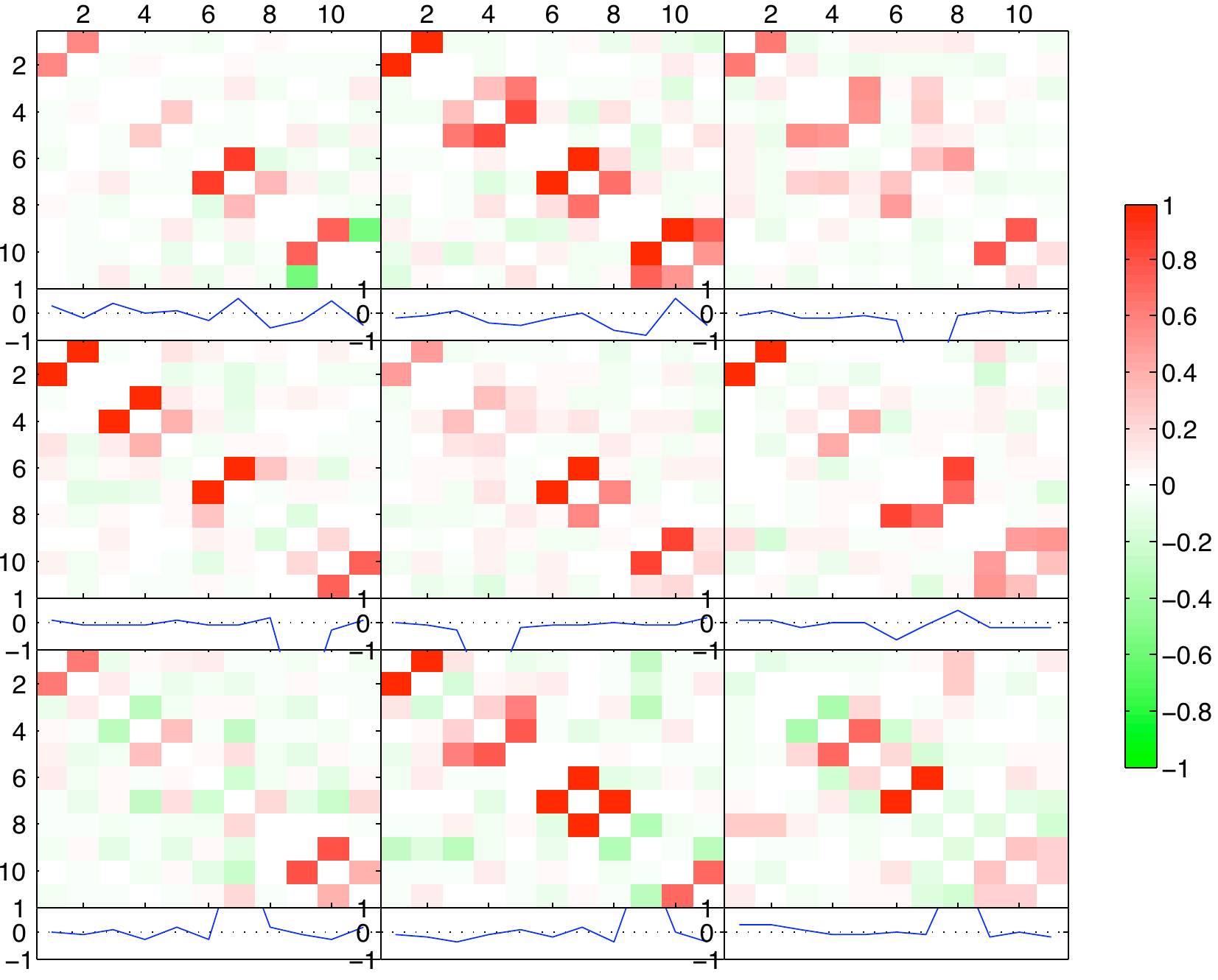}
   \caption[Interaction maps between proteins for different conditions.]{Interactions $J_{ij}$ (color map)  and biases $h_i$ (blue line) for all  external conditions $\mathcal{C}=1\dots 9$, proceeding top down, left to right, computed (both quantization and maximum entropy reconstruction) separately for each condition. All interactions $J_{ij}$ are drawn on the same scale, with red color indicating positive and green color indicating negative couplings. Conditions 1 and 2 represent cells exposed to the naturally occurring stimulatory chemical signals; other conditions represent  environments where ``intervention'' chemicals -- which are supposed to lock the activity states of certain nodes to either ``on'' or ``off'' -- have been added to the stimulatory chemicals of condition 1.} 
   \label{fig_pn2}
\end{figure}

Figure \ref{fig_pn2} shows reconstructed interaction maps $J_{ij}$ and biases for each condition's data quantized and analyzed separately\footnote{Our data was intrinsically continuous and was discretized into 2 levels. The biases $h_i$ that we compute simply reflect the overall bias of the, e.g., node $\tilde{\sigma}_i$ to tend towards -1 vs 1. This is not really a property of the data, but of where we draw the discretization thresholds; this is in contrast to the interactions $J_{ij}$ which truly reflect the interactions in the data. }.  Interestingly, both condition 1 and 2 exhibit a similar pattern of interactions, with those of condition 1 being a subset of condition 2; moreover they also agree with the conventional map of interactions in Fig.~\ref{f_mapk}, except for the interaction between 10 and 11 (p38, JNK) in condition 2. A possible explanation for this interaction is the cross-talk in the MAPKK pathway upstream of p38 and JNK: unobserved biomolecules that couple pairs of observed proteins would induce effective interactions between them. 
In general, the interaction matrices are sparse, and most of the small coupling constants can be set to zero with minimal change to the distribution (not shown)\footnote{One starts with the smallest couplings and proceeds towards bigger ones by setting them to zero and calculating the Jensen-Shannon distance between such ``pruned'' and the original distributions. 
For conditions 1 and 2, if all exchange interactions but for the ``skeleton'' around the diagonal are set to 0, the Jensen-Shannon distance will be around $0.015$, i.e.\ one would need on the order of 70 samples to distinguish the full maximum entropy from the pruned distribution.}.

Note again that we are looking only at fluctuations around a naturally stimulated steady state. 
These fluctuations are much smaller then those induced by intervening chemicals, which is presumably why we detect only a subset of full interactions.


\begin{figure}[!h]
  \begin{center}    
   
      \includegraphics[width=\linewidth]{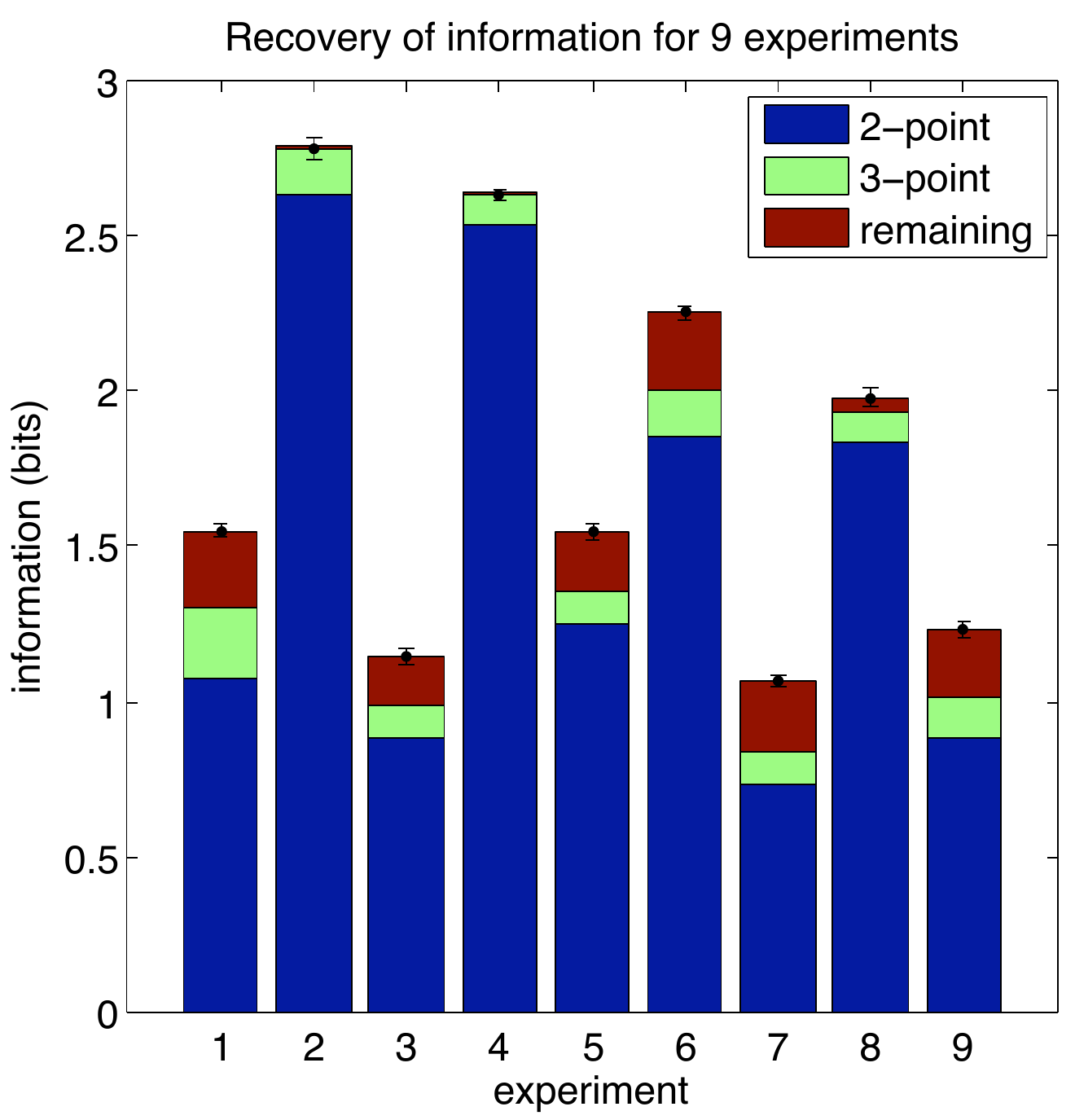}      \caption[Information capture by Ising models.]{How much of the total multi-information in the distribution of activation levels does the pairwise model capture? For each of the 9 conditions on the horizontal axis, the data is discretized  into 2 levels and a pairwise (or three-body) Ising model is constructed. The bar height is the total multi-information of the distribution [Eq (\ref{multii})]. The blue (green) segment represents the information of the second (third) order, $I^{(2,3)}[p(\vec{\sigma})]$, of Eq (\ref{c2_multiicon}). The error bars are entropy estimation errors from the \emph{direct} estimation obtained by 100 repeated reestimations \citep{minoam}.
   } 
   \label{fig_info}

  \end{center}
\end{figure}

How much of the complexity of the true distribution is captured by the maximum entropy approximation? To answer this question we look at the fraction of the multi-information of the real distribution that is captured by the pairwise model. As Fig.~\ref{fig_info} demonstrates, in case of condition 2 it recovers almost all of the 2.8 bits of total information; for condition 1, however, the fraction is around 70 percent out of the total of 1.5 bits\footnote{There might be larger systematic errorbars on experiments 1, 6 and 9 because the distribution seems considerably more uniform than for other experiments and we are low on samples.}.

\begin{figure}[!h]
  \begin{center}    
          \includegraphics[width=2.75in]{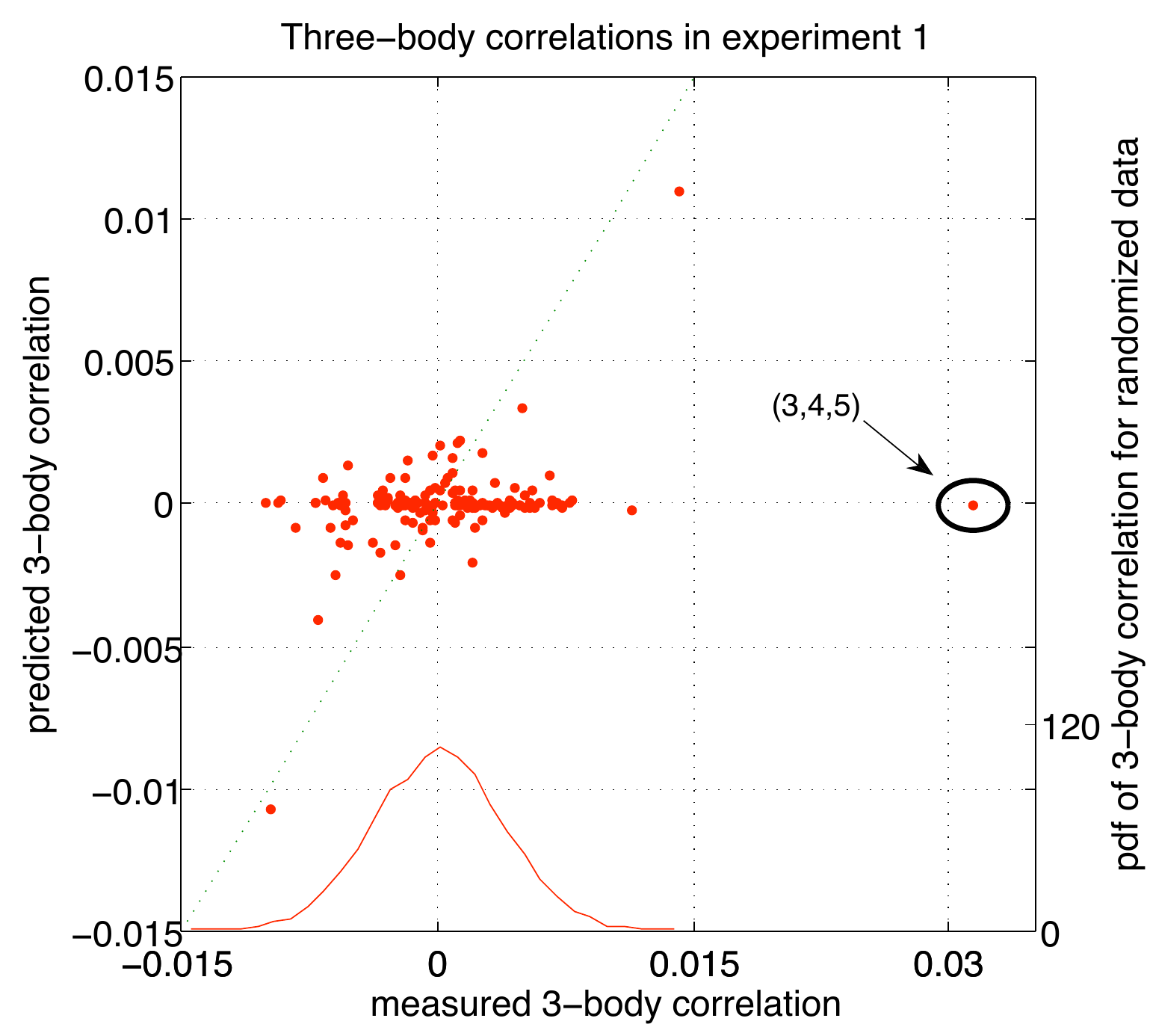}
          \caption[Verification of the model by 3-point correlations.]{Comparison between three-point connected correlations measured from the data (horizontal axis) and predicted by the pairwise Ising model (vertical axis). Due to small sample effects, there are error bars in the correlation estimates; the distribution in red is the distribution of the three-point connected correlations in a shuffled dataset, and the spread in that distribution arises purely from small-sampling effects. We see that three-point correlations are small with the exception of several elements, and all but one agree with the model prediction. The mismatch is the three-body correlation between biomolecules 3, 4 and 5 (see main text).} 
   \label{fig_threepoint}
  \end{center}
\end{figure}

A test of the pairwise model involves comparing the predictions about connected three-point correlations $\langle (\sigma_i-\bar{\sigma}_i)(\sigma_j-\bar{\sigma}_j)(\sigma_k-\bar{\sigma}_k)\rangle$ with values estimated from the data, as shown in Fig.~\ref{fig_threepoint}. Thee-point statistics have not been constrained by construction in our model, and are therefore a real prediction of the maximum-entropy distribution. As expected, the match between predictions and measurement is good in condition 2 (not shown), while for  condition 1 we see a single three-point predicted correlation deviating strongly from its measured value. The corresponding biomolecules are $\sigma_2$, $\sigma_3$ and $\sigma_4$, namely PLC$\gamma$, PIP2 and PIP3, and they are suspected to form a feedback loop [Fig \ref{f_mapk}]. To ascertain that it is not only the observed correlation, but actually a true triplet interaction between the molecules that generates the discrepancy, we can build a new maximum-entropy model consistent with three-point marginals. The corresponding distribution has the following form:
\begin{equation}
H=-\sum_i h_i\tilde{\sigma}_i-\frac{1}{2}\sum_{ij}J_{ij}\tilde{\sigma}_i\tilde{\sigma}_j-\frac{1}{6}\sum_{ijk}J_{ijk}\tilde{\sigma}_i\tilde{\sigma}_j\tilde{\sigma}_k
\label{eq_triplet}
\end{equation}

When we solve for the unknown $\{h_i,J_{ij},J_{ijk}\}$ in the distribution of Eq (\ref{eq_triplet}), the largest three-point interaction term is $J_{345}$. Moreover, in order to convincingly show that it really is $J_{345}$ that fixes the offending three-point correlation (as opposed to all other triplet degrees of freedom in Eq (\ref{eq_triplet})), we construct yet another maximum entropy model: a pairwise Ising system that in addition to all pairwise correlations constrains exactly one three-point marginal, $p(\sigma_3,\sigma_4,\sigma_5)$, and has a single three-point coupling, $J_{345}$. The agreement between prediction and observations is then restored up to third-order in correlations, at the cost of one additional underlying interaction. Experimentally it is also known that PLC$\gamma$ hydrolyses its substrate PIP2 to produce PIP3; furthermore it is suspected that PIP3 can recruit PLC$\gamma$. The example analysis presented in this lecture sweeps a lot of checks and details under the rug in order for the lecture to remain straightforward; for details see Ref \cite{phd}.
%


 In summary, we believe that the maximum-entropy network reconstruction procedure offers a viable alternative to Bayesian network reconstruction. The theoretical foundation  provides a way of decomposing the total information of a given distribution into a sum of positive terms [Eq~(\ref{c2_multiicon})], each of which indicates the extent to which maximum-entropy models incorporating successively higher order marginals recover the total complexity \citep{schneidman+al_03}. A failure to account for the total information with a simple model is diagnostic of complexity being unaccounted for in the model; to pinpoint the problem, one compares the prediction and measurement of next order correlations; hopefully, the failure is localized and not distributed through the network. If this is the case and fixing the failure requires the introduction of a single new interaction, we might believe that we have learned something new about the system. In the presented example of the MAP cascade, the pairwise model accounted very well for data in condition $\mathcal{C}=2$, and less well in $\mathcal{C}=1$; but even in the latter case, an addition of a single combinatorial three-point interaction has lead to a considerably improved model. More importantly, the analysis approach is not good for a single system only, but is a principled framework that can include systematically more and more complexity until the data can be accounted for.

In general, principled methods for inferring network structure from the data are not yet widely used; part of the reason is that the power of these methods ultimately derives from the quality of the measurements. If the experimental noise levels swamp the intrinsic noise of the regulatory process, then the benefits of observing the correlated fluctuations in the system are lost. In signaling pathways and genetic regulation the network reconstruction procedures deployed to date have mostly been technical demonstrations that the approach is possible or feasible, and its correctness was judged by comparison to a manually curated ``gold standard'' network, assembled from the literature. 

\subsection{Relation to neuroscience}
In the case of the neural networks in the retina, true new insight about how the network works was provided by maximum-entropy models \cite{eladn,gtkacikn}. Recordings of neural activities there have consistently shown that any pair of neurons in close physical proximity in the retina has activities that are weakly, but statistically significantly correlated (i.e. the average correlation coefficients in pairs of neural spike trains were $\sim 0.05$). It was usually assumed that this meant that correlations are a small effect and can safely be neglected, leading to an interpretation that separate retinal ganglion cells are independent encoders of information about the light in the visual field. However, when looking at larger and larger groups of neurons together (not just pairs), it became increasingly clear that the assumption of neural independence leads to worse and worse predictions about how groups should behave in comparison to the experiments; at groups comprising $N=10$ neurons the mismatch was extremely large. 

Ref.~\cite{eladn} proceeded to fit maximum entropy models with pairwise interactions to groups of up to 15 neurons. The resulting models accounted for data very well, leading to a reinterpretation of the neural behavior, where a dense network of weak interactions induces strong effects on how groups or populations of neurons behave. These analyses were later extended to groups of 40 neurons, and analogies could be made between the behavior of these inferred networks and the theoretical models of frustrated collective behaviors (spin glasses) in physics \cite{gtkacikn}.
\section{Towards a possible network design principle}
\label{lec7}
In the last lecture, we will seriously  consider the idea that the \emph{function of biological  regulatory networks is to transmit information}. Armed with the mathematical concept of information introduced in Section~\ref{lec4}, we will be able to formalize the notion that a transcription factor present at concentration $c$ drives a set of downstream genes, $\{g_i\}$, and that the expression levels of these genes therefore jointly carry information about $c$. We will argue that there exist certain regulatory wiring diagrams for the network $c\rightarrow \{g_i\}$ along with the associated regulatory parameters (such as interaction strengths), which increase, or even maximize, the information transmission between the transcription factor and its regulatory targets. If we believe that the ability to transmit information is under positive selection, then evolution might drive real regulatory networks towards such maxima. We end by proposing that such an optimization principle could be a good candidate for a ``design principle''  for biological information processing networks. We will use measurements from early \emph{Drosophila} development to illustrate these ideas.

\subsection{Early morphogenesis in \emph{Drosophila melanogaster}}
After fertilization, interesting processes are taking place in the ellipsoidally shaped egg, about half a millimeter in length, of the fruit fly. A single nucleus undergoes 14 rounds of nuclear divisions, before large-scale spatial rearrangements, easily observed under the microscope, start happening approximately 3 hours after fertilization. During these first hours, all nuclei are floating in a shared pool of cytoplasm, in a structure that is known as a syncytium; only at division cycle 14 do individual nuclei cellularize. 

In a groundbreaking series of genetic experiments, researchers have shown that during these early developmental stages  when all nuclei look identical and no differentiated structures are visible, important cell-fate determination steps are already taking place. Looking along the long axis of the ellipsoidal egg, known as the AP (anterior-posterior) axis, one can see about 100 rows of nuclei at cell cycle 14. These nuclei express  proteins (mostly transcription factors) that will confer cell fate: nuclei belonging to various spatial domains of the embryo express specific combinations of genes that will lead these nuclei to become precursors of different tissues. Stainings for relevant transcription factors  have shown a  remarkable degree of precision with which the spatial domain boundaries are drawn in each single embryo, and a stunning reproducibility in positioning of these domains between embryos. Although probably a slight overgeneralization, we can say that at the end of cell cycle 14, along the AP axis, each row of nuclei reliably and reproducibly expresses a gene expression pattern that is characteristic of that row only -- in other words, the nuclei have unique \emph{identities} encoded by expression levels of developmental TFs along the long axis of the embryo.

Decades of research have focused on the following questions about early development, with the hope that what is learned in \emph{Drosophila} will shed light on how development and cell differentiation proceed in general:
{\bf (i)} How are the spatial domains established? What are the inputs that break the initial symmetry where all nuclei start out in the same state with the same genetic material? {\bf (ii)} What is the wiring diagram of a network that takes the input information and processes it to the point where the nuclei have their identity encoded in the expression pattern of late developmental genes? {\bf (iii)} What is responsible for the precision and reproducibility of cell fate determination? What are the limits to this precision? Are there mechanisms that confer robustness to certain environmental fluctuations, such as natural variation in temperature or physical size of the egg? {\bf (iv)} What are the molecular mechanisms of regulation and cross-regulation implicated in the developmental network? {\bf (v)} What are the mechanisms that allow the signals to propagate spatially in the developing egg and that coordinate the response of different nuclei, such that e.g. the expression domains are not ``noisy'' or jagged?

The answers to some of these questions are certainly known qualitatively, although we are still trying to make the models and measurements fit quantitatively. Briefly, the mother breaks the symmetry by depositing sources of so-called  maternal \emph{morphogens}, or diffusible transcription factors, at key points of the embryo. For example, the source mRNA for the bicoid protein that was discussed in Section~\ref{lec3}, is positioned at the anterior pole of the embryo. There, mRNA is translated into protein, which diffuses along the AP axis and is continuously degraded, establishing a steady-state concentration profile along the AP axis which is well-fit by an exponential:
\begin{equation}
c(x) = c_0e^{-x/\lambda}
\end{equation}
where $x$ is the distance on the AP axis measured from the anterior pole and $\lambda = 0.25 L$ (and $L$ is the total length of the embryo).

This spatial gradient is a chemical coordinate system: it is thought that each nucleus can read off the local concentration of bicoid (and other morphogens), and based on these inputs, drive the expression of the second layer of developmental genes (called the gap genes, which we denote by $g_i$); these in turn lead to ever more refined spatial patterns of gene expression that ultimately generate the cell fate specification precise to a single-nuclear row.
\subsection{Posing the question}
Let's attempt to put together all that we have learned up to now about gene regulation, noise in gene regulation and information theory. On one hand we can make a simple back-of-the-envelope calculation: If there are 100 distinguishable states of gene expression along the AP axis responsible for 100 distinct rows of nuclei, some mechanism must have delivered $I\approx \log_2(100)\approx 7$ bits of information to the nuclei. That's the minimum amount of information needed to make a decision about the cell fate along the AP axis. Similar patterning mechanisms also act along the other axes of the embryo, and if each of the 6000 nuclei at cell cycle 14 were uniquely determined, these systems together would have to deliver about 13 bits of information.

We also know something about the information flow in genetic regulatory networks, and we can start by asking not how much information the nuclei need, but how much can be delivered. We have studied how bicoid regulates one of the gap genes, hunchback, in Sections~\ref{lec2} and \ref{lec3}. We will first take a look at this single regulatory element and ask if, given the measured noise, the element is used optimally as part of the regulatory network. Then we will generalize by assuming that $c\rightarrow \{g_i\}$, i.e. that bicoid input $c$ regulates a set of (possibly interacting) gap genes $g_i$. The noise in gene expression was discussed in Section~\ref{lec3} for the regulation of hunchback by bicoid; if that can be taken as typical for other elements of the network, we can indeed compute $I(c;\{g_i\})$ and ask if that quantity can approach the $I\approx 7$ bits of information that is needed on theoretical grounds.

\subsection{Information transmission from bicoid to huncbback}
\begin{figure}
\includegraphics[width =  \linewidth]{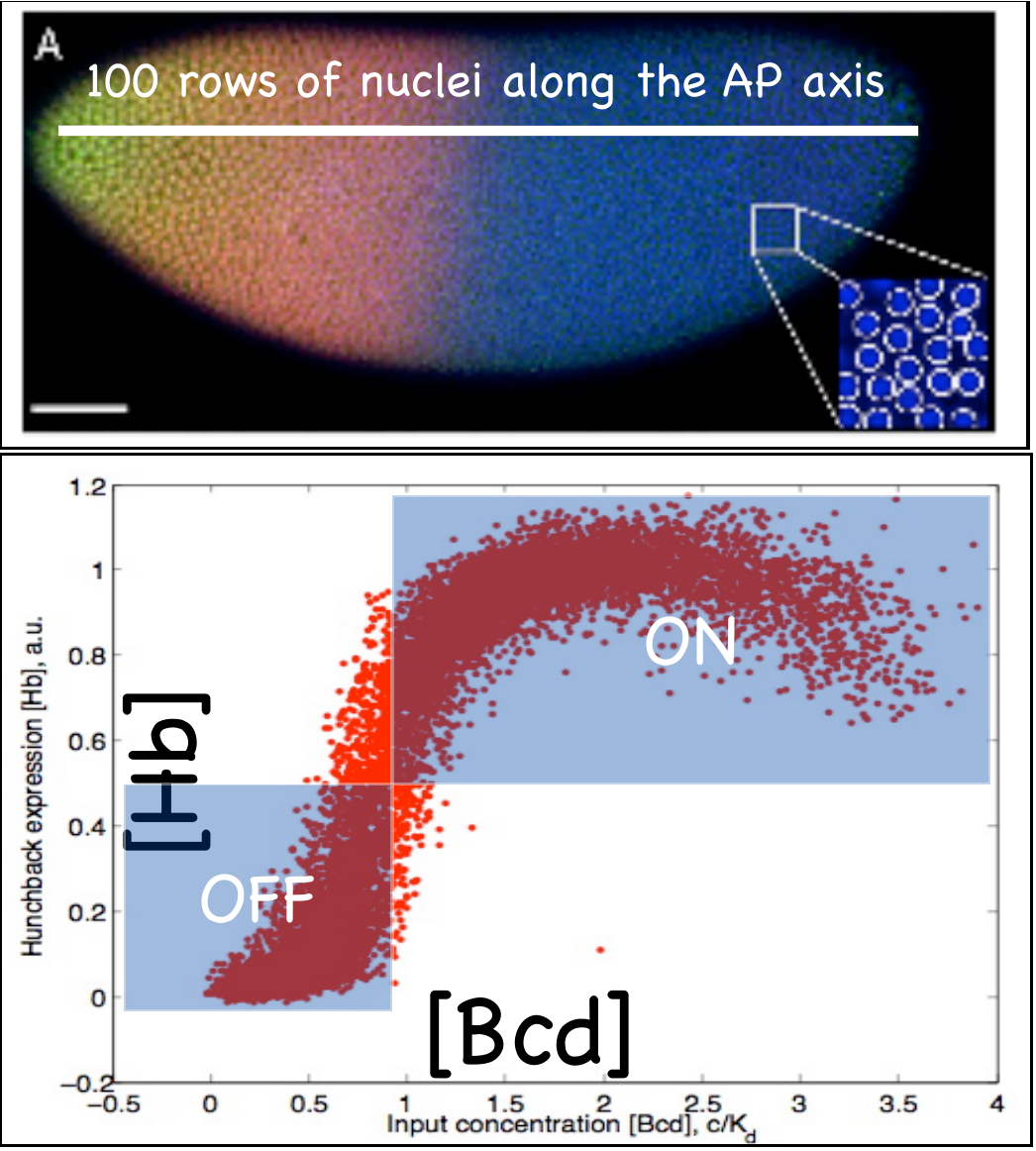}
\caption{\emph{Drosophila melanogaster} embryo at cell cycle 14. Nuclei stained in blue (see inset), bicoid stained in green and hunchback stained in red, data reproduced from Ref \cite{thomas}. At this stage, about 6000 nuclei are present in the embryo, of which about a quarter are visible under a single microscope view. Each nucleus provides a joint quantitative readout proportional to bicoid and hunchback intensities; the data is shown in scatter plot below. Usually hunchback was understood as having a single precise boundary that separates the domain of high expression (``on'') from the domain of low expression (``off''). We would like to use information theory to make this statement precise and find out if the bicoid/hunchback regulatory element really is just a binary switch. }
\label{f-droso-ap}
\end{figure}

By simultaneously observing the concentrations of bicoid ($c$) and hunchback ($g$) across the nuclei of an embryo, one can sample the joint distribution $P(c,g)$, see Fig.~\ref{f-droso-ap}. Usually it was assumed that hunchback provides a sharp, step-like response to its input, bicoid; mathematically, this would mean that the bcd/hb input/output relation is switch-like, with an ``on'' and an ``off'' state, yielding information transmission capacities of about 1 bit. However, is this really the case?

Using our estimation methods from Section~\ref{lec4} we can measure directly how much information bicoid $c$ and hunchback $g$ carry about each other. The result $I_{\rm expt}(c;g)=1.5\pm 0.1$ bits, where the error bar is computed across 9 embryos. This is an experimentally determined quantity, and the errors (apart from the estimation bias) are related mostly to our ability to fairly sample the distribution $P(c,g)$ across the ensemble of nuclei. Our sampling is not complete because a single microscope view only records about a quarter of all nuclei, but we believe that that sampling is not very biased. Another point to have in mind is that the computation of $I(c;g)$ reflects all statistical dependency in the probabilistic relation $c\rightarrow g$: both the direct regulation, as well as any possible indirect regulation through an unknown intermediary $x$, e.g. $c\rightarrow x \rightarrow g$. If, however, $g$ is regulated also by an input $y$ independent of $c$, that is $\{c,y\}\rightarrow g$, and our experiment does not record $y$, then we might be assigning some variability (or noise) to $g$, although that noise really would be a systematic regulatory effect caused by $y$. In this last case, we would measure a smaller value of $I(c;g)$ than would underestimate the real precision in the system; the true value would only be revealed upon recording the unobserved regulator $y$ and computing $I(\{c,y\};g)$.

Having these caveats in mind, our first finding is that the information transmission of 1.5 bits between bicoid and hunchback that we measure from the data is larger than 1 bit, which would be needed if bicoid/hunchback transformation were a simple binary switch. To our knowledge this was one of the first times that a quantitative measure of ``regulatory power'' was computed for a genetic regulatory element that was measured in a high-precision experiment. 

Given the noise in gene expression, can we put an upper bound of how much information could have maximally been transferred between bicoid and hunchback? To do this, let's start by writing:
\begin{equation}
P(c,g)=P(g|c)P_{TF}(c).
\end{equation}
As shown in Sections~\ref{lec3},\ref{lec4}, the term $P(g|c)$ describes the input/output properties of the regulatory element. From experiment, we can determine the mean response $\bar{g}(c)$ of the regulatory element and the noise in the response, $\sigma_g^2(c)$. These quantities have been plotted in Figs.~\ref{f-noise2},\ref{f-noise5}; if the noise is Gaussian (and to a good approximation, it is), these two measurements determine $P(g|c)$ fully.

To ask about the maximum achievable information transmission given the measured input/output relation $P(g|c)$, we write the Lagrangean
\begin{equation}
\mathcal{L}[P_{TF}(c)]= I(c;g) - \Lambda\int dc\;P_{TF}(c),
\end{equation}
where $\Lambda$ is a Lagrange multiplier that will enforce the normalization of $P_{TF}(c)$, while 
\begin{equation}
I(c;g)=\int dc\;P_{TF}(c)\int dg P(g|c) \log_2\frac{P(g|c)}{P(g)}
\end{equation}
is the mutual information, and $P(g)=\int dc\;P_{TF}(c)P(g|c)$. We can now look for the optimal distribution of inputs, $P_{TF}(c)$, which must satisfy:
\begin{equation}
\frac{\delta\mathcal{L}[P_{TF}(c)]}{\delta P_{TF}(c)}=0. \label{variation}
\end{equation}
One way to solve this variational problem is numerically. For details see Refs \cite{ggpre,ggpnas}; here we only report on the results. 

\begin{figure}
\includegraphics[width =  \linewidth]{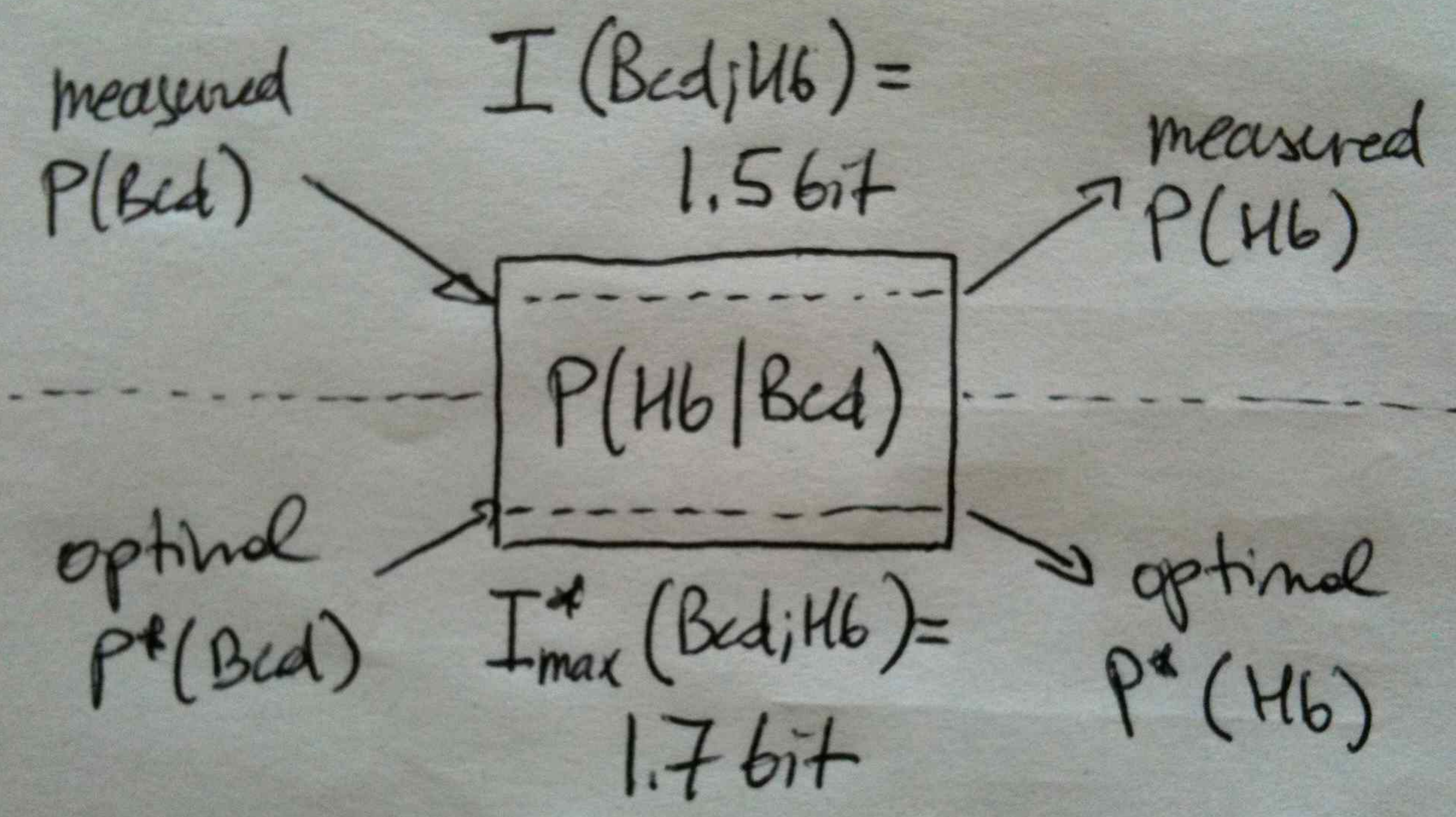}
\caption{The real (measured) information transmission and the maximal information transmission (channel capacity) in the bicoid/hunchback regulatory system. The input/output relation $P(g|c)=P(\mathrm{Hb}|\mathrm{Bcd})$ is measured and held fixed. To estimate the true information transmission of 1.5 bits, the experimentally sampled $P_{TF}(\mathrm{Bcd})$ is used to construct the joint $P(c,g)$. To find the channel capacity, $P_{TF}(\mathrm{Bcd})$ is varied until the information-maximizing choice is found numerically, denoted as $P^*(\mathrm{Bcd})$; this yields 1.7 bits of capacity. The optimal choice for the input distribution also predicts the optimal distribution of outputs, shown in Fig.~\ref{f-optinfo}.}
\label{f-infoinfo}
\end{figure}

\begin{figure}
\includegraphics[width =  \linewidth]{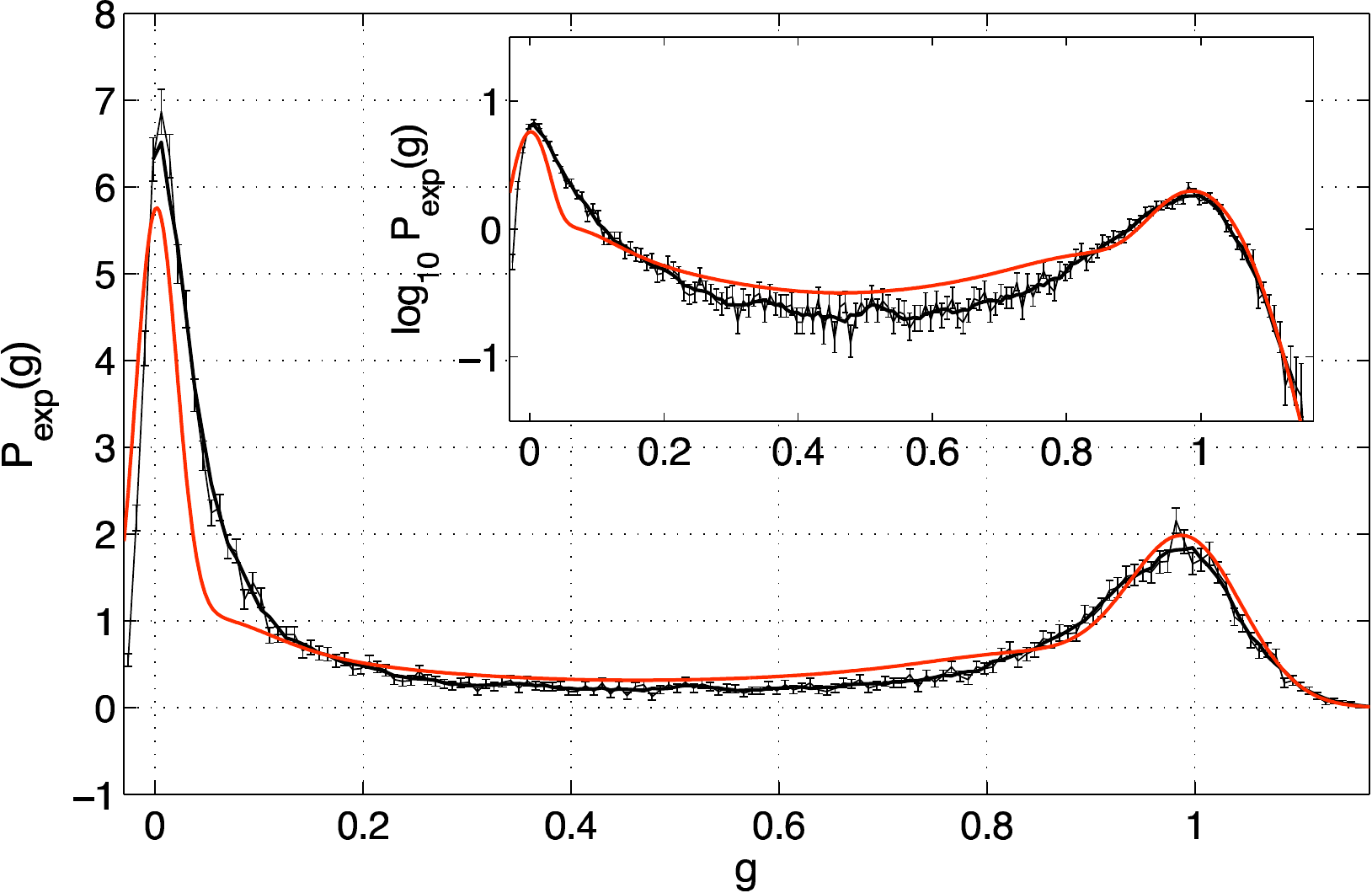}
\caption{The measured (black) and predicted optimal (red) distribution $P(g)$ of hunchback expression levels across an ensemble of nuclei in the \emph{Drosophila} embryo. The expression level $g$ goes from 0 (no induction, posterior) to 1 (full induction, anterior). A considerable fraction ($\sim 30\%$) of nuclei express intermediate levels of hunchback, and the noise in the system is low enough that this intermediate expression level could constitute a separate signaling level from 0 and 1; this would be consistent with the observed information of $1.5$ bits that intuitively corresponds to $2^{1.5}\sim 3$ distinguishable levels of gene expression. The inset shows the same plot on the logarithmic scale.}
\label{f-optinfo}
\end{figure}

We find that holding $P(g|c)$ fixed as determined from the data on bicoid/hunchback relationship, and optimizing $P_{TF}(c)$ numerically, yielded the maximal channel capacity of $I^*(c;g)=1.7$ bits, see Fig.~\ref{f-infoinfo}. Additionally the optimal $P_{TF}^*(c)$ predicts the optimal distribution of hunchback expression levels observed across the ensemble of nuclei, through $P^*(g)=\int dc\; P(g|c) P^*_{TF}(c)$, and the optimally predicted distribution matches the measured distribution very well [Fig.~\ref{f-optinfo}].
The value found for the maximal information transmission (channel capacity) shows that the real biological system is operating close to what is achievable given the noise, that is $I_{\rm expt}(c;g)/I^*(c;g)\approx 90\%$. The high value is somewhat unexpected given that we know that hunchback is regulated also by other inputs, and that bicoid also regulates other targets. Nevertheless this finding is a good motivation to consider taking  maximization of information transmission seriously as a possible design principle.

Can we comment on the values in the range of $I\sim 1.5-1.7$ bits? It turns out that the bicoid gradient is read out directly by 4 gap genes: hunchback, kruppel, giant and knirps. If each would independently be able to encode  $\sim 1.5$ bits, then together this genes could convey $I(c;\{g_i\})\sim 6-7$ bits of information about bicoid and would thus achieve the amount needed for AP patterning. In this case, we would be able to claim consistency with the back-of-the-envelope calculation that \emph{requires} at least this amount of information for the AP specification. Before reaching such a conclusion, however, we need to resolve the following issues: {\bf (i)} The readout (gap) genes $\{g_i\}$ are probably not independent, but have some  redundancy, which will mean that they convey less than the sum of their individual information values about $c$; such redundancy, as we find below, can be alleviated by proper network wiring; {\bf (ii)} The next layer of developmental cascade after the gap genes is not regulated \emph{solely} through the gap genes, but receives inputs from maternal morphogens directly; therefore, the gap genes are not a single bottleneck through which the information can flow; {\bf (iii)} especially at the poles of the embryo, gradients other than bicoid provide spatial information about the AP position; {\bf (iv)} our formulation of the problem assumes steady state gap gene readout from a stable gradient; it is not clear that such steady state is really reached in the timeframe necessary for nuclear specification. 

Further experiments and theory will be needed to successfully address these, and possibly other, issues. Despite these concerns, we hope that the discussion provides motivation for looking at quantities like $I(x;c)$ -- the information that the morphogen gradient encodes about the physical location $x$; at $I(c;\{g_i\})$, and at $I(x;\{g_i\})$ -- the information that later developmental genes (like gap genes) carry about the physical location. Information processing inequalities also constrain the relationships between these (directly measurable) quantities, providing an implicit check of whether we have missed some unobserved regulatory pathway. Before proceeding, we note that experiments that probe these quantities are not easy, because they require us to measure simultaneously the expression levels of a number of genes, nucleus by nucleus, in order to estimate both the mean response $\bar{g}_i(c)$

\subsection{The small-noise approximation}
Analytically, the problem of Eq~(\ref{variation}) is tractable in the so-called small-noise limit, which we present here  and use to explore the optimal architecture of small regulatory networks.

Having seen that in at least one biological system the information transmission can come close to the channel capacity (maximum achievable transmission given noise), we would like to elevate this finding to a principle: let us find network wiring diagrams and interaction parameters that transmit the most information from input TFs to the regulated output genes.

We will consider networks where a single transcription factor at concentration $c$ can regulate a set of $K$ target genes $\{g_i\}$, which may be interacting in a feed-forward network. For now, we will not consider feedback loops that can cause multistable behavior. It is clear that without any constraint, the information transmission can trivially be increased by decreasing the noise, and in biochemical networks noise can be decreased arbitrarily by increasing the number of signaling molecules, both on the input side ($c$) and on the output side ($\{g_i\}$). The crucial idea is therefore \emph{to optimize information subject to biophysical constraints, i.e. subject to using a fixed number of signaling molecules}.

With these assumptions in mind, we sketch the derivation of information transmission in the following text; for details see Refs \cite{ggpre, ggpre2, awpre}. For additional work on information transmission in biochemical networks see Refs \cite{zivetal, tw1, alex1}.

The dynamics of gene expression for genes $\{g_i\}$ is given by 
\begin{equation}
\tau\frac{dg_i}{dt}=f_i(c;\{g_j\}) - g_i + \xi_i,
\end{equation}
where $\tau$ is the protein lifetime, $\xi_i$ is the Langevin noise term (explained in Section~\ref{lec3}), and $f(c,\{g_j\})\in [0,1]$ is the regulatory (input/output) function, describing the activation rate of gene $g_i$, given the input $c$ and the expression levels of all the other genes. Various regulation functions were discussed in Section~\ref{lec2}; for combinatorial regulation, the most flexible one that we have examined was the Monod-Wyman-Changeaux (MWC) regulation function: 
\begin{eqnarray}
f_i(c;\{g_j\})& =& \frac{1}{1+e^{F_i(c,\{g_j\})}},\\
F_i(c,\{g_j\})&=& -n^i_c \log(1+c/K^i_c)- \nonumber \\
&-&\sum_j n^i_j \log(1+g_j/K^i_j) + \tilde{L}^i. \label{lgvn}
\end{eqnarray}
In this model, every regulatory input to $g_i$ contributes a term to the ``free energy'' $F$, and each such term is parametrized by $n^i_j$, the number of binding site for $g_j$ in the promoter of $g_i$, and $K_j^i$, related to the energy of binding to that binding site; as before, $\tilde{L}$ is the free energy offset between the ``on'' and ``off'' states when no transcription factor is bound. If we want to avoid feedback and multistability, we can always renumber the genes such that each gene $g_i$ only depends on the input $c$ and other genes $g_j$ where $j<i$.

The regulation in a network of a single input $c$ and $K$ target genes $g_i$ is then described by unknown constants $\{\tilde{L}^i, K^i_j, n^i_j, n^i_c, K^i_c\}$. When $n^i_j\rightarrow 0$, the regulation of gene $j$ by gene $i$ is absent, that is, in the wiring diagram the arrow from $g_j$ to $g_i$ disappears.

Before proceeding, we need also to compute the noise in this regulatory network. The noise in $g_i$ is given by two contributions: the output noise from generating a finite number of proteins of $g_i$, and the input diffusive noise because $g_i$ is regulated by $c$ and other $g_j$. The noise in our setup with $K$ target genes is fully determined by a $K\times K$ covariance matrix:
\begin{equation}
C_{ij}(c)=\langle(g_i - \bar{g}_i(c))(g_j-\bar{g}_j(c))\rangle,
\end{equation}
which can be computed from Eqs~(\ref{lgvn}), as shown in Refs~\cite{ggpre2,awpre}. 

In addition to computing this matrix, we find that there is a \emph{single dimensionless parameter $C$} describing the dynamic range of the input, $c\in [0,C]$, that will control the shape of the optimal solutions\footnote{This is true if all genes $\{g_j\}$ have the same parameters (such as diffusion constant and degradation times), an approximation that we make.}. $C$ is the maximal concentration for the input $c$, expressed in ``natural units of concentration,'' $c_0=N_{\rm max}/D a \tau$, i.e. the maximum number of independent molecules of the output $N_{\rm max}$, divided by the relevant diffusion constant, typical size of the binding site $a$ and the integration protein lifetype $\tau$. Large values for $C$ mean that the output noise is dominant over the input noise, while a small dynamic range and therefore small $C$ means that the input diffusive noise in $c$ is the dominant noise in the system.

With the covariance matrix in hand, the distribution of outputs given the input $c$ is a multivariate Gaussian:
\begin{equation}
P(\{g_j\}|c)=\frac{e^{-\frac{1}{2}\sum_{i,j=1}^K (g_i-\bar{g}_i(c)) C_{ij}^{-1}(g_j-\bar{g}_j(c))}}{(2\pi)^{K/2} \sqrt{|C|}}. \label{gnoise}
\end{equation}

In the language introduced in Section~\ref{lec4}, this is the ``encoding'' distribution. Suppose that we now  ask the ``decoding'' question: having seen the values of gap genes $\{g_i\}$, what is the most likely value of $c$ that produced them, and what is the variance in $c$? If the noise is small, $P(c|\{g_j\})$ will also be Gaussian, which can be found from Eq~(\ref{gnoise}) and the Bayes' theorem:
\begin{equation}
P(c|\{g_j\})\propto e^{-\frac{1}{2}\frac{(c-c^*(\{g_j\}))^2}{\sigma_c^2(\{g_j\})}},
\end{equation}
where $c^*(\{g_j\})$ is the most likely value for $c$ that gives rise to the observed $\{g_j\}$, and 
\begin{equation}
\frac{1}{\sigma_c^2(c)}=\sum_{ij}\frac{d\bar{g}_i}{dc} C^{-1}_{ij}\frac{d\bar{g}_j}{dc};
\end{equation}
$\sigma_c$ is the effective noise level in the input that accounts for all the noise in the system\footnote{In small noise approximation one can reassign the noise from the input to the output and vice versa through the mean input/output relation, as shown in Fig.~\ref{f-noiseprop}. }.

Information $I(c;\{g_j\})$ is 
\begin{eqnarray}
I(c;\{g_j\}) &=& S[P_{TF}(c)] - \langle S[P(c|\{g_j\})]\rangle_{P_{TF}(c)},
\end{eqnarray}
where the distribution of inputs, $P_{TF}(c)$ is unknown. We want to find the maximal information transmission given the known noise, therefore, we look for the maximum of $I$ with respect to $P_{TF}(c)$, just as we did in Eq~(\ref{variation}), while insisting that $P_{TF}(c)$ be normalized. We find that
\begin{equation}
P^*_{TF}(c)=\frac{1}{Z}\frac{1}{\sigma_c(c)},
\end{equation}
that is, the system should optimally use those input levels $c$ more frequently that have proportionately smaller effective noise. Using this optimal choice, the information, in bits, will be:
\begin{equation}
I(c;\{g_j\})=\log_2\frac{Z}{\sqrt{2\pi e}} \label{infofinal}
\end{equation} 

This is as far as we can puch analytically; $I(c;\{g_j\})$ still depends on the parameters $\{\tilde{L}^i, K^i_j, n^i_j, n^i_c, K^i_c\}$ that determine the wiring diagram of the network and the strengths of the regulatory arrows. The last remaining task is, therefore, to numerically optimize Eq~(\ref{infofinal}) with respect to these parameters, and examine the structure of optimal solutions.
\subsection{Optimal network architectures}
We can finally ask what are the optimal input/output curves for $K$ genes $\{g_i\}$, regulated by the single input $c$, if we do or do not allow for mutual interactions between the outputs. These results are a function of $C$, the dynamic range of the input.

\begin{figure}
\includegraphics[width =  \linewidth]{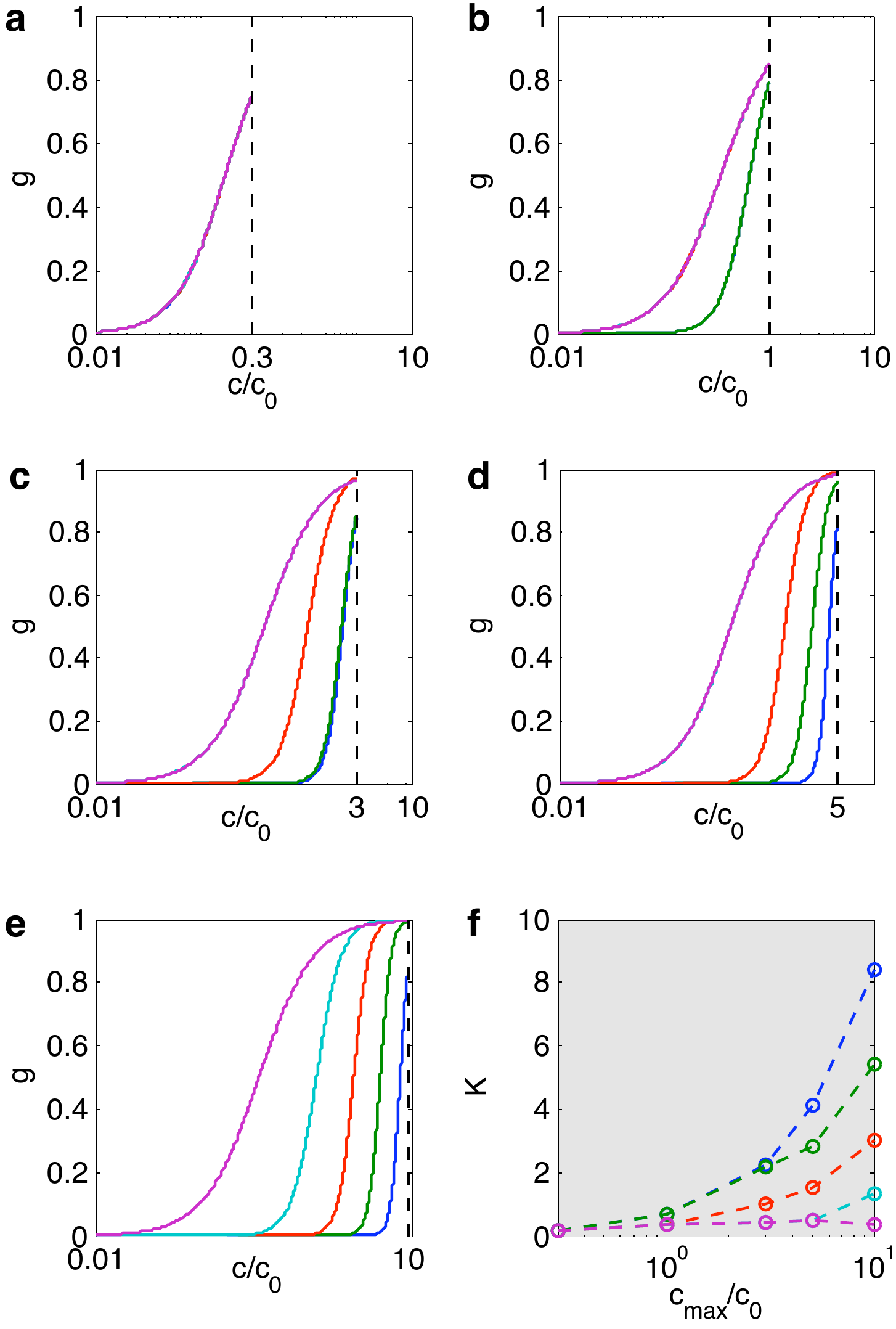}
\caption{The optimal input/output relations for $K=5$ genes, $\{g_1(c),\dots,g_5(c)\}$ (shown in various colors), regulated independently by a common input, $c$. The first 5 panels show optimal solutions depending on the dynamic range of the input, $C$, that is, when $c\in [0,C]$. As $C$ is increased, the totally redundant solution, where $\bar{g}_1(c)=\dots=\bar{g}_5(c)$, slowly becomes non-redundant and transitions into the tiling solution at high $C$, where each $g_i$ independently covers a subrange of concentrations for the input $c$. The last panel shows the optimal values for the dissociation constants, $K_i$, of all 5 genes, as a function of $C=c_{\rm max}/c_0$. }
\label{f-5genes}
\end{figure}

Figure~\ref{f-5genes}  shows the example solutions for $K=5$ noninteracting genes as a function of $C$. We see that there are two regimes: at low $C$, the optimal solutions  for all 5 genes  have exactly the same parameters, and therefore their input/output curves overlap perfectly. Why is this behavior optimal, if at first glance all the genes appear completely redundant? At low $C$, the input noise is dominant, and the best strategy is to have all $K=5$  genes read out the input $c$ and lower the input noise by averaging: using $K$ readouts should lower the effective noise by a factor of $\sqrt{K}$.

At high $C$ another strategy, called the \emph{tiling} solution, becomes optimal: here, each gene $g_i$ changes its expression considerably over some limited range of inputs, and various genes $g_i$ encode various non-overlapping input ranges; in other words, each $g_i$ ``reports'' on its own range of inputs, while the other $g_j$ have either not switched on yet, or are already saturated. We can explore the transition from redundant to tiling solutions in detail, and we can carefully study the scaling of information capacity $I(c;\{g_i\})$ with the number of genes $K$ in each solution \cite{ggpre2}.

Although interesting from a theoretical perspective, the redundant and tiling solutions are not what is actually observed in the real regulatory networks in \emph{Drosophila}. In particular, when $\{g_i\}$ are independent, the only possible input/output relations are sigmoid; there are no stripe forming solutions, where $g_i$ would turn on at some concentration $c$ and turn off at some higher concentration. Can such solutions emerge if the activating and repressing interactions between the output genes are allowed?

\begin{figure}
\includegraphics[width =  \linewidth]{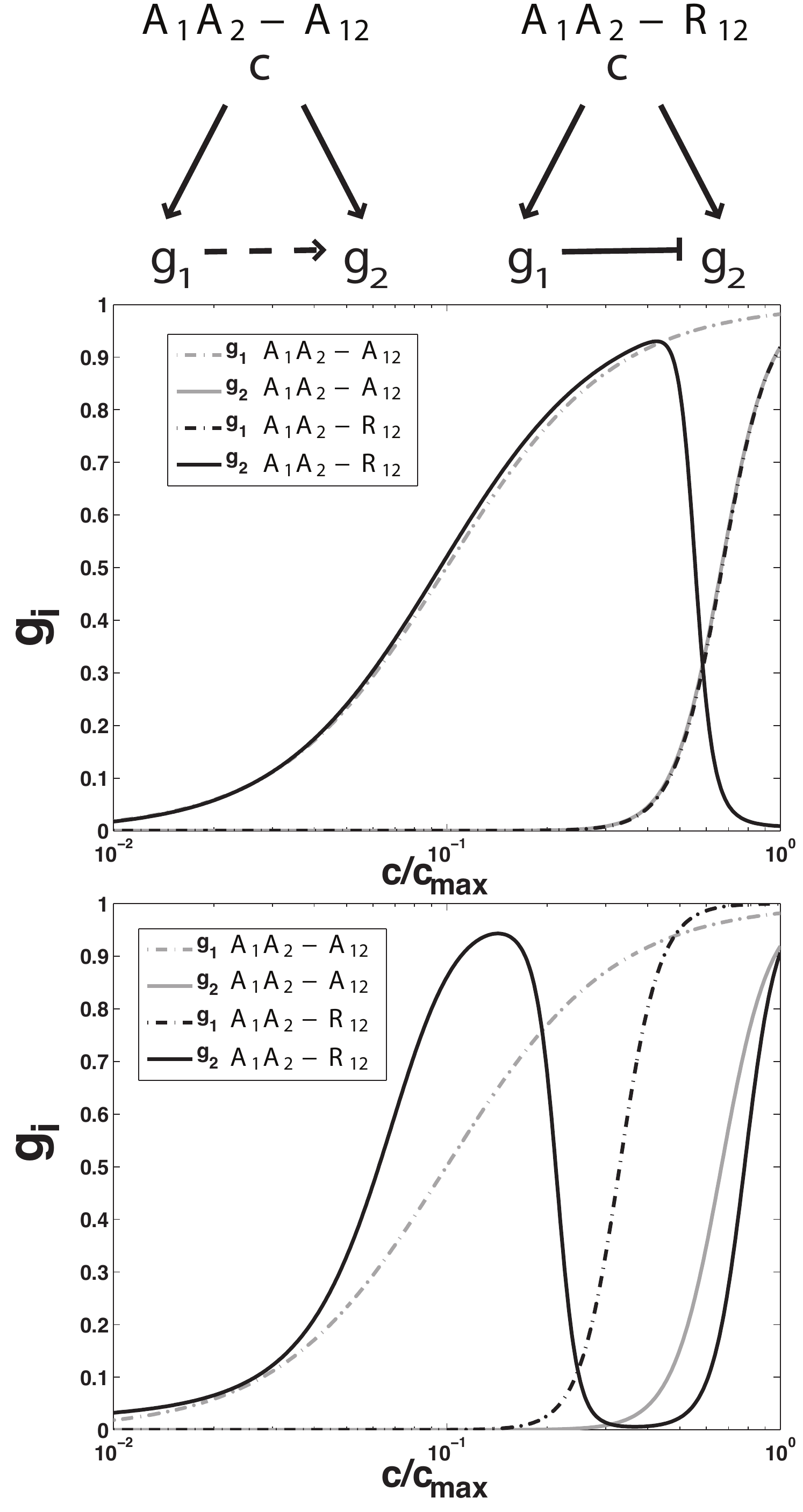}
\caption{The optimal input/output relations for two genes $g_1,g_2$ regulated by a common input $c$, with cross-regulatory feed-forward interactions and Hill model of regulatory functions. In case the activating arrow is allowed between $g_1$ and $g_2$, the optimal solution (gray lines) is not different from a non-interacting system, where $c$ independently regulates $g_1$ and $g_2$: both the input/output curves as well as the information transmission values are the same. In the case where $c$ activates $g_1$ and $g_2$, but $g_1$ can repress $g_2$, qualitatively new input/output shapes  can be optimal (black lines). Here, the combinatorial regulation of $g_2$ by $g_1$ and $c$ makes the apparent input/output relation $g_2(c)$ behave non-monotonically and produce a stripe. }
\label{f-combgenes}
\end{figure}

Indeed we find that this is the case, as shown in Fig.~\ref{f-combgenes}. If the interactions between two output genes $\{g_1(c),g_2(c)\}$ are allowed, the information maximizing wiring diagram includes ``lateral repression'' between the two genes that are jointly activated by a common input. This also generates apparent input/output curves that are non-monotonic: $g_2$ as a function of $c$ is seen to exhibit a stripe of activation. Further work has confirmed that such stripe-like patterns optimize information transmission. Interestingly, a similar pattern of interconnections (``lateral inhibition'') is known to occur in neural networks involved in early sensory processing. The function of such connections is to decrease the redundancy in the outputs -- with no interconnections in the tiling solution, when the gene with the highest $K_d$ is saturated and fully active, we \emph{know} that all the other genes are also fully on and saturated: they are therefore providing redundant information. In other words, when there is no interactions, the only patterns of activation (in a simplified picture when the genes are binary) are $000,001,011,111$ for a case of 3 genes. Patterns such as $010$ or $110$ cannot be accessed if there is no lateral interactions. If they exist, however, these patterns can be generated and they can encode additional useful information about their input $c$, increasing information transmission.

Our understanding of information transmission in transcriptional networks is far from complete. Nevertheless, the richness of solutions and network topologies that emerges from a single optimization principle in a one-parameter ($C$) problem is very encouraging, as is the qualitative matching to the stripe-like solutions in early \emph{Drosophila} development. Further efforts need to be invested into understanding multi-stability, feedback loops and autoregulation, and in the incorporation of other biologically realistic detail. Hopefully, this (or some other) design principle will in the future enable us to understand the wiring of biological networks and derive it from a mathematical measure of their function, rather than reconstructing it back  from painstaking molecular disassembly of the network into its component parts.
\section*{Conclusions}
\label{conc}
Biology presents an interesting challenge to physicists: many symmetries and simplifications applicable in ordered (but dead) systems are absent in biology, and this complexity of life can be intimidating. On the other hand, biological systems have evolved for function, and as we make progress in formalizing this notion mathematically, we hope to gain new insights and predictive power.  Assembling real physical models of biological information processing networks and connecting them to the genotypes on one hand, and to their function and selection on the other, will require tools from physics, biology, population dynamics, computer science, information theory and other disciplines, and this cross-disciplinary nature should make such research attractive to new students. I hope these lecture notes provide one interesting entry point to this new and exciting field. 
\section*{Acknowledgements}
I would like to thank the organizers of QECG 2010 Summer School, especially prof. Jonathan Miller, for their kind invitation. In addition, I am grateful to colleagues who have contributed to the research presented in these lectures: Vijay Balasubramanian, Michael Berry, William Bialek, Curt Callan, Thomas Gregor, Justin Kinney, Phil Nelson, Elad Schneidman, and Aleksandra Walczak. 

\end{document}